# Integrated photonics on thin-film lithium niobate

**Di Zhu,**[1,*] **Linbo Shao,**[1] **Mengjie Yu,**[1] **Rebecca Cheng,**[1] **Boris Desiatov,**[1] **C. J. Xin,**[1] **Yaowen Hu,**[1] **Jeffrey Holzgrafe,**[1] **Soumya Ghosh,**[1] **Amirhassan Shams-Ansari,**[1] **Eric Puma,**[1] **Neil Sinclair,**[1,2] **Christian Reimer,**[3] **Mian Zhang,**[3] and **Marko Lončar**[1,*]

[1]*John A. Paulson School of Engineering and Applied Sciences, Harvard University, 29 Oxford Street, Cambridge, Massachusetts 02138, USA*
[2]*Division of Physics, Mathematics and Astronomy, and Alliance for Quantum Technologies (AQT), California Institute of Technology, 1200 E. California Boulevard, Pasadena, CA 91125, USA*
[3]*HyperLight Corporation, 501 Massachusetts Avenue, Cambridge, Massachusetts 02139, USA*
dizhu@g.harvard.edu, loncar@seas.harvard.edu

Lithium niobate (LN), an outstanding and versatile material, has influenced our daily life for decades: from enabling high-speed optical communications that form the backbone of the Internet to realizing radio-frequency filtering used in our cell phones. This half-century-old material is currently embracing a revolution in thin-film LN integrated photonics. The success of manufacturing wafer-scale, high-quality, thin films of LN on insulator (LNOI), accompanied with breakthroughs in nanofabrication techniques, have made high-performance integrated nanophotonic components possible. With rapid development in the past few years, some of these thin-film LN devices, such as optical modulators and nonlinear wavelength converters, have already outperformed their legacy counterparts realized in bulk LN crystals. Furthermore, the nanophotonic integration enabled ultra-low-loss resonators in LN, which unlocked many novel applications such as optical frequency combs and quantum transducers. In this Review, we cover—from basic principles to the state of the art—the diverse aspects of integrated thin-film LN photonics, including the materials, basic passive components, and various active devices based on electro-optics, all-optical nonlinearities, and acousto-optics. We also identify challenges that this platform is currently facing and point out future opportunities. The field of integrated LNOI photonics is advancing rapidly and poised to make critical impacts on a broad range of applications in communication, signal processing, and quantum information. © 2021 Optical Society of America.

**Table of Contents**

## 1. Introduction

Integrated photonics holds great promise for realizing low-cost and scaled optical solutions for communication, sensing, and computation. Moreover, the miniaturization and integration of photonic structures enable new design knobs and functionalities that are inaccessible in their bulk counterparts. Many material systems have been investigated and adopted for integrated photonics. Good examples include silicon (Si) [1], indium phosphide (InP) [2], silicon nitride (SiN$_x$) [3], gallium arsenide (GaAs) [4], aluminum nitride (AlN) [5,6], and silicon carbide (SiC) [7]. Despite great progress, these material platforms cannot simultaneously support ultra-low propagation loss, fast and low-loss optical modulation, and efficient all-optical nonlinearities.

Studied since the 1960s, lithium niobate (LiNbO$_3$, LN) is one of the most versatile and attractive materials for photonics, owing to its exceptional electro-, nonlinear-, and acousto-optic properties, as well as its wide transparency window and relatively high refractive index. For instance, LN electro-optic modulators are ubiquitous in fiber-optic communications, while periodically poled lithium niobate (PPLN) is widely used for wavelength conversion and photon pair generation. Despite its great potential, LN had generally fallen behind the competing integrated photonic platforms, mainly due to the significant difficulties of material integration and processing. Traditional integrated LN devices are based on low-index-contrast waveguides, normally formed by titanium (Ti) indiffusion or proton exchange in bulk LN. These devices have weak mode confinement, large device footprints, and reduced nonlinear efficiencies. As a result, LN devices have largely remained bulky, discrete components. This is in stark contrast to other major integrated photonic platforms, typically based on thin films supported by low index cladding, which benefit from high-index-contrast and sub-wavelength optical confinement.

Followed by foundational works laid around the early 2000s [8–10], high-quality thin-film LN wafers prepared from ion slicing and wafer bonding have recently been made commercially available. With breakthroughs in fabrication techniques, ultra-low-loss, high index-contrast, nanophotonic LN waveguides have been realized [11–20]. Within the past few years, a complete set of integrated optical components have been developed on the thin-film LN platform with unprecedented performances. Some notable examples include compact and ultra-high-performance modulators [21–25], broadband frequency comb sources [26–28], as well as record-efficiency wavelength converters [29–31] and photon-pair sources [32–34]. This versatile platform can host a wide variety of devices, as illustrated in Figure 1, including various optical and optomechanical cavities, tunable filters, electro-optic modulators, acousto-optic modulators, microwave-to-optical transducers, nonlinear wavelength converters, frequency combs, non-classical light sources, detectors, and quantum memories. With such a rich component toolbox, we expect thin-film LN to become a material platform of choice for realizing multi-functional, high-performance integrated photonic circuits for both classical and quantum applications.

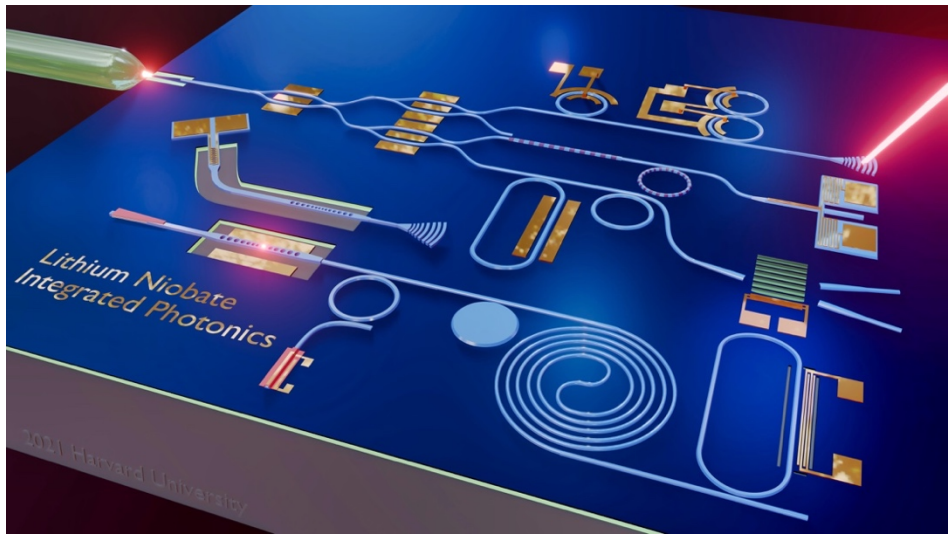

Figure 1. Integrated photonic devices on thin-film lithium niobate. The illustrated components include tapered waveguides and grating couplers, a spiral delay line, ring and microdisk resonators, photonic crystal cavities, an electro-optic frequency comb source, periodically poled waveguide and ring, a heterogeneously bonded laser source, an amorphous Si photodetector, Mach Zehnder and in-phase/quadrature electro-optic modulators, acousto-optic modulator, single- and coupled-ring modulators, a microwave-to-optical transducer, and an integrated single-photon detector as well as an emitter. Note that this figure only serves as a visual illustration of the wide range of devices that can be hosted on the platform. When integrating different categories of devices on a single chip, one may need to address practical fabrication issues to ensure process compatibility.

This Review aims to present a comprehensive introduction of integrated photonics based on thin-film LN, from basic principles to the state of the art. In Section 2, we start by introducing the material properties of LN and basic passive optical components such as waveguides and cavities. In Section 3, we discuss electro-optics, from its most ubiquitous application in electro-optic modulators to novel development of frequency combs, quantum transducers, and synthetic space photonics. Section 4 covers the diverse all-optical nonlinearities in thin-film LN, such as $\chi^{(2)}$-based wavelength conversion, Kerr comb generation, Raman lasing, and supercontinuum generation. In Section 5, we survey recent advances in integrated piezo-optomechanics (acousto-optics), which bridges optics, electronics, and mechanics. Section 6 describes recent efforts on heterogeneous integration that combines thin-film LN with other mature material platforms or unique functional components for both classical and quantum applications. In Section 7, we discuss challenges and opportunities, and provide an outlook. We hope this Review, in conjunction with some other recent reviews on this topic [35–42], can give readers a starting point to delve into this exciting and rapidly advancing field.

## 2. Material and integrated passive optics

In this section, we give an overview of the material properties of LN and summarize how thin-film LN is prepared. We then review waveguides and cavities on the thin-film LN platform, which are the basis for various functional and active devices discussed in later sections.

### 2.1. Material properties of lithium niobate

The study of material properties of LN traces back to the 1960s, and summaries of its physical properties can be found, for example, in Ref. [43,44]. The crystal structure of LN belongs to the $3m$ point group: it exhibits three-fold rotation symmetry about the $c$-axis, commonly defined as the $Z$-axis, and mirror symmetry about three planes that are 60 degrees apart (Figure 2). In this review and most LN integrated photonics references, the $X$ and $Y$ axes are defined such that the LN crystal is mirror-symmetric about the $YZ$ plane, as shown in Figure 2(b). LN is a ferroelectric material [45]. Figure 2(a) depicts the relative positions of the lithium (Li) and niobium (Nb) atoms with respect to the oxygen (O) octahedra. Along $+c$-axis, the octahedra is filled with Li, Nb, vacancy, Li, Nb, vacancy, and so on. A strong external electric field can relocate Li and Nb (cations), and move Li to the adjacent vacancy. This results in an inversion of ferroelectric domain orientation, which is the basis of periodic poling as used in quasi-phase matching.

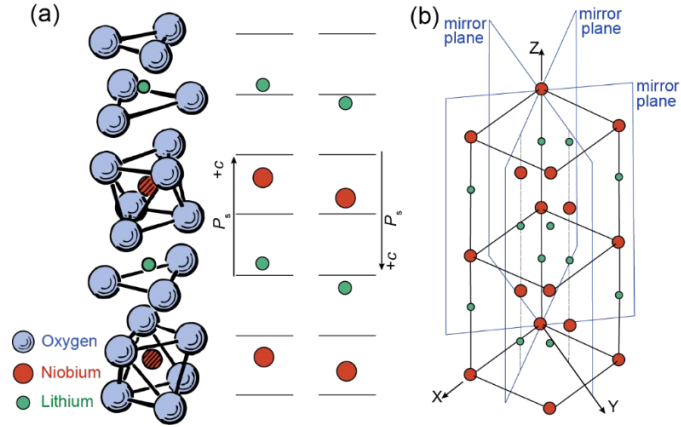

Figure 2: Crystal structure of lithium niobate. (a) Positions of lithium (Li) and niobium (Nb) atoms with respect to oxygen (O) octahedra in the ferroelectric phase. Poling moves the positions of Li and Nb, resulting in the inversion of the spontaneous polarization and domain orientation. (b) Standard definition of X, Y, Z axes, where Y aligns with a mirror plane. Adapted from [43,45].

In Table 1, we list some key material properties of LN and compare them to other popular materials used in integrated photonics. LN has a broad transparency window, from 350 nm to 5 µm, covering the visible, near-infrared and mid-infrared wavelength range. Its relatively large refractive index (~2.2 at 1550 nm) allows high-index-contrast waveguides to be formed on top of most amorphous and crystal substrates (e.g., SiO$_2$ or sapphire). Its high Curie temperature (~1210 °C) ensures a stable ferroelectric phase to render it compatible with a wide range of fabrication processes and operation conditions. Unlike Si and SiN$_x$, LN is a non-centrosymmetric crystal and possesses large second-order nonlinear coefficients ($d_{33}$= -27 pm/V). Moreover, it allows ferroelectric domain engineering, which makes it a leading material for nonlinear optical frequency conversion and generation (e.g., second harmonic generation, difference/sum frequency generation, or spontaneous parametric down-conversion). Crucially, LN's exceptional Pockels coefficient ($r_{33} \approx 31$ pm/V) has made it the material of choice for electro-optic (EO) modulators, which is at the core of telecommunication networks. LN also has a third-order nonlinear coefficient ($n_2 = 1.8 \times 10^{-19}$ m²/W) that is comparable to other commonly used optical materials, making Kerr comb generation possible. Considering acoustic applications, the speed of sound on the surface of bulk LN or inside thin-film LN is ~4 km/s, resulting in a higher acoustic index than SiO$_2$ and sapphire. This means that the thin-film LN platform is not only able to guide photons, but also phonons. Combining this with large piezo-electric coefficients (for example, $d_{22} = 21$ pC/N and $d_{33} = 16$ pC/N for Rayleigh waves, and $d_{15} = 74$ pC/N for shear waves) and photoelastic coefficients (e.g., $p_{13} = 0.133$ and $p_{44} = 0.146$) of LN, it is possible to realize a wide variety of novel piezo-optomechanical devices, in addition to more traditional surface-acoustic-wave (SAW) microwave filters. At last, LN has thermo-optic ($\frac{dn_e}{dT} = 3.34 \times 10^{-5}$ K$^{-1}$ at 1523 nm and 300 K [46]), pyroelectric (~95 μCm$^{-2}$K$^{-1}$ [47]), and photorefractive effects. They may be leveraged to provide additional functionalities and engineering knobs, such as wavelength tuning of modulators and resonators, but may also cause detrimental effect and limit device operation (especially for photorefraction, see Section 7.1).

**Table 1: List of key properties of some materials that are commonly used in photonics. Linear optical refractive indices ($n_0$) are given at 1550 nm, with *o/e* denoting ordinary/extraordinary axis.**

| Material | Point group symmetry | Optical index | RF dielectric Constant | $\chi^{(2)}$(pm/V) | $n_2 = \frac{3\chi^{(3)}}{4\varepsilon_0 c n_0^2}$ (m²/W) | EO coefficient (pm/V) | Photoelastic coefficients (1) | Piezoelectric strain coeff. (pC/N) |
|---|---|---|---|---|---|---|---|---|
| LiNbO$_3$ | 3m | 2.21 (o) 2.13 (e) | $\varepsilon_{11,22} = 44$ $\varepsilon_{33} = 27.9$ (clamped, high-frequency response) | $d_{31} = -4.3$ $d_{33} = -27.0$ $d_{22} = 2.10$ @1064nm | $1.8 \times 10^{-19}$ @1550nm | $r_{13} = 9.6$ $r_{22} = 6.8$ $r_{33} = 30.9$ $r_{51} = 32.6$ | $p_{11} = -0.026$ $p_{12} = 0.09$ $p_{13} = 0.133$ $p_{14} = -0.075$ $p_{31} = 0.179$ $p_{33} = 0.071$ $p_{41} = -0.151$ $p_{44} = 0.146$ | $d_{31} = -1.0$ $d_{22} = 21$ $d_{33} = 16$ $d_{15} = 74$ |
| LiTaO$_3$ | 3m | 2.18 (o) 2.17 (e) | $\varepsilon_{11,22} = 38.3$ $\varepsilon_{33} = 46.2$ | $d_{31} = 0.85$ $d_{33} = 13.8$ | - | $r_{13} = 6.2$ $r_{22} = 0$ $r_{33} = 28.5$ $r_{51} = 8.4$ | $p_{11} = -0.081$, $p_{12} = 0.081$ $p_{13} = 0.093$ $p_{14} = -0.026$ $p_{31} = 0.089$ $p_{33} = -0.044$ $p_{41} = -0.085$ $p_{44} = 0.028$ | $d_{31} = -3.0$ $d_{22} = 9.0$ $d_{33} = 9.0$ $d_{15} = 26$ |
| Si | m3m | 3.49 | 11.7 | 0 | $5 \times 10^{-18}$ @1550nm | 0 | $p_{11} = -0.094$ $p_{12} = 0.017$ $p_{44} = 0.051$ | 0 |
| SiO$_2$ | Amorphous | 1.44 | 3.9 | 0 | $3 \times 10^{-20}$ @1550nm | 0 | $p_{11} = 0.121$ $p_{12} = 0.270$ | 0 |
| α-Quartz | 32 | 1.53 (o) 1.54 (e) | 3.8 | $d_{11} = 0.584$ @1318 nm | $2.6 \times 10^{-20}$ @1318nm | $r_{11} = 0.230$ | $p_{11} = 0.16$ $p_{12} = 0.27$ $p_{13} = 0.27$ $p_{14} = -0.03$ $p_{31} = 0.29$ $p_{33} = 0.10$ $p_{41} = -0.047$ $p_{44} = -0.079$ | $d_{11} = 2.3$ $d_{14} = -0.67$ |
| SiN$_x$ | Amorphous | 2.00 | 7 | 0 | $2.5 \times 10^{-19}$ @1550nm | 0 | | 0 |
| AlN | 6mm | 2.12 (o) 2.16 (e) | 9 | $d_{31} = 1.6$ $d_{33} = 4.7$ @1550nm | $2.3 \times 10^{-19}$ @1550nm | $r_{13} = 0.67$ $r_{33} = -0.59$ | $p_{11} = -0.1$ $p_{12} = -0.027$ $p_{13} = -0.019$ $p_{33} = -0.107$ $p_{44} = -0.032$ $p_{66} = -0.037$ | $d_{31} = -2.0$ $d_{33} = 5.0$ $d_{15} = 4.0$ |
| ZnO | 6mm | 1.93 (o) 1.94 (e) | 10.4 | $d_{33} = 1.23$ @1064nm | $6.88 \times 10^{-15}$ (ZnO:Ni thin film) @1064nm | $r = 0.5$ (ZnO:Mn thin film) $r = 2.6$ (single crystal) | $p_{11} = 0.222$ $p_{12} = 0.099$ $p_{13} = -0.111$ $p_{31} = 0.0888$ $p_{33} = -0.0235$ $p_{44} = 0.0585$ | $d_{31} = -5.1$ $d_{33} = 12.4$ $d_{15} = -8.3$ |
| GaAs/ AlGaAs | $\bar{4}3m$ | 3.37 | 12.88 | $d_{36} = 170$ @1064 nm | $2.6 \times 10^{-17}$ @1550nm | $r_{41} = 1.43$ | $p_{11} = -0.165$ $p_{12} = -0.140$ $p_{44} = -0.072$ | $d_{14} = 2.6$ |
| Diamond | m3m | 2.39 | 5.66 | 0 | $4 \times 10^{-20}$ | 0 | $p_{11} = 0.125$ $p_{12} = -0.325$ $p_{44} = 0.11$ | 0 |

## 2.2. Thin-film lithium niobate

While traditional diffusion-based photonic devices in bulk LN have played an important role in optical and RF signal processing, they have several important limitations, predominately originating from the large optical mode due to the low index contrast. For example, EO modulators realized using this approach require large drive voltages and challenging RF-optical dispersion engineering. Furthermore, it is impossible to make tight bends in these waveguides, which prevents realization of, for example, low-loss ring or race-track resonators. These limitations, along with the desire to integrate many different functionalities on the same LN chip, motivated the production of thin-film LN-on-insulator (LNOI) wafers.

Historically, the realization of high-quality LN thin films has proven difficult. Various methods have been explored, including chemical vapor deposition [48], RF sputtering [49], pulsed laser deposition [50], sol-gel [51], and molecular beam epitaxy [52]. However, these approaches failed to produce highly crystalline materials, and lattice matching requirements severely limit the substrate options for epitaxial growth [35].

The development and refinement of the "Smart Cut" technology [53], commonly used to generate SOI wafers, has emerged as a the standard technique for producing high-quality LN thin films [35,54]. The use of crystal ion slicing to prepare thin-film LN started in late 1990s [8] and early 2000s [10], including a patent that was filed in 1998 and expired in 2018 [9]. Instead of making a thin film using deposition or growth methods, the film is split from a high-quality bulk substrate through ion slicing and thermal annealing, and is bonded to an insulating substrate, typically through adhesive (e.g. using benzocyclobutene, BCB) [55] or direct wafer bonding [10]. However, while both BCB and direct bonding techniques have succeeded in producing single-crystalline LNOI, the direct bonding method is preferred for many applications, as the higher annealing temperatures allowed by the process are necessary for recovering nonlinear and electro-optic properties of the ion-sliced material [35].

The Smart Cut fabrication process for LNOI is illustrated schematically in Figure 3. From a high-quality LN substrate, a cleavage plane is first defined at the desired film thickness via high-dose implantation of $He^+$ (or $H^+$) ions. In parallel, the substrate wafer, usually LN or Si with an oxide layer, is prepared. After the LN substrate is bonded to the wafer, thermal annealing is used to split the substrate along the cleavage plane, defining the thin film. Additional annealing reduces ion-implantation induced crystal defects, and polishing improves surface smoothness. It is important to note that most LNOI fabrication is performed using congruent rather than stoichiometric LN, meaning that the crystals are lithium deficient [44].

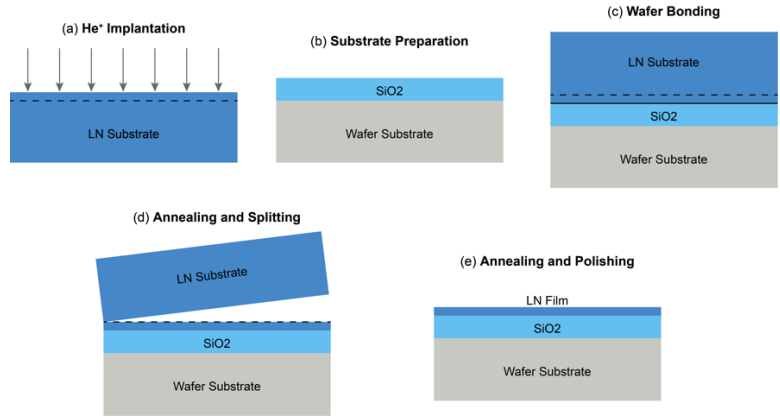

Figure 3: Schematic of the Smart Cut process for a LNOI wafer.

Through the use of ion slicing with direct wafer bonding, large-area, single-crystalline LN thin films on SiO$_2$-on-LN substrates were demonstrated [56,57]. This fabrication method can also be employed to produce other high-quality ferroelectric thin films, as well as more exotic iterations of thin-film LNOI, such as doped and stoichiometric LN wafers. Currently, LNOI wafers with sizes up to 6 inches have been made commercially available from companies such as NanoLN, Partow Technologies, NGK Insulators, and SRICO.

*2.3. Waveguides*

Waveguides are the backbone of integrated photonic circuits. In this section, we summarize how waveguides are formed in thin-film LN (Figure 4). Throughout the text, we use "bulk" to refer diffusion-based waveguides in bulk LN (Figure 4(a)). For thin-film devices, we use the term "monolithic" to describe devices realized by direct etching (Figure 4(b)). This is to differentiate from "hybrid" devices based on rib-loading or heterogeneous bonding (Figure 4(c) and (d)), which involve additional materials other than the LNOI substrate and any cladding materials. Furthermore, terms such as thin-film lithium niobate (TFLN) and thin-film lithium niobate on insulator (TFLNOI) are often used in the literature. They are, in most cases, interchangeable to lithium niobate on insulator, or LNOI.

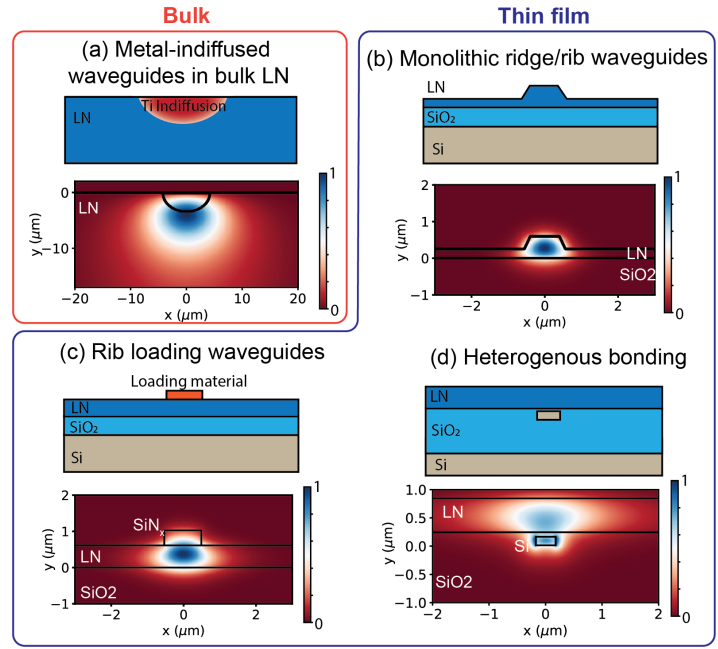

Figure 4: Major waveguide types in LN and their typical mode distribution. (a) Metal-indiffused waveguide in bulk LN (similar for proton-exchanged waveguides). The boundary of the waveguide core in the lower panel indicates diffusion length; actual index variation extends beyond that. (b) Monolithic ridge/rib waveguides in thin-film LN. (c) Rib-loaded waveguides. (d) Heterogeneously bonded thin-film LN on photonic integrated circuits made from other materials.

2.3.1. Weakly guiding diffused waveguides in bulk LN

Traditional LN waveguides are often based on titanium (Ti) in-diffusion or proton-exchange [36]. Ti in-diffused waveguides are formed by depositing Ti strips on bulk LN substrates followed by thermal annealing. Ti diffuses into LN crystals with a typical diffusion length of a few micrometers. The presence of Ti ions causes an increase in the optical refractive index ranging from 0.001 to 0.04, depending on Ti density. Proton-exchange waveguides are formed by immersing LN into a liquid source of hydrogen. At moderate temperatures (between 150-400 °C), protons ($H^+$) replace lithium ions ($Li^+$), causing a refractive index change (only in the extraordinary direction, $\Delta n_e$) around 0.1 or less. Further improvement of the waveguide quality could be achieved by a reverse-proton exchange (RPE) process, which re-diffuses $Li^+$ ions into the top layer of the crystal, leading to a more symmetric optical mode. Such index perturbation-based waveguides can also be formed by ion implantation [58] and femtosecond laser writing [59]. A major drawback of these waveguides is the low index contrast, resulting in weak optical confinement (mode area of 10 to 100 $\mu m^2$) and large bending radius (millimeter). They are therefore unsuitable for dense integration, unable to realize micro-resonators, and challenging to perform dispersion engineering.

2.3.2. Monolithic ridge/rib waveguides on thin-film LN

The availability of thin-film LNOI facilitates the fabrication of monolithic ridge waveguides with large index contrast (roughly an order of magnitude larger than ion-diffused waveguides) and strong mode confinement. Figure 4(b) shows the cross-section of a typical LNOI rib waveguide with slanted sidewalls and a slab (due to limitations in the dry etching process as discussed later). The slab is desirable in, for example, electro-optical modulators, since it increases the microwave field inside LN. Furthermore, it can help reduce overlap of the optical mode with etched surfaces, and thus reduce optical losses.

To avoid dry etching, LN can be processed by mechanical means. Ridge waveguides using optical-grade diamond blade dicing [60–63] have been demonstrated (Figure 5(a)). Diced waveguides can achieve high aspect ratios (>10 [62]) and low optical loss (<1 dB/cm [14]), but are challenging for realization of complex shapes such as couplers, crosses, and tapers. Chemical-mechanical (chemomechanical) polishing (CMP) has also been employed to fabricate ridge waveguides on thin-film LNs (Figure 5 (b)) [12,64,66–68]. A typical process [12] involves (1) Cr deposition, (2) patterning the Cr mask through femtosecond laser ablation, (3) removing uncovered LN through CMP, and (4) Cr removal. The CMP process yields waveguides with sub-nanometer surface roughness and propagation losses as low as 0.027 dB/cm [12]. Arrays of beam splitters and MZIs have also been demonstrated using this process [66,67], making future large-scale photonic integrated circuits (PICs) possible. One problem with CMP-based LN waveguides is that they have shallow sidewalls, limiting the minimum bending radius. LN can also be wet etched (e.g., using $HF:HNO_3$ solution) [69], and LNOI waveguide has been demonstrated using ion-beam-enhanced KOH etching [70,71]. Furthermore, waveguiding in thin-film LN using proton-exchange [72–74] and helium-ion-implantation [75] have been reported. These waveguides have relatively low index contrast and mode confinement in the lateral direction despite their relatively small mode volume compared to their bulk counterparts. In addition, proton-exchange tends to degrade EO and nonlinear coefficients [73,76]

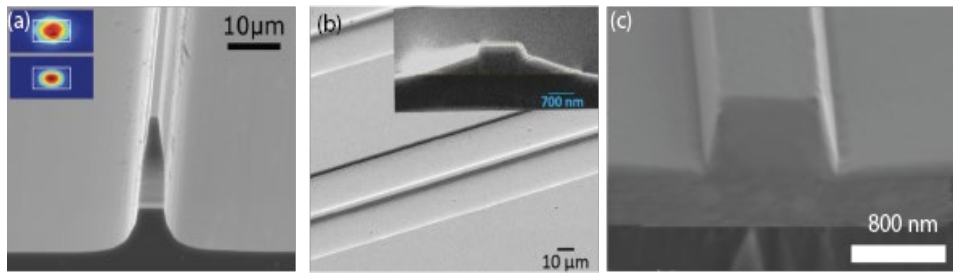

Figure 5: Monolithic LNOI ridge/rib waveguides. (a) Precision-diced waveguide [inset shows measured (top) and simulated (bottom) fundamental TE mode)] [60], (b) chemical-mechanical polished waveguides (inset shows waveguide cross-section) [64], (c) dry etched waveguide [65]. (a) Reproduced from [60], © 2015 Optical Society of America; (b) reproduced from [64] under open access license; (c) reproduced from [65], © 2019 Optical Society of America (OSA).

Compared to mechanical dicing/polishing and wet etching, dry etching is much preferred for integrated photonics. It is anisotropic, well-controlled in etch depth, able to transfer complex 2D patterns, and compatible with multi-layer processing. However, dry etching of LN had been notoriously difficult. Unlike most other integrated photonics platforms such as Si and $SiN_x$, a proper reactive ion-etching recipe was not available for LN. Fluorine-based reactive-ion etching (RIE) can remove LN effectively by forming volatile, fluorinated niobium (Nb) species [77]. However, it forms lithium fluoride (LiF), which is non-volatile, and result in severe redeposition problems [78]. This effect is less critical in proton-exchanged waveguides due to low Li contents [77,79,80] but problematic for thin-film LN waveguides. LiF is highly resistive to further etching, and thus increases sidewall roughness and scattering loss. Early on, focused ion beam milling has also been used to pattern LN [81,82]. They are suitable for defining small and sparse patterns such as photonic crystals, however this technique is not time-efficient for larger patterns or wafer-scale fabrication.

Currently, pure physical etching with $Ar^+$ plasma (or $Ar^+$ milling) [83] is among the most popular dry etching approaches for LNOI. It can be carried out with different tools, such as inductively coupled plasma (ICP), electron-cyclotron resonance (ECR), or ion-beam etch (IBE) systems. There are a few challenges associated with pure physical etching. First, it has low etch selectivity with respect to the available lithography resists (usually around 1:1), resulting in limited etch depth. Though shallow etched LNOI can also support guided modes [84], deep etched waveguides would have tighter mode confinement. Commonly used hard masks include amorphous Si [85] and Cr [86]. Various photo or electron-beam resists are also used as etch masks, such as hydrogen silsesquioxane (HSQ) [11,15,87], Zeon electron-beam positive-tone resist (ZEP) [27,88], CSAR [87], SU8 [55], and even positive-tone UV-lithography photoresist [89]. The other challenge with $Ar^+$ etching is that it forms nonvertical sidewalls due to redeposition. Unlike LiF redeposition (which is in the form of particles and clusters), LN redeposition from $Ar^+$ etching can be smooth and does not introduce large scattering loss. Furthermore, it can be removed using wet chemistry, after the dry etch process is finished, to further reduce optical losses. Nonetheless, the purely mechanical nature of the etch results in "over-cut" waveguide profile with trapezoidal cross-section. The sidewall angle, typically in 40º–80º range, ultimately limits the minimum feature size and spacing between neighboring structures. Lower vacuum pressure and higher plasma power generally produce steeper sidewalls. Some processes introduce fluorene gases such as $CHF_3$ [86], aiming to improve the sidewall angle. Also, wet chemical cleaning can help remove redeposition and other contaminants, and reduce sidewall roughness [89–91].

Besides etching, waveguide fabrication also requires lithography and post-processing. Lithography can be done either by photo or electron-beam lithography. Since LN is insulating, it commonly experiences significant charging during electron-beam lithography, which can be addressed using a spin-coated aqueous, conductive solution (e.g. Espacer$^{TM}$), or a thin layer of a metallic element (e.g. Cr). Post-processing may include several steps such as acid/solvent cleaning, thermal annealing, polishing, and oxide cladding.

With the optimization of lithography, etching, and post-etch cleaning, the propagation loss of dry-etched LNOI waveguides has improved from > 6 dB/cm [55] in 2007 to 0.027 dB/cm in 2017 [11]. Devices with optical loss below 0.1 dB/cm can now be routinely produced by several groups, making LNOI a reliable and scalable low-loss platform for integrated photonics. However, these values are still beyond the material absorption rate measured in whispering gallery resonators made from bulk congruent LN (<0.4 dB/m [92,93]). The major loss mechanism in dry-etched LNOI waveguide is the contribution from sidewall roughness. Various methods have been reported to reduce surface roughness, including CMP [13,94] and gas clustered ion beam smoothening [19]. Alternatively, one can design the waveguide cross-section so that the mode is more confined in the core and less exposed to the sidewalls using a wide and thin waveguide [95]. Finally, we note that the impact of ion implantation, used for ion slicing, on the intrinsic material losses is not fully understand at the moment.

### 2.3.3. Hybrid waveguides

Another class of waveguides on thin-film LN is the rib-loaded waveguide, which avoids the necessity of LN etching. In this category, a strip of material is deposited/patterned on top of the unetched LN thin film, as shown in Figure 6 (a)-(c). The patched material usually has a similar or higher refractive index as LN but is easier to deposit and etch. This way, an effective rib waveguide is formed, where part of the optical mode is confined in the LN slab and can access its unique material properties. Various loading materials have been used, including $SiN_x$ [17,25,99–102], $TiO_2$ [96,103], chalcogenide glass [104], $Ta_2O_5$ [105], and Si [97,106]. Recently, it has been demonstrated that the loading ridge can also be low-index material (e.g., photo or electron-beam resist) [107,108]. Different from standard rib-loaded waveguides, this approach is based on the concept of bound state in the continuum (BIC), where waveguiding is achieved by destructive interference of all leakage channels [109]. BIC waveguides, however, only guide light in specific polarization and wavelengths.

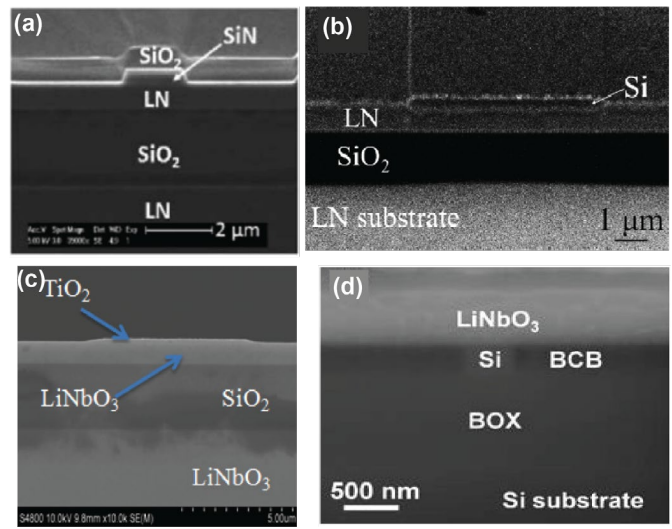

Figure 6: Hybrid waveguides on LN. (a)-(c) Rib-loaded waveguide in LNOI using $SiN_x$ [17], Si [96], and $TiO_2$ [97], respectively. (d) LN thin film bonded on SOI waveguide [98]. Reproduced with permission [17,96–98], © 2016, 2015, 2017, 2012 Optical Society of America.

It is also possible to bond LN to PICs made from other mature material platforms, such as $Si_3N_4$ [110] and Si [23,98,111–114]. A subtle but important difference between the rib-loading and bonding approach is that LN is the main host material in the former but a stamping/subsidiary material in the latter. These hybrid approaches represent a promising path towards heterogeneous integration of LN with other PIC platforms, complementing their missing functionalities such as EO modulation, second-order nonlinearity, and the ability of quasi-phase matching. One disadvantage, however, is that only a portion of the optical mode is in the LN which results in weaker nonlinear and EO interactions.

Figure 7 summarizes the propagation losses of different types of LNOI waveguides reported in the literature. Low-loss waveguides with propagation losses below 1 dB/cm were made possible since 2015 using multiple methods, and a few *dB/m* can now be achieved using dry-etching and chemo-mechanical polishing.

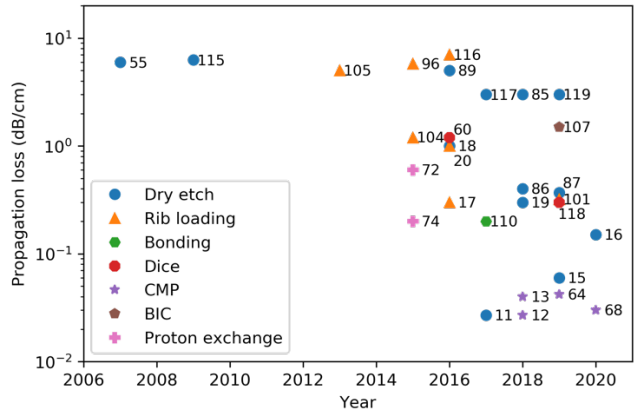

Figure 7: Propagation loss of various waveguides on thin-film LN. Reference numbers of some representative works are labeled. Reference: [11–13,15–20,55,60,64,68,72,74,85–87,89,96,101,104,105,107,110,115–119]

### 2.3.4. Fiber-to-chip coupling

The tight optical confinement of LNOI waveguides drastically reduces device footprint and enhances nonlinear optical interactions. It, however, presents a significant challenge in optical packaging due to the mode mismatch between the waveguides and standard optical fibers. In order to create an efficient interface between the optical fibers and the chip, two common schemes, namely, end-fire coupling (also called butt-coupling) and grating coupling have been widely studied (see Figure 8).

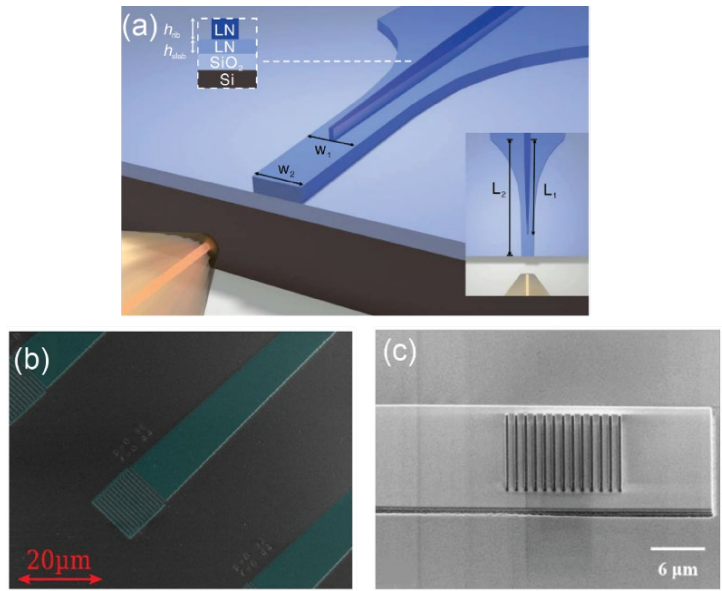

Figure 8 Fiber-to-chip coupling. (a) End-fire coupling through a mode converter with two-step tapering. (b) Thin-film LN grating coupler. (c) Hybrid amorphous Si grating on LN waveguide. (a)-(c) Reproduced from [120], [121], and [122], © 2019, 2019, and 2018 Optical Society of America.

End-fire coupling can be done by aligning a lensed fiber to a diced/polished or etched facet of the waveguide. This approach is broadband and can work for any polarizations. However, due to the mismatch between the spot size of the lensed fiber (~2 μm diameter) and waveguide mode (~1 μm), typical fiber-to-chip coupling efficiencies are between -4 to -6 dB [32]. To mitigate this problem, a mode converter in the form of an inverse taper can be used to expand the waveguide mode at the facet [120,123–128]. Since a typical LNOI rib waveguide has a profile that consists of a trapezoidal ridge on top of a slab, single-step tapering will only reduce the ridge width, which pushes the optical mode to the slab instead of expanding it symmetrically into the cladding, and in fact reduces coupling efficiency. Thus, two-step tapering is needed. This is accomplished by tapering down both the rib and slab that gradually pushes the mode to the oxide cladding. Using this approach, fiber-to-chip coupling loss of 1.7 dB per facet has been achieved [120] (see Figure 8(a)). More recently, fiber-to-chip coupling loss as low as 0.54/0.59 dB per facet for TE/TM mode has been demonstrated by adding a SiON cladding waveguide to the LNOI adiabatic taper [128].

Grating couplers allow vertical coupling at arbitrary positions of the chip with relatively large alignment tolerance. However, their operation bandwidth is narrow, and they are sensitive to polarization. Grating couplers are therefore convenient for rapid testing of large arrays of linear optical devices (including optical modulators) at wafer scale, but not well suited for characterization of most nonlinear devices that involve multiple wavelengths and polarizations. There have been a number of reports on LNOI grating couplers [121,129–134], with a current record coupling efficiency as low as -3.5 dB per coupler [121] (see Figure 8(b)). Two unique problems of LNOI grating couplers are: (1) the anisotropy of LN causes orientation-dependent coupling, (2) nonvertical sidewall angle in dry etching limits minimum pitch size and fill-factor of the gratings. On the other hand, bonded LNOI has the unique advantage that it allows buried metal layers, which can serve as an efficient optical reflector to minimize downward coupling and improve coupling efficiency [129,130]. Besides monolithically etched LN gratings, another promising approach is to pattern hybrid gratings on top of LNOI waveguide using, for example, amorphous Si [16,122]. Si gratings have larger refractive index and allow finer and sharper structures, alleviating the limitations imposed by LN etch. This approach has demonstrated coupling efficiency as high as ~-3 dB per coupler with a 1-dB bandwidth of 55 nm [122] (see Figure 8(c)).

Beyond what has been demonstrated on the LNOI platform, approaches that have been explored on other material platforms such as adiabatic tapered fibers [135,136], apodized [137], angled-etched [138], or multi-layer gratings [139] may be adapted to further improve fiber-to-chip coupling efficiencies. In addition, for hybrid systems, it is possible to adiabatically couple the optical mode from LN to, for example, Si [22] or $SiN_x$ [110] waveguides, and leverage the rich optical packaging toolbox developed for those platforms to achieve efficient coupling [140].

Fiber-to-chip coupling is a ubiquitous problem shared by all high-index-contrast integrated photonic platforms. As LNOI devices move forward from proof-of-concept demonstrations to mission-critical products, we anticipate this optical packaging problem can be collectively addressed by the academic community as well as commercial sectors.

### 2.4. Cavities

Optical microcavities are another set of essential components for integrated photonics. They enable wavelength filtering and strongly enhance nonlinear-, electro-, and acousto-optic interactions. In this section, we briefly review the state-of-the-art of optical cavities on thin-film LN, including microdisks, rings and racetracks, and photonic crystals (see Figure 9).

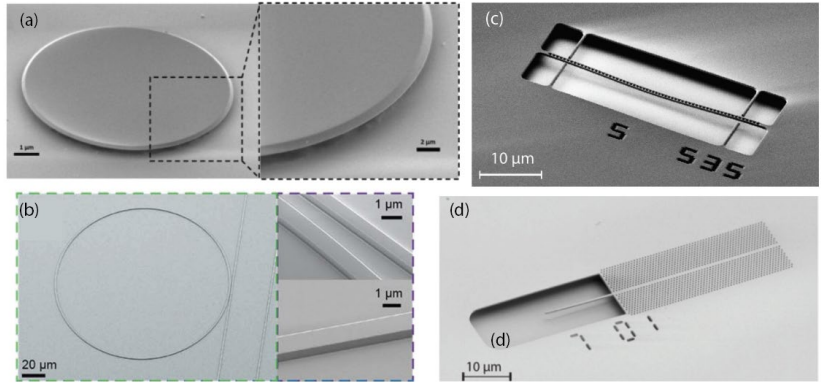

Figure 9: Optical microcavities based on thin-film LN. (a) Microdisk [141], (b) microring [11], (c) 1-D photonic crystal cavity [142], and (d) 2-D photonic crystal cavity [143]. (a)-(c) Reproduced from [141], [11], and [142], © 2014, 2017, 2017 Optical Society of America; (d) Reproduced from [143] © 2019 WILEY-VCH Verlag GmbH & Co. KGaA, Weinheim.

### 2.4.1. Microdisk

On-chip microdisks support whispering gallery modes with ultra-high quality-factors (Q-factor) and compact mode volumes. Thanks to their high-quality factors, they have been widely used for nonlinear optics, cavity quantum electrodynamics and optomechanics.

Fabrication of microdisk resonators in thin-film LN have been achieved by dry-etching (highest Q-factor ~$10^6$) [18,141,144,145], femtosecond laser ablation followed by FIB sidewall milling and high-temperature annealing (highest Q-factor ~$10^7$) [146–151], direct FIB milling (Q-factor ~$10^5$) [152], or CMP (highest Q-factor ~ $10^7$) [153–155]. On LNOI substrates, the LN microdisk can be easily suspended/undercut by removing the buried oxide using HF (see Figure 9(a)). It is also possible to fabricate suspended microdisks on bulk LN by chemically etching a buried layer of damaged lattice induced by ion implantation [156]. Besides single-crystal LN, microdisks made of polycrystalline LN deposited by pulsed laser deposition on a Si substrate have also been reported, showing a Q-factor of $3.4\times10^4$ that is likely limited by the polycrystal itself [157].

It is worth noting that mechanically polished, millimeter-scale disk resonators made from pristine bulk LN crystals have shown Q-factors as high as $2\times10^8$ [93], an order of magnitude higher than what has been demonstrated in thin-film LN. This indicates that the current cavity Q-factors using the LNOI platform are likely still largely limited by surface scattering. Another reason could be the induced damages resulting from ion-slicing process used in thin-film LN production, which may result in higher material-limited loss.

LNOI microdisks have enabled various demonstrations ranging from electro-optic modulators [145] and nonlinear wavelength converters [141,150–153,157–165] to cavity optomechanics [166]. Some of these demonstrations will be discussed in later sections.

### 2.4.2. Microring and racetrack

Microring and racetrack resonators are formed by waveguide loops and support cleaner resonance modes than microdisks due to the suppression of higher-order radial modes. They follow the same fabrication procedure as waveguides, and their intrinsic Q-factors directly reflect the propagation losses of the waveguide. Therefore, microring/racetrack resonators are often used as an accurate way of measuring waveguide losses, especially when the loss is low. Excluding bending loss, the intrinsic Q-factor follows $Q_i = \frac{2\pi n_g}{\lambda \alpha}$, where $n_g$ is the group velocity, $\lambda$ is the free-space wavelength, and $\alpha$ is the propagation loss. The free-spectral range (*FSR*) of the resonance can be used to calculate $n_g = \lambda^2/(FSR \cdot L)$, where $L$ is the cavity length. In addition, the intrinsic $Q$ of the resonator is related to the measured loaded $Q$ factor by $Q_i = 2Q_l/(1 \pm \sqrt{T_0})$, where $T_0$ is the normalized transmission, and + and - correspond to under- and over-coupled regimes, respectively. At critical coupling, $T_0 = 0$ and $Q_i = 2Q_l$.

On the LNOI platform, microring and racetrack resonators have been fabricated by dry etching (see Figure 9(b)) [11,15,28,55,85,88,167], rib loading [104,105], CMP [12], as well as using BIC waveguides [107]. The state-of-the-art quality factors of these resonators are listed in Table 2, with the highest intrinsic Q-factor achieved to date ~$10^7$. Furthermore, resonators have enabled low-voltage EO modulators [55,85,171], broadband EO frequency comb generation [26], ultra-efficient parametric wavelength conversions [30,31,94,168,169,172,173], and Kerr comb generation [27,28,174,175]. Coupled rings combined with EO modulation have also been explored to form a photonic molecule [176], and have found applications in frequency shifting [177] as well as microwave-to-optical photon conversion [178,179]. Details of these applications will be discussed in later sections.

### 2.4.3. Photonic crystal

Photonic crystal (PhC) cavities simultaneously possess high Q-factor and wavelength-scale mode volume, which enhance optical nonlinearities and reduce the physical footprint. In 2013, Diziain et al. demonstrated second-harmonic generation in suspended 2D photonic crystals fabricated from bulk LN by FIB and wet etching of an ion-damaged buried layer [180]. On LNOI substrates, both 1D and 2D photonic crystal cavities have been demonstrated (see Figure 9(b) and (d)) [142,143,170,181,182]. With small mode volume and large local field intensity, photorefractive effects (discussed in Sec. 7) are enhanced to the extreme, enabling photon-level tuning of cavity resonances [183]. As a cavity, it also allows resonant EO modulation [184], with the advantage of reduced electrode size and thus higher RC-limited operating frequency. In addition, optomechanical crystals [142,170,182] that confine both optical and mechanical modes have been demonstrated on suspended LN thin films.

Table 2. State-of-the-art Q-factors of some representative optical cavities in the thin-film LN platform, noted by their cavity type and fabrication method. Intrinsic Q-factors are cited when available.

| Cavity type | Fabrication method | Q-factor ($10^6$) | Year | Ref. |
|---|---|---|---|---|
| Microdisk | Dry etched | 1.70 | 2017 | [159] |
| Microdisk | Femtosecond laser + FIB | 9.61 | 2019 | [151] |
| Microdisk | CMP | 14.6 | 2018 | [153] |
| PPLN microdisk | Dry etching, piezo-force-microscopy poling | 8.0 | 2020 | [162] |
| Microring | Rib-loaded | 0.13 | 2015 | [104] |
| Microring | Dry etch | 10 | 2017 | [11] |
| Microring at visible wavelength | Dry etch | 11 | 2019 | [15] |
| Microring | BIC | 0.577 | 2019 | [107] |
| Microring | CMP | 11.4 | 2018 | [12] |
| PPLN ring | Dry etched, Z-cut poling | 1.8 | 2020 | [168] |
| PPLN racetrack | Dry etched, X-cut poling | 0.37 | 2019 | [30] |
| Ring-shaped WGR | Dry etch followed by CMP | 3.0 | 2017 | [94] |
| PPLN Ring-shaped WGR | Dry etch followed by CMP, Z-cut poling | 1.3 | 2018 | [169] |
| Photonic crystal | Dry etch | 0.109 | 2017 | [142] |
| Piezo-optomechanical crystal | Dry etch | 0.47 | 2020 | [170] |

## 3. Electro-optics

The electro-optic (EO) effect is perhaps the most attractive property of LN. This effect directly mixes optical and RF fields, enabling optical modulation, sideband generation, and frequency shifting in the GHz range. Low-loss ridge waveguides with closely placed microwave electrodes, strong field confinement and high-quality resonators in thin-film LN drastically improve EO interaction strength. In recent years, we have seen a proliferation of works on EO modulators in LN, based on both non-resonant and resonant optical structures, pushing towards wider microwave bandwidth and lower half-wave voltages. Such efficient EO interaction has also enabled EO frequency combs, which are of interest for applications in spectroscopy and topological photonics. In quantum applications, EO-based frequency conversion serves as a promising candidate to link microwave photons (e.g. in superconducting quantum systems) and optical photons (for long-haul networks). In addition, time-modulation enables study of synthetic dimensions in frequency domain. This section covers EO optical modulators, EO frequency comb sources, photonic molecules, cavity EO for quantum transduction, as well as EO-modulation-based synthetic photonics.

### 3.1. Basic formulation of electro-optic modulation

We first summarize some basic formulation to describe the linear EO effect. The linear EO strength in a material is described using the Pockels tensor, $\bar{\bar{r}}$. It relates the applied DC/RF electric field to the change in the inverse permittivity tensor, or impermeability tensor, by $\Delta(\frac{1}{n^2})_{ij} = \sum_k r_{ijk} E_k$, where $i, j, k$ can take values of 1 to 3, corresponding to $x$, $y$ and $z$ crystal directions. Because of the symmetry property, $r_{ijk} = r_{jik}$, a contracted index notation is usually used: $r_{ijk} \equiv r_{Ik}$, with $I = 1, \ldots, 6$ [185]. The correspondence between $I$ and $ij$, as well as the matrix form of $r_{Ik}$ with symmetry reduction for LN (point group $3m$), are listed in Eq. (1) below.

$$r_{Ik} = \begin{bmatrix} 0 & -r_{22} & r_{13} \\ 0 & r_{22} & r_{13} \\ 0 & 0 & r_{33} \\ 0 & r_{42} & 0 \\ r_{42} & 0 & 0 \\ -r_{22} & 0 & 0 \end{bmatrix} \text{ with } I = \begin{cases} 1 & for\ ij = 11 \\ 2 & for\ ij = 22 \\ 3 & for\ ij = 33 \\ 4 & for\ ij = 23\ or\ 32 \\ 5 & for\ ij = 13\ or\ 31 \\ 6 & for\ ij = 12\ or\ 21 \end{cases} \quad (1)$$

In integrated photonics, modulation strengths are often calculated using coupled-mode theory from index perturbations. Thus it is often more convenient to directly express the permittivity change as a function of the applied electric field: $\Delta \epsilon_{ij} = -\sum_k \epsilon_{ii} \epsilon_{jj} r_{ijk} E_k / \epsilon_0$ [186,187].

In LN modulators, the strongest and most utilized EO coefficient is $r_{33}$ (~31 pm/V), which induces an index change along the $Z$ crystal axis when the $E$ field is applied along $z$ as well. Note that LN is anisotropic and $Z$ is the extraordinary axis ($n_{33} = n_e = 2.14$ and $n_{11} = n_{22} = n_o = 2.21$ at 1550 nm). The induced permittivity change along z is $\Delta \epsilon_z = -n_e^4 \epsilon_0 r_{33} E_z$, and the refractive index change is approximately $\Delta n \approx -1/2\ n_e^3 r_{33} E_z$. Most of the devices reviewed in this section use $r_{33}$.

Another noteworthy but less exploited EO coefficient is $r_{51} = r_{42} \approx 30$ pm/V. It causes deformation of index ellipsoid in the $xz$ or $yz$ crystal plane when DC/RF field is applied along $x$ or $y$, respectively. The permittivity change under such modulation follows $\Delta \bar{\bar{\epsilon}} = -\epsilon_0 r_{42} n_o^2 n_e^2 E_y \begin{bmatrix} 0 & 0 & 0 \\ 0 & 0 & 1 \\ 0 & 1 & 0 \end{bmatrix}$ or $-\epsilon_0 r_{51} n_o^2 n_e^2 E_x \begin{bmatrix} 0 & 0 & 1 \\ 0 & 0 & 0 \\ 1 & 0 & 0 \end{bmatrix}$. This effect can induce polarization rotation, as used in Pockels cells. In bulk LN waveguides, it has been used for single-sideband modulation and magnetic-free optical isolation by mimicking an EO-driven rotating waveplate [188]. Solc-type polarization control and wavelength filtering have also been demonstrated utilizing this off-diagonal EO coefficient in PPLN waveguides [14,189]. In cavity-EO, this off-diagonal EO coefficient can couple two resonances with different polarizations [190,191].

*3.2. Non-resonant electro-optic modulators*

EO modulators translate electrical signals to light and are essential components in modern telecommunication networks. LN has been the material of choice for modulators because of its exceptional EO property, low optical and microwave losses, and excellent temperature stability. Integrated EO modulators in bulk LN have been studied extensively for decades [192–194]. Though basic principles remain similar, migrating these modulators into the thin-film platform brings unique advantages and engineering opportunities. In this section, we review non-resonant optical modulators based on thin-film LN waveguides, with specific focus on traveling-wave modulators, including their figures of merit, design strategies, performance tradeoffs, and latest experimental demonstrations.

3.2.1. Common modulator configurations

Figure 10 shows typical EO modulators in X or Y-cut LN (with Z crystal axis in plane and perpendicular to the waveguides). In phase modulators (see Figure 10(a)), an applied voltage induces an index change in the waveguide and causes a phase shift of the transmitted light. When driven with a strong RF signal, phase modulators generate sidebands, and can be used for frequency comb generation [195] and frequency shifting [196,197]. Figure 10(b) shows a typical intensity modulator based on a Mach-Zehnder interferometer embedded in a coplanar waveguide (CPW) electrode. This design features a push-pull configuration, where the two arms naturally experience electric fields with opposite polarities and thus opposite phase shifts. This is the most common configuration in thin-film LN modulators. In addition, combining two MZMs can realize an in-phase/quadrature (IQ) modulator (see Figure 10(c)) [16]. IQ modulators can encode information in both amplitude and phase, and are important for coherent transmission systems. Besides MZM, Michelson interferometer modulator (MIM) has also been studied [198,199], as illustrated in Figure 10(d). Here, we have been focused on X/Y-cut modulators. In principle, Z-cut EO modulators can also be implemented by placing electrodes on top of the waveguides. But so far they are less common in thin-film LN devices, because it is more difficult to efficiently deliver the electric field on the sub-micrometer waveguides while maintaining low microwave and optical loss.

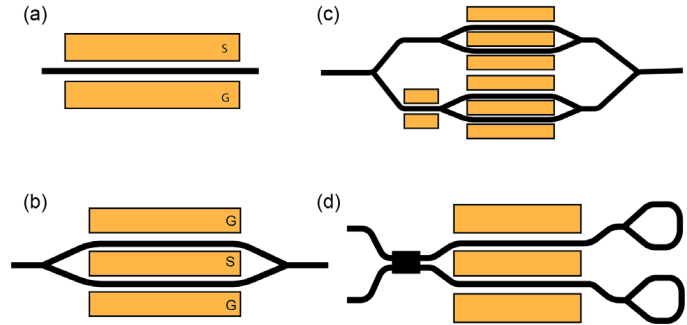

Figure 10: Four common configurations of EO modulators. (a) EO phase modulator, where an applied voltage induces an index change of the waveguide and thus causes a phase shift in the output light. (b) Mach-Zehnder intensity modulator, consisting of a pair of phase modulators embedded in a Mach-Zehnder interferometer. (c) In-phase/quadrature (IQ) modulator, where both amplitude and phase of the output light can be controlled. (d) Michelson interferometer modulator (MIM), where optical signals are reflected back at the end of both arms.

3.2.2. Performance metrics

The key metrics for EO modulators include half-wave voltage ($V_\pi$), EO bandwidth (*BW*), and insertion loss (*IL*). For intensity modulators, the extinction ratio (*ER*) is also a critical parameter. In addition, linearity and dynamic range are of interest in analog applications. Here, we go through each of them and discuss their design rules and state-of-the-arts. Some of these topics have also been discussed in Ref. [200,201]. Figure 11 shows a representative MZM on thin-film LN, with measurement result on $V_\pi$, ER, and EO bandwidth.

*Half-wave voltage ($V_\pi$).* In phase modulators, the half-wave voltage refers to the voltage required to achieve $\pi$-phase shift. In intensity modulators, it refers to the voltage required to switch the output between "on" and "off" states (see Figure 11(b) as an example). Benefiting from the push-pull configuration, an MZM has a $V_\pi$ that is half of a phase modulator. In addition, shorter wavelength requires lower $V_\pi$. To compare modulator designs fairly, in the rest of this section we will give $V_\pi$ values referenced to MZMs operating near 1550 nm when citing literature (i.e., $V_\pi$ reported in phase modulators will be divided by half). The $V_\pi$ of a modulator depends inversely on its length $L$ in the ideal case, and $V_\pi L$ is a commonly quoted figure for comparing EO interaction strength.

LNOI waveguides feature compact mode areas and allow small electrode gaps. Typical X/Y-cut modulators can support CPW gaps as small as ~5 μm. Further reducing the gap typically leads to excess optical absorption and microwave loss, and requires more precise alignment in fabrication. Measurements of DC/low-frequency $V_\pi L$ in monolithic (dry-etched) LNOI MZMs (at telecom wavelength) typically range from 2.1 to 2.5 V-cm [16,21,22], and can reach, for example, 1.8 V-cm with 3.5 μm electrode gap [85]. In hybrid LN modulators, since the optical mode is only partially confined in LN, $V_\pi L$ tends to be slightly larger [23,25,104,105,107,116,202], but can also be engineered to be comparable to monolithic ones: 2.1 V-cm has been demonstrated in SiN$_x$-loaded LN MZM [203].

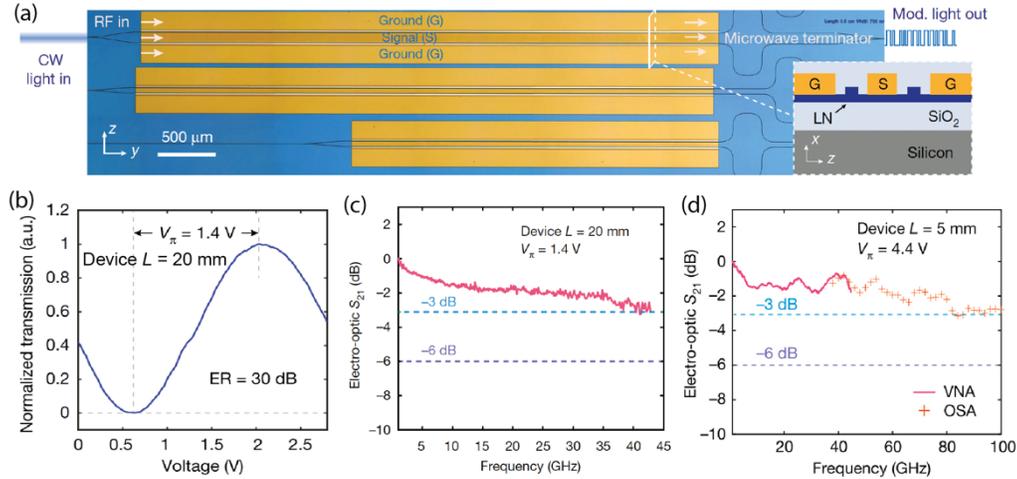

Figure 11: Traveling-wave lithium niobate electro-optic Mach-Zehnder modulator. (a) Device layout and cross-section. (b) DC response of the modulator, showing a $V_\pi$ of 1.4 V in a 20 mm device with an ER of 30 dB. (c), (d) Electro-optic frequency response. Shorter modulators can achieve larger EO bandwidth but require larger $V_\pi$. Reproduced from [21], © 2018 Springer Nature Limited.

Modeling of $V_\pi$ requires the calculation of microwave and optical mode overlap - both can be simulated using standard electromagnetic mode solvers. Mathematical formulations can be found in, e.g., Ref. [201,204]. In general, maximum optical-microwave mode overlap inside LN along the correct crystal direction (Z for $r_{33}$) is desired. It is worth mentioning that the electric field in the LN *cannot* be simply approximated as voltage across the electrode over the gap size, $V/g$. Since the dielectric constant of LN ($\epsilon_{r,z} = 28$ and $\epsilon_{r,x/y} = 44$ for clamped, high-frequency response) is drastically different from its cladding material (e.g., $\epsilon_r = 3.9$ for SiO$_2$), the electric field is non-uniform in the gap, as can be seen in Figure 12. Intuitively, when one draws a line from the ground to the signal (passing through the LN ridge), we have $V = -\int \bar{E} \cdot d\bar{l}$. At material boundaries, the perpendicular $D_\perp = \epsilon E_\perp$ field, not the $E$-field, is continuous. Therefore, in high $\epsilon$ medium, $E$ field is small. In other words, voltage drops preferentially across the low (microwave) index material instead of the high index LN. In fact, this is a critical problem in designing modulator electrodes. To enhance the $E$-field in the LN waveguide, it is beneficial to have partially etched waveguides and place the electrodes directly on the slab. In general, a device with thicker slab, wider waveguide, and a dielectric cladding would have lower $V_\pi$. Furthermore, a cladding material with high RF index but low optical index is highly desirable for this purpose. As a practical example, Xu et al. use 1 μm waveguide when routing the optical signals while tapering up to 4 μm wide waveguides in the phase modulation region to reduce $V_\pi$ [16]. It is worth mentioning that wider waveguides also allow wider electrode gap (given the same $V_\pi$), which can reduce microwave loss of the electrode and increase bandwidth (as discussed in the following subsections).

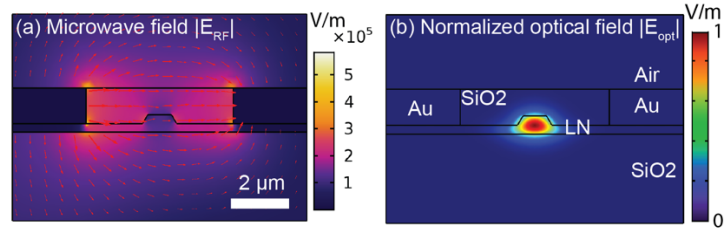

Figure 12: Microwave (a) and optical (b) electric field distribution in a monolithic X-cut EO modulator. The electrode has a gap of 5 μm. The microwave field is normalized so that the voltage across the two electrodes is 1 V. Due to the large microwave permittivity of LN, the electric field in the LN ridge is weaker than that in the cladding.

*EO bandwidth.* Modulator responses are frequency dependent. Definitions of EO bandwidth vary within the literature, but commonly used standards include 3 dB and 6 dB bandwidth (i.e., the frequency where modulation efficiency drops by 3 or 6 dB from the reference frequency). The reference frequency can be chosen at or near DC [21], but may also be taken at a higher frequency (e.g. 1 GHz) [16,205]. Choosing a non-DC reference frequency has the following rationales. First, DC and low frequency responses are difficult to measure and sometimes unstable. It is observed that DC response of LN modulators sometimes tends to be inefficient, unstable, and drift over time due to photorefraction, photoconduction, and surface charge accumulation [206,207]. In commercial LN modulators, it is common to have separate DC bias electrode and feedback circuit to stabilize the operation. Thermal-optic phase shifters can be used for DC control as well. They more stable and compact [16], but at the expense of static power dissipation. Second, some modulator designs have a rapid roll-off at low frequencies, often due to impedance mismatch. This narrow-band roll-off may not be a concern for some applications; instead, the flatness and modulation response at high frequency may be more important.

In non-traveling-wave modulators, the EO bandwidth is mainly restricted by the *RC* (resistance-capacitance) limit of the electrode. Though LN possesses one of the largest EO coefficients among other materials, the absolute index modulation under a reasonable voltage is still weak. For instance, an electric field of $10^5$ V/m can only induce an index change of $1.4 \times 10^{-5}$. To accumulate a $\pi$ phase shift, the modulator requires millimeter- or even centimeter-scale length given a reasonable driving voltage. Electrodes with such length will have a large capacitance. At frequencies $\omega > 1/RC$, the capacitor (which is the electrode) has low impedance so the voltage would not drop across it. Conventionally, $R$ is at 50 Ω, determined by the source and cable impedance. For example, a typical few-micron-gap CPW electrode has a capacitance on the order of 100 pF/m, resulting in a *RC*-limited, 3-dB EO bandwidth of about 15 GHz in a 2 mm-long modulator [85].

To achieve an EO bandwidth of tens to hundreds of GHz, traveling-wave modulators are necessary, where the electrode is a transmission line. In traveling-wave modulators, three key design considerations are particularly important: (1) velocity matching, (2) impedance matching, and (3) microwave loss.

Velocity matching allows microwave and optical signals to co-propagate at the same speed so that the modulation accumulates constructively. A natural question is "which velocity to match?" [208] Considering the case of a phase modulator, where a single-tone microwave signal modulates continuous-wave (CW) light, it is tempting to think that their phase velocities should match so that their phase fronts align when propagating. However, once the optical field is modulated, new frequencies (sidebands) are generated, and the centroid of the modulated optical signal travels at its group velocity ($v_{g,opt} = d\omega_{opt}/d\beta_{opt}$, where $\omega_{opt}$ and $\beta_{opt}$ are the angular frequency and propagation constant of the optical signal, respectively). *Therefore, the phase velocity of the microwave should match the group velocity of the optical signal* [209,210]. This can also be understood using a phase-matching picture: (1) energy conservation requires the frequency difference between the original and modulator light to be equal to the microwave frequency, $\Delta\omega_{opt} = \omega_{RF}$; (2) similarly, momentum conservation requires $\Delta\beta_{opt} = \beta_{RF}$. As a result, $d\omega_{opt}/d\beta_{opt} = \omega_{RF}/\beta_{RF}$ is needed for efficient modulation, i.e., optical group velocity matches microwave phase velocity. For practical applications such as optical communication, instead of a single-tone or narrowband signal, the microwave drive is usually a complex waveform. It is natural to think that the group velocity of the microwave signal should match that of light. While the concept of group velocity well describes the propagation speed of the envelope of an optical wavepacket (carrier frequency at hundreds of THz and bandwidth of tens of GHz), it is less proper for the electrical signals, which usually span from DC to tens or hundreds of GHz (e.g., for information encoded in square waves). It is worth noting that while the optical group and phase velocities/indices differ significantly in thin-film LN waveguides (usually >15% difference), microwave transmission lines have relatively small dispersion so their group and phase velocities/indices are similar (usually < 1% difference and will asymptotically approach a constant at higher frequencies). In practice, velocity mismatch matters more at higher frequencies. It is therefore appropriate to match the microwave effective index to the optical group index towards the highest target modulation frequency. In bulk LN modulators, the microwave index is mainly restricted by the permittivity of LN, which is very high ($\epsilon_{r,z} = 28$ and $\epsilon_{r,x/y} = 44$) and hard to match the optical index ($\epsilon_{r,z} = 4.6$ and $\epsilon_{r,x/y} = 4.9$). In thin-film LN modulators, as the LN's participation in the microwave field is small, the effective index of the transmission line is easier to engineer and can be effectively tuned by the thickness of the cladding or buried oxide, as well as metal thickness.

Impedance matching ensures proper RF interface between the modulator and driving electronics. It is conventional to design the characteristic impedance ($Z_0$) of the transmission line to be 50 Ω to match standard 50 Ω RF electronics, including source, amplifier, cable, termination, etc. However, impedance matching is often compromised (usually with $Z_0$ < 50 Ω) due to practical design constraints. Impedance mismatch leads to two main issues. (1) It causes microwave reflection. Reflections between the two ends of the transmission line can form standing waves (resonances) and cause ripples in EO frequency response, which is especially prominent when the transmission line is short or low-loss. In addition, reflections at the source end prevents efficient power delivery, with a reflected power of $|(Z_L - 50\ \Omega)/(Z_L + 50\ \Omega)|^2$. (2) Given the same RF power, lower characteristic impedance results in smaller voltage, following $P = V^2/Z_0$. Combining input reflection and voltage scaling, the impedance mismatch gives an overall $V_\pi$ penalty of $1/[2Z_0/(Z_0 + 50\ \Omega)]$, or equivalently, an EO S21 offset of $20\log[2Z_0/(Z_0 + 50\ \Omega)]$. For instance, $Z_0$ = 40 Ω will cause 12.5% increase in $V_\pi$, or equivalently, a 1 dB offset in EO S21. The effect of EO offset, as well as resonance ripples, can be seen in the simulated EO S21 in Figure 13 (iii), where $Z_0$ is set to 40 Ω. Note that the rapid drop-off near DC is because at low frequency, the transmission line is much shorter than the wavelength, so the source directly sees an input impedance of 50 Ω (neglecting DC resistance) given by the load/termination. This drop-off is even sharper for longer modulators. In the presence of the sharp drop-off, the definition of 3-dB EO bandwidth varies sensitively to the choice of reference frequency. Also, $Z_0$ is frequency dependent and tends to be higher at lower frequencies due to increased line inductance. In thin-film LN modulators, because of the narrow CPW gap (typically around 5 μm), the center conductor needs to be very narrow (~ 10 μm) to achieve 50 Ω impedance, which induces significant ohmic microwave losses. In many cases, it is more beneficial to increase the center conductor width (e.g., 20-30 μm). This will reduce $Z_0$ to around 35 - 40 Ω, but can significantly reduce the transmission line loss (by a few dB/cm) and may outweigh the penalty paid by impedance mismatch. Under this tradeoff, one needs to strike a balance when designing a modulator depending on specific bandwidth and $V_\pi$ requirement.

Given that velocity matching can be achieved relatively easily as compared to its bulk counterpart, transmission line loss is the main limiting factor for high-frequency operation of LNOI modulators. Major loss factors include metal loss (ohmic heating), dielectric loss, and surface roughness [211]. In general, metal loss dominates and increases as $\sqrt{f}$ due to skin effect (skin depth $\delta_s = \sqrt{1/(\pi f \sigma \mu_0)}$, where $f$ is frequency, $\sigma$ is the conductivity, and $\mu_0$ is the vacuum permeability). Dielectric loss depends on the material's loss tangent and scales linearly with frequency. The choice of the substrate and cladding/buried oxide material critically affect the dielectric loss. In addition, metal roughness effectively increases the travel path of current and thus increases loss, but their effects have not been studied in depth in the context of thin-film LN modulators.

In standard CPWs, the tuning knobs are very limited—gap size, center conductor width, metal thickness, and dielectric above and below the metal. This results in a small engineering margin. More importantly, due to the small gaps used in LNOI modulators (typically ~ 5 μm), the microwave mode is highly confined and causes significant current crowding near the metal edge. As a result, typical losses of CPWs in LNOI modulators are around 0.7 – 1.1 dB cm$^{-1}$ GHz$^{-1/2}$. Recently, segmented CPWs have been applied to LNOI modulators [205]. The segmented CPW consists of a standard CPW and microstructured segments extending out from the main CPW in the gap region. The segments prevent current from flowing on the edge of the metal near the gap region and significantly reduces microwave loss (0.26 dB cm$^{-1}$ GHz$^{-1/2}$ demonstrated in [205]). Similar designs have been used in III-V modulators [212], mainly to improve velocity matching, because the segments add capacitance and reduce the phase velocity. For LNOI modulator, lower-permittivity carrier material (such as quartz) may be needed instead of standard Si to compensate the slow-wave effect and maintain velocity matching.

Considering all three factors mentioned above, the frequency response (EO S21) of a EO modulator follows [213]

$$m(\omega) = \left|\frac{2Z_{in}}{Z_{in}+Z_L}\right|\left|\frac{(Z_L+Z_0)F_+ + (Z_L-Z_0)F_-}{(Z_L+Z_0)e^{\gamma_m L}+(Z_L-Z_0)e^{-\gamma_m L}}\right|, \qquad (2)$$

where $\omega$ is the microwave frequency, $Z_{in} = Z_0 \frac{Z_L+Z_0\tanh(\gamma_m L)}{Z_0+Z_L\tanh(\gamma_m L)}$ is the input impedance of the transmission line, $F_\pm = (1-e^{\pm\gamma_m L - j\frac{\omega}{c}n_{g,opt}L})/(\pm\gamma_m L - j\frac{\omega}{c}n_{g,opt}L)$ accounts for the forward/backward propagating waves, $\gamma_m = \alpha_m + j\frac{\omega}{c}n_m$ is the complex microwave propagation constant ($n_m$ being the microwave effective index, $\alpha_m$ being the loss rate), $n_{g,opt}$ is the optical group index, $L$ is the modulator length, and $c$ is the speed of light in free-space. Here, we assume both the source and load (termination) have impedance $Z_L$ (in most cases at 50 Ω). The EO frequency response (EO S21) in dB is $20\log(m(\omega))$. It can be seen that a thorough characterization of the electrical response of the microwave transmission line already allows us to accurately predict the EO response, as long as the optical group index is known. As a rule of thumb, in the case of perfect velocity and impedance matching, 3 dB EO bandwidth corresponds to ~6.4 dB electrical bandwidth (often referred as EE bandwidth) of the transmission line. This is essentially a loss limited bandwidth. Figure 13 shows some simulated EO S21 curves based on Eq. (2) to illustrate the effects of metal loss, impedance mismatch, velocity mismatch, and modulator length. These effects have been discussed individually above.

Due to microwave loss and velocity mismatch, shorter modulators have larger EO bandwidth (see Figure 11(c) vs. (d), as well as Figure 13(i) vs. (v)), but the shorter length also results in larger $V_\pi$. The bandwidth-length product may be considered as an engineering trade-off in modulator design [214,215]. In addition, since $V_\pi L$ is roughly a constant given a certain electrode and waveguide design, $BW/V_\pi$ (referred as voltage-bandwidth limit) becomes a valuable figure of merit when comparing modulator performance.

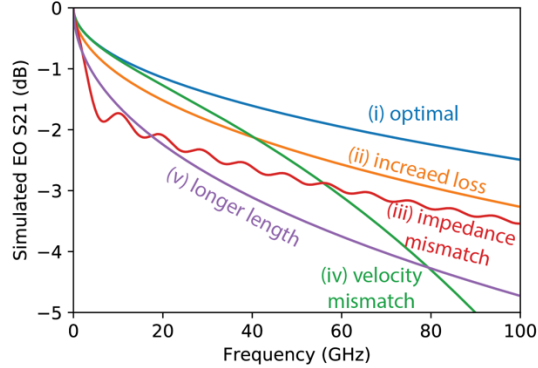

Figure 13: Illustration of how EO bandwidth is affected by transmission line loss, impedance mismatch, velocity mismatch, and modulator length. (i) Simulated EO S21 of a MZM of which the transmission has a characteristic impedance of 50 Ω, propagation loss of 0.75 dB cm$^{-1}$GHz$^{-1/2}$, perfectly matched phase velocity $n_m = n_{g,opt} = 2.28$, and modulation length of 7 mm. (ii) an increased transmission line loss (0.9 dB cm$^{-1}$ GHz$^{-1/2}$) accelerates the roll-off in EO response. (iii) Impedance mismatch (40 Ω) causes an offset and form ripples in EO response. (iv) velocity mismatch ($n_{g,opt} - n_m = 0.2$) leads to rapid roll-off at high frequencies; (v) longer length (1.4 cm) reduces EO bandwidth, but it is worth mentioning that the actual $V_\pi$ is still lower than that of shorter modulators.

**Insertion loss (*IL*).** The *IL* can be divided into two parts: fiber-to-chip coupling loss and on-chip insertion loss. Fiber-to-chip coupling has been discussed in Section 2.3.4. On-chip insertion loss includes (1) metal absorption on the electrode, (2) waveguide propagation and bending loss, and (3) insertion loss of the splitters/combiners (usually in the form of y-splitter [15,21] or multimode interference (MMI) couplers [16]). With state-of-the-art dry etching processes, waveguide propagation loss is typically < 0.3 dB/cm (with 0.027 dB/cm measured in Ref [11], see Section 2.3); bending loss can be mitigated by increasing bending radius; splitter insertion loss as low as 0.054 dB has been demonstrated using dry-etched MMI in LN [16]. Excess optical loss mainly comes from absorption in the metallic electrode, which becomes worse when there is misalignment between the metal and optical layer. Metal absorption thus causes a trade-off between $V_\pi$ and on-chip insertion loss: small electrode gap and wide waveguide reduces $V_\pi \cdot L$ but increases optical absorption in the metal.

***Extinction ratio (ER).*** The *ER* is defined to be the output intensity ratio between the "on" and "off" states. It is largely determined by the design and fabrication of the MZI. Specifically, unbalanced loss in the two MZI arms will decrease *ER*, and imperfections in the splitter/combiner will decrease *ER*. In addition, polarization impurity in the MZM, for example from waveguide bending [216], will also affect the *ER*. Typical measured *ER*s of LNOI MZMs are around 20 to 30 dB (tested at low frequency) [16,21,23,203]. *ER* > 40 dB has been demonstrated in 3 mm MZM [22], In addition, by cascading two MZIs, *ER* of 53 dB has been demonstrated in Ref. [119].

**Linearity and dynamic range.** In the small signal regime when biased at the quadrature point, the modulated optical intensity responds linearly to the applied electrical voltage. However, at large microwave drives, the nonlinearity of the modulator becomes significant and affects the signal integrity in RF photonics systems and analog fiber-optic datalinks. The linearity can be quantified using, for example, 3rd-order intermodulation distortion (IMD3) spurious-free dynamic range (SFDR). Given two RF input at frequencies $f_1$ and $f_2$ (both within the modulator bandwidth), the intermodulation products include $mf_1 \pm nf_2$ ($m$ and $n$ are integers). IMD3 includes contributions from $2f_1 - f_2$ and $2f_2 - f_1$. These components are the most challenging distortion products since they are directly adjacent to the input tones. IMD3 SFDR refers to the input RF power range between the two points where the fundamental modulation output (RF power from photodetector) is beyond the system noise floor (lower limit) and where the IMD3 becomes larger than the noise floor (upper limit) (see Figure 14(a)). Alternatively, it can be seen as the signal-to-noise ratio of the fundamental output when the IMD3 reaches the system noise floor. Mathematically, it is defined as $SFDR = (P_{int}/NF)^{2/3}$, where $P_{int}$ is the output power (at the detector) where fundamental and IMD3 intercepts, and $NF$ is the noise floor (in units of W/Hz or dBm/Hz). As a result, SFDR has the unit of dBHz$^{2/3}$. All these definitions are equivalent since the fundamental has a slope of 1 and IMD3 has slope of 3. Figure 14(a) illustrates the concept and Figure 14(b) shows measured IMD3 SFDR data from a thin-film LN modulator [22]. It is clear that SFDR not only depends on the linearity of the modulator, but is also affected by noises. SFDR of a photonic link depends on laser noise, noise figure of optical and RF amplifiers, noise of the detector, as well as linearity and insertion loss of the modulator. SFDR is relatively under-studied in the current literature of LNOI modulators. Based on the limited number of reports [22,25,202,217], measured IMD3 SFDRs using LNOI modulators are in the range of 90 - 100 dBHz$^{2/3}$, which has not outperformed the best demonstrate in bulk LN modulators (> 120 dBHz$^{2/3}$ reported [218]). As the lasers, detectors, and amplifiers used in these reports are different, it is difficult to make fair comparisons of the modulator linearity across literatures. Future increase of optical pump power and reduction of coupling loss in LNOI modulator-based datalinks may substantially increase the measured SFDR.

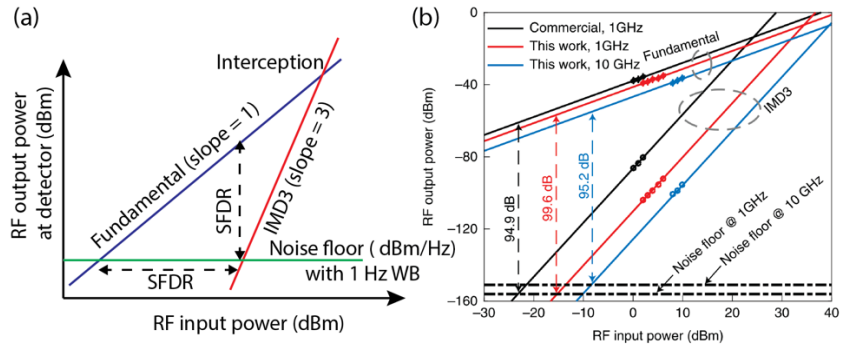

Figure 14: Third-order intermodulation distortion (IMD3) spurious-free dynamic range (SFDR). (a) Illustration of IMD3 SFDR. (b) Measured IMD3 SFDR in a thin-film LN modulator [22]. (b) Reproduced from [22] with permission**,** ©The Author(s), under exclusive licensed to Springer Nature Limited 2019.

### 3.2.3. State of the art

Though thin-film LN-based EO modulators were first demonstrated as early as 2005 [219], they have since undergone relatively slow development. This is mainly due to the lack of low-loss micro/nano-structured waveguides. The first breakthrough came around 2013 when rib loading was proven to be an effective solution to avoid LN etching [105]. Since then, various loading materials have been exploited, and rib-loaded modulators have become a major approach [25,116,203,220,221]. Meanwhile, other approaches such as proton-exchange in thin-film have also been explored [73,222]. The second breakthrough happened around 2016 when dry-etched LNOI modulators reached propagation loss below 1 dB/cm [20]. Along with this monolithic approach, the field has seen rapid progresses [85,119,198,223–226], including the demonstration of >100 GHz bandwidth modulators operating at CMOS-compatible voltages [21], and coherent modulators that support data rate up to 320 Gbit/s [16]. Moreover, heterogeneous integration by bonding thin-film LN (either etched [22,199] or unpatterned [23]) on SOI and $SiN_x$ PICs have also proven to be a promising direction. This direction is now being investigated not only by academics, but also by silicon photonic foundries and national laboratories/facilities [202,227]. It may lead to scalable production and multi-function integration for thin-film LN modulator technologies.

Table 3 summarizes the performance of some representative EO modulators based on thin-film LN, including both monolithic and hybrid approaches, and covering Mach-Zehnder intensity modulators, phase modulators, and Michelson interferometer modulators.

Table 3: Representative non-resonant EO modulators based on thin-film LN. MZM: Mach-Zehnder modulator; PM: phase modulator; IQ modulator: in-phase/quadrature modulator; MIM: Michelson interferometer modulator. $V_\pi$ and $V_\pi \cdot L$ refer to low-frequency (near-DC) values, and those for PMs are divided by two to match MZMs. All modulators were measured near 1550 nm.

| Modulator type | Platform/Method | $V_\pi \cdot L$ (V-cm) | $V_\pi$ (V) | L (mm) | 3 dB EO BW (GHz) | ER (dB) | Propagation loss (dB/cm) | Year | Ref. |
|---|---|---|---|---|---|---|---|---|---|
| MZM | Monolithic | 2.2 | 1.5 | 15 | 20 | - | 0.5 | 2017 | [228] |
| MZM | Monolithic | 2.2/2.3/2.8 | 4.4/2.3/1.4 | 5/10/20 | 100/80/45 | 30 | 0.3 | 2018 | [21] |
| MZM | Monolithic | 3.12 | 2.6 | 12 | 56 | - | - | 2019 | [225] |
| MZM[a] | Monolithic | 2.7 | 1.35 | 20 | 175[b] | 20 | < 0.5 | 2020 | [205] |
| MZM[c] | Monolithic | 1.8 | 9 | 2 | 15 | 10 | 3 | 2018 | [85] |
| PM | Monolithic | 4.7 | 4.7 | 10 | 40 | - | 1 | 2016 | [20] |
| PM | Monolithic | 1.748 | 1.9 | 9.2 | 47[d] | - | 7 | 2018 | [24] |
| PM | Monolithic | 3.5 | 1.75 | 20 | >45 | - | 0.5 | 2019 | [195] |
| IQ modulator | Monolithic | 2.47 / 2.325 | 1.9 / 3.1 | 13 / 7.5 | 48 / 67 | 25 | 0.15 | 2020 | [16] |
| MZM | Etched LN on SOI | 2.55 / 2.22 | 5.1 / 7.4 | 5 / 3 | 70 | 40 | 0.98 | 2019 | [22] |
| MZM | Ta2O5 rib loading | 4 | 6.8 | 6 | - | 20 | 5 | 2013 | [105] |
| MZM | Chalcogenide rib loading | 3.8 | 6.3 | 6 | 1 | 15 | 1.2 | 2015 | [104] |
| MZM | SiNx rib loading | 3 | 2.5 | 12 | 8 | 13.8 | 7 | 2016 | [116] |
| MZM | SiNx rib loading | 3.1 | 3.88 | 8 | 33 | 18 | - | 2016 | [25] |
| MZM | SiNx rib loading | 1.925 | 0.875 | 24 | - | 30 | <2.25 | 2020 | [203] |

| | | | | | | | | |
|---|---|---|---|---|---|---|---|---|
| MZM | LN on SiNx | 6.67 | 3.34 | 5 | 30.55 | > 20 | 1.6[e] | 2020 | [202] |
| MZM | LN on SOI | 6.7 | 13.4 | 5 | > 106[f] | 30 | 0.6 | 2018 | [23] |
| MIM | Etched LN on SOI | 1.2 | 12 | 1 | 17.5 | 30 | 0.3[g] | 2019 | [199] |
| MIM | Monolithic | 1.4 / 1.52 | 14 / 7.6 | 1 / 2 | 12 | 27.6/20 | 4.6 | 2019 | [198] |

[a]Segmented CWP electrode.
[b]Measured 1.8 dB EO roll-off at 50 GHz; extrapolated 3 dB EO bandwidth to be ~175 GHz.
[c]Not traveling wave.
[d]Estimated from electrical bandwidth; Measurement up to 500 GHz without significant roll-off from index mismatch.
[e]Estimated based on [229].
[f]Limited by measurement instrument; extrapolated to > 200 GHz.
[g]Estimated from calculation.

### 3.3. Resonant electro-optic modulators

The non-resonant EO modulators described above can achieve low $V_\pi$ and high bandwidth, but require long devices (centimeters for few-volt $V_\pi$) to do so. For many applications in high-density integrated photonics, more compact modulators are required. One straightforward way to accomplish this is to use resonant EO modulators, such as ring modulators and photonic crystal cavity modulators, illustrated in Figure 15(a) - (e). In their most basic form, such modulators allow the resonance frequency to be changed by an applied voltage, which can change the transmission of light at a frequency near resonance. This section reviews progress on such resonant modulators in thin-film LN.

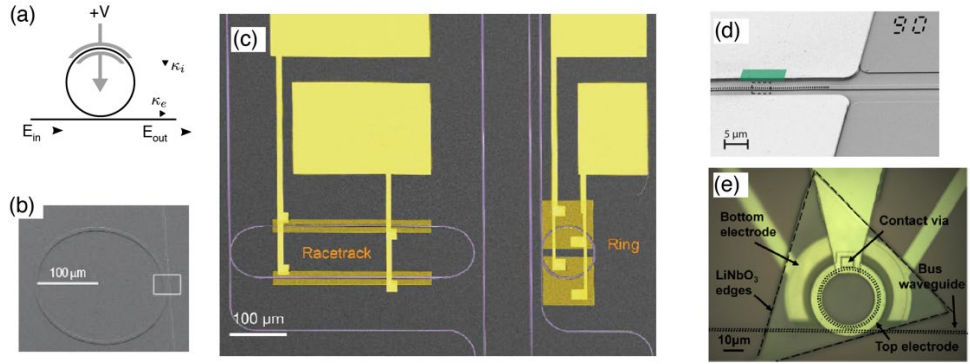

Figure 15: Resonant electro-optic modulators in thin-film lithium niobate. (a) Schematic diagram of a typical ring modulator. The input light ($E_{in}$) couples to a ring resonator with internal and external coupling rates ($\kappa_i$ and $\kappa_e$ respectively). The gate voltage enables modulation of the output light ($E_{out}$). (b) An early Z-cut ring modulator, with electrodes placed above and below the device layer [55]. (c) Ring and racetrack modulators in X-cut lithium niobate [85]. (d) A photonic crystal cavity modulator [184]. (e) A ring resonator in a hybrid LN/silicon platform [113]. (b) Reproduced from [55], © 2007 Nature Publishing Group; (c) Reproduced from [85], © 2018 Optical Society of America; (d) reproduced from [184], licensed under a Creative Commons Attribution 4.0 International License; (e) reproduced from [113], © 2014 Optical Society of America.

One challenge when working with resonant EO modulators is that a voltage applied to the modulator can change both the amplitude and phase of the transmitted light, depending on the coupling ratio $\eta = \kappa_e/\kappa_i$, where $\kappa_i$ and $\kappa_e$ are the intrinsic (due to loss in the cavity) and extrinsic (due to coupling to external waveguides) power loss rates, respectively. The transfer function for monochromatic light with frequency $\omega$ through a modulator like that shown schematically in Figure 15(a) with a resonance frequency $\omega_r = \omega_0 + \chi V$ that shifts linearly with applied voltage at a rate $\chi$ is given by [230]

$$t = \frac{E_{out}}{E_{in}} = \frac{(\kappa_i - \kappa_e) - 2i\Delta}{(\kappa_i + \kappa_e) - 2i\Delta}, \quad (3)$$

where $\Delta = \omega - \omega_r$. This optical transfer function does not represent either pure phase or amplitude modulation, but in certain regimes it can act as such. As an example that is particularly relevant for high-density photonic circuits, consider the phase modulation case where $\kappa_e \gg \kappa_i$ and the input light is resonant with the zero-voltage resonance frequency $\omega = \omega_0$. In this regime the power transmission is nearly lossless $T = |t|^2 \approx 1$ and the output light is phase shifted relative to the input by $\delta\phi = \pi - \frac{4\chi \delta V}{\kappa_e}$ for small voltage signals $\delta V$. Note that for larger voltages of order $|V| \gtrsim \frac{\kappa_e}{4\chi}$, the phase response becomes nonlinear and saturates near the bounds $\delta\phi = 0$ and $\delta\phi = 2\pi$. If purely linear phase response is required, it can be achieved with non-resonant phase modulators described in the previous section.

Resonant modulators can achieve strong modulation despite their small size because the optical field circulates and interacts with the microwave field produced by the electrodes many times. Typical resonant modulators are made using a cross-sectional waveguide and electrode geometry that is similar to that of a non-resonant modulator. In such a case, a useful way to compare the performance of a resonant and non-resonant modulator is to consider the effective interaction length of the modulator $l_{\text{eff}} = l_f \alpha$, given by the product of the optical field attenuation length $l_f = \frac{2c}{\kappa n_{\text{eff}}}$ inside the resonator, where $\kappa = \kappa_i + \kappa_e$ is the total power decay rate, and the effective electrode coverage fraction $\alpha$. For some types of resonant modulators like ring resonators, the electrode geometry is constrained such that the electric field cannot be applied along the optimum $z$ crystal axis [204], which also reduces the effective electrode coverage fraction $\alpha$ below the simple geometric result. For thin-film lithium niobate, low loss resonators with $Q=10^7$ at wavelengths near 1.5 μm can be achieved, corresponding to EO interaction lengths of order 1 m—about 10-fold longer than typical non-resonant EO modulators.

However, the compactness and high modulation efficiency of resonant modulators comes at the cost of reduced EO bandwidth. The optical field within the resonator can only respond to modulation at a rate set by the total decay rate $\kappa$ of resonator [231]. For example, for a resonator with $Q = 10^7$ at wavelengths near 1.5 μm, $\kappa = 2\pi \times 20$ MHz.

The design of EO modulators based on resonance frequency modulation requires a difficult tradeoff between modulation efficiency (set by the effective EO interaction length) and bandwidth (set by the resonator linewidth). However, for many applications such as optical switch networks [232] and optical beam steering [233], even the relatively low bandwidth provided by high-Q resonators is sufficient. For these applications, compact resonant modulators in thin-film LN offer a promising alternative to techniques like thermal tuning, with lower power dissipation and higher bandwidth. Note that this tradeoff can be tailored by using multiple coupled optical resonators, such as in coupled resonator optical waveguides [234] and the photonic molecules described in Section 3.5. Additionally, by modulating the resonator-waveguide coupling instead of the resonance frequency, the bandwidth limitation can be reduced [231].

Over the past two decades, resonant EO modulators in thin-film LN have been widely studied. Table 4 presents a selection of recent works on single-resonator modulators, and compares relevant figures of merit for these devices. Both monolithically dry-etched thin-film LN resonators [55,85,171,176,184] and hybrid ones based on rib loading [101,104,105] or heterogenous bonding [111–113] have been explored. Typical resonant EO modulators in thin-film lithium niobate have resonances that can be electrically tuned at low frequencies by 1-20 pm/V. This voltage tuning rate figure of merit is determined in part by the same electrode design considerations as non-resonant modulators. For example, more closely spaced electrodes improve the voltage tuning rate. This effect is exploited by [112], in which the doped Si waveguide is used as part of the electrode. Additionally, a large effective electrode coverage fraction $\alpha$ improves the voltage turning rate. Modulators based on photonic crystal cavities like that in [184] have near-optimal effective electrode coverage fraction $\alpha$ because they do not require waveguide bends. Besides ring, racetrack, and photonic crystal cavities, Bragg reflectors may also be used to form resonant EO modulators [235] or tunable filters [236].

Table 4: Representative resonant EO modulators based on thin-film LN.

| Resonator type | Platform/Method | Voltage tuning rate (pm/V) | Q-factor (loaded) | Crystal cut | Ref | Notes |
|---|---|---|---|---|---|---|
| Ring | Monolithic | 1.05 | 4×10$^3$ | Z | [55] | |
| Ring | LN on SOI | 1.7 | 1.7×10$^4$ | Z | [111] | |
| Ring | Ta$_2$O$_5$ rib loading | 4.5 | 7.4×10$^4$ | X | [105] | |
| Ring | LN on SOI | 12.5 | 1.1×10$^4$ | Z | [112] | Si waveguide used as conductive electrode |
| Ring | LN on SOI | 3.3 | 1.4×10$^4$ | X | [113] | |
| Ring | Chalcogenide rib loading | 3.2 | 1.2×10$^5$ | Y | [104] | |
| Ring | Monolithic | 2.15 | 2.8×10$^3$ | Z | [171] | |
| Photonic crystal cavity | Hybrid Si on bulk LN | ~2 | 1.2×10$^5$ | X | [114] | Nonlinear shift of resonance with voltage observed attributed to the field effect in Si. |
| Racetrack | Monolithic | 7 | 5×10$^4$ | X | [85] | |
| Ring | Monolithic | 4 | 2×10$^6$ | X | [176] | Single ring properties in dual ring device |
| Ring | SiN$_x$ rib loading | 1.8 | ~9×10$^4$ | X | [101] | |
| Photonic crystal cavity | Monolithic | 16 | 1.34×10$^5$ | X | [184] | |

## *3.4. Electro-optic frequency combs*

### 3.4.1. Non-resonant electro-optic comb

Optical frequency combs (OFC) have enabled numerous applications in science and engineering. The ability to achieve strong optical confinement at the micro- and nano-scale on low-loss optical platforms have allowed the realization of chip-scale OFCs. One way to generate OFCs is through Kerr nonlinearity, which will be discussed in Section 4.5. Another way is through the linear electro-optic (Pockels) effect, by phase-modulating an optical signal in a material with a large $\chi^{(2)}$ value. A comprehensive review of EO frequency combs, their principles, and applications can be found in Ref. [237]. These non-resonant EO-comb sources offer several promising features, including intrinsic mutual coherence between all the comb lines, convenient tuning of the repetition rate and center frequency, and favorable spectral flatness. The latter would be of great importance for optical communication [238], optical arbitrary waveform generation [239], and radio-frequency (RF) photonics [240].

An essential parameter for generating a frequency comb using an electro-optic phase modulator is the modulation index ($\beta = \pi \frac{V_p}{V_\pi}$) where $V_p$ and $V_\pi$ are the peak voltage for the driving microwave field and half-wave voltage of the modulator, respectively. By feeding a continuous-wave laser at the frequency of $\omega_c$ to a phase modulator, and modulating it with a microwave field given by $V(t) = V_p sin(\omega_m t)$, the optical output field can be written as

$$E(t) = E_0 e^{i\omega_c t} e^{i\beta \sin(\omega_m t)} \tag{4}$$

Taking the Fourier-transform of the time-domain signal and applying Jacobi-Anger expansion [241] allow us to express the output field in the frequency domain as

$$E(\omega) = E_0 \sum_{n=-\infty}^{+\infty} J_n(\beta)\delta(\omega - n\omega_m - \omega_c), \qquad (5)$$

where $J_n$ is the Bessel functions of the first kind, and $\delta$ is the Dirac delta function. The output spectrum consists of equidistant spectral elements with spacing $\omega_m$ around the center frequency of the input field, $\omega_c$. $J_n$ defines each comb line's amplitude while the modulation index can tune the overall shape of the spectrum.

Figure 16(a) shows an experimentally measured frequency comb generated from a LNOI phase modulator. This modulator was fabricated on an X-cut thin-film LN and has a length of 2 cm with $V_\pi$ of 3.5-4.5 V in the frequency range between 5-40 GHz [195]. By driving it at substantial RF input voltage (~$4V_\pi$), a relatively broad comb with over 40 spectral elements was generated with 30 GHz spacing. The ability to make tightly confined waveguides with low optical loss and efficient overlap of the optical field with microwave field (defined by proper waveguide etch depth and electrode gaps) are the key reasons to achieve such a performance.

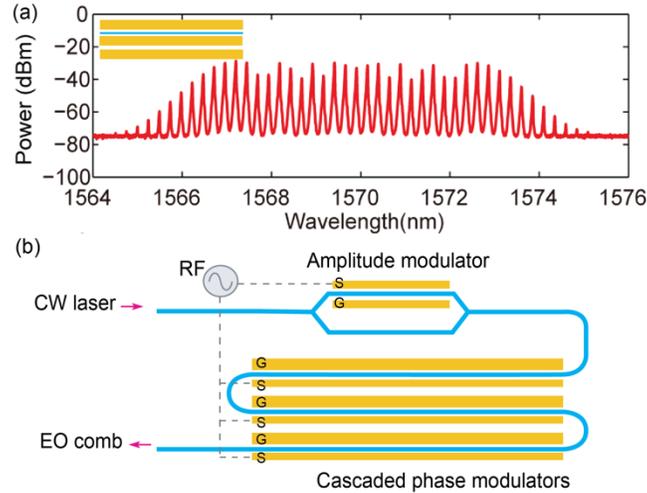

Figure 16: Non-resonant EO comb. (a) A frequency comb generated by driving a single LNOI phase modulator [195]. (b) A scheme to generate flat EO comb by cascading multiple EO amplitude and phase modulators. (a) Reproduced from [195] © 2019 IEEE.

The limited microwave power applied to individual phase-modulators restricts the number of generated comb lines in the non-resonant frequency-comb sources. This issue can be addressed by cascading more devices to achieve broader combs. It is also known that the addition of a Mach-Zehnder intensity modulator in series with the phase-modulator can improve the spectral flatness of the generated spectrum, such as the configuration illustrated in Figure 16(b). Inspired by the bulk examples [242] and cascading a phase modulator with an amplitude modulator in an X-cut LNOI platform, authors in Ref. [243] could trim their output comb spectrum to have a spectral flatness of 0.89 dB although for a limited number of comb lines.

### 3.4.2. Resonant electro-optic comb

As indicated in the previous section, the small spectral bandwidth could potentially limit the applications of non-resonant EO combs. By placing a phase-modulator in an optical cavity, phase modulation efficiency can be enhanced by the number of passes (determined by the Finesse of the cavity) of the light through the modulator, which could increase the comb bandwidth. The key idea here is to realize that the driving microwave field of the phase-modulator must be resonant with the free spectral range (FSR) of the optical cavity to maximize the efficiency of the cascaded side-band generation. This idea has been studied for almost 50 years now [244–247].

With the advent of ultra-low-loss LNOI waveguides and resonators [11], resonant-based EO combs have been realized on-chip with unprecedented performance [26]. In Ref [26], injecting a continuous-wave (CW) laser into a high-$Q$ optical race-track cavity (Figure 17(a)) (FSR~10 GHz), which is modulated at a microwave frequency equal to the FSR, a frequency comb spectrum consisting of 900 lines in the telecommunication region was generated (Figure 17(b)). A unique property of these comb sources is their ability to generate the spectrum at multiples of the optical FSR (defined by the size of the race-track cavity) by changing the microwave driving frequency, as shown in Figure 17(c). Moreover, the dispersion engineering of the LNOI waveguides would help to overcome the fundamental limitation of the comb bandwidth for resonant EO-combs based on bulk elements. In resonant EO-combs, a flat dispersion is desirable; otherwise, the mismatch between the microwave frequency and the cavity FSR introduces a spectral cut-off to the comb spectrum, which can be numerically modelled [248]. As opposed to Kerr combs, this process of comb generation is not sensitive to the group velocity dispersion at the pump frequency; therefore, it could potentially extend to a wider range of optical pump wavelengths. This flexibility allows pumping at different wavelengths and generating multiple combs simultaneously. A few areas of improvements can be foreseen in the near future. First, current efficiency and output power of LNOI EO comb are relatively low compared to Kerr combs. It is shown that employing an intermediate cavity between the input coupler and the EO-comb cavity can increase the output comb power[248]. Second, the bandwidth can be future improved using resonators with higher Q factors. With these improvements, LNOI resonant EO combs will become a particularly useful element for applications in telecommunications and microwave photonics.

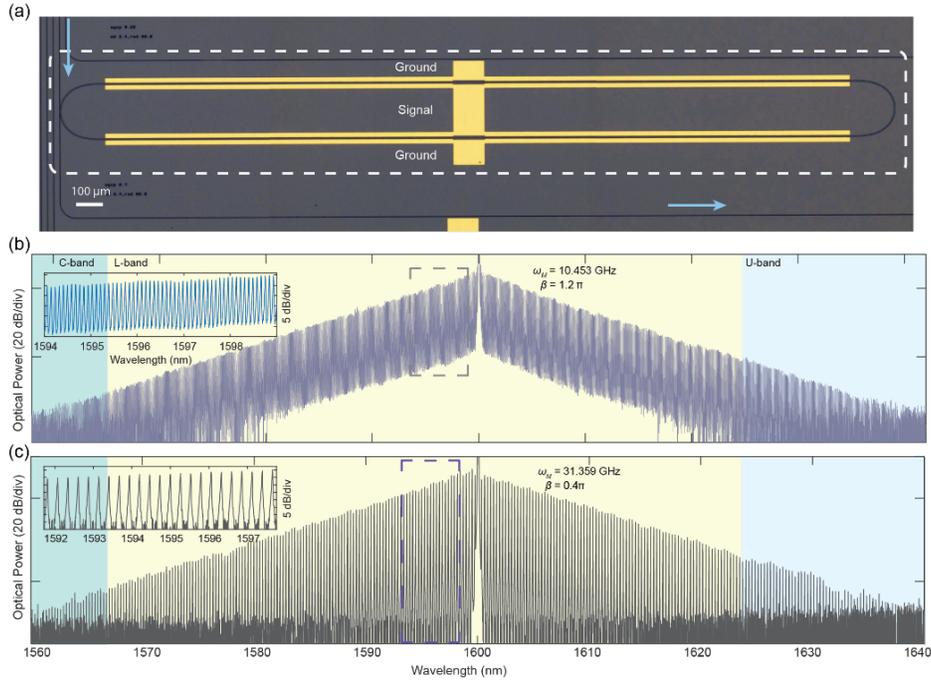

Figure 17: Resonant electro-optic frequency comb sources. (a) Micrograph of a resonant integrated electro-optic frequency comb source fabricated on an X-cut lithium niobate on insulator wafer. (b) The output measured spectrum with a line spacing of 10 GHz spanning from 1560-1640 nm. The left inset shows a magnified portion of the comb spanning 4 nm. The right inset shows the measured transmission spectrum for an optical resonance which experiences broadening by increasing the microwave driving power. (c) EO comb generated by driving the 10 GHz FSR resonator with a 30 GHz microwave. (a)-(b) Reproduced from [26], © The Author(s), under exclusive license to Springer Nature Limited 2019.

### 3.4.3. Spectroscopy

Frequency combs are powerful tools for precision spectroscopy [249]. Dual-comb spectroscopy [250] is an emerging technique based on Fourier transform spectroscopy [251] with no moving parts. In this technique, one comb with a repetition rate $f_{rep}$ interrogates the sample of interest (e.g., a gas cell), and beats with a second comb (as a local oscillator) with a repetition frequency of $f_{rep}+\delta f$ on a fast photo-diode. The time-domain interference is recorded, and the Fourier transform of time trace reveals a spectrum consisting of the beat notes of the two combs in the radio-frequency (RF) domain. This technique holds promise in several aspects as it offers broadband measurements with a resolution given by the comb line spacing and a frequency accuracy provided by an atomic clock. One of the requirements for dual-comb spectroscopy is the need for mutual coherence time between the two comb sources. In most demonstrations, the two combs are generated from separate sources (e.g., mode-locked lasers), and establishing a mutual-coherence could be technically challenging. In dual-comb setups with EO comb sources, the mutual coherence can be achieved by driving with phase-locked microwave sources and shared optical pump laser [252]. These EO comb sources can also provide high agility, flexible operating frequency, and tunable repetition rate. Benefiting from LNOI technology, an on-chip dual-comb spectrometer in the telecommunication region with a resolution of ~10 GHz were demonstrated using a pair of cavity-based EO comb microring sources [250]. Using a single CW laser source to feed both microrings resulted in a coherence time of $5\times10^{-3}$ seconds. An acetylene gas-cell at atmospheric pressure was interrogated in this dual-comb setup, and its transmittance and dispersion were measured. A residual (difference between the measured spectrum and the HITRAN database [253]) of 10% and a standard deviation of 3.4% was reported. As it was discussed earlier, EO-comb sources can be operated over different center frequencies due to the flexible operating frequency and natural phase-locking established by the microwave driving field. Therefore, one can inject two CW lasers that are spectrally distant and generate two separate combs in a single device. Using this concept, for the first time, the authors demonstrated a spectrally-tailored dual-comb interferometer which could probe two different regions at telecommunication wavelengths that are 6.6 THz apart. Such a spectrometer can potentially probe transitions that are spectrally far apart with an optimized signal to noise ratio.

Compared to other dual-comb spectrometers based on Kerr combs [254,255], this platform has the potential of increasing the resolution i.e. the repetition rate of the comb sources with more versatility. By cascading a cavity-based and a non-resonant EO-comb, a 3-GHz comb was achieved on an X-cut LNOI platform [256]. The low-repetition-rate output spectrum was achieved by densifying the original 10-GHz spectrum with a filling factor set by the non-resonant EO-comb. Potentially the phase-modulator can increase the spectral density with any integer division of the original comb spacing.

### *3.5. Coupled-resonator-based modulators*

EO modulation in coupled resonators brings rich physics and leads to novel and useful device concepts. In this section, we review the basic principles and recent progress of coupled-resonator-based EO modulators in the thin-film LN platform.

By forming hybrid modes, coupled resonators host resonance modes with energy splittings that are different from their intrinsic FSR. Dynamical EO modulation of these resonators can induce optical transition between these energy levels. The concept of creating a photonic two-level system was first demonstrated in 1998 [257]. Such artificial photonic molecules produce various optical phenomena and allow for non-trivial light control. In general, a photonic two-level system requires two optical resonance modes that are strongly coupled, so photons can transition from one mode to another before being dissipated or lost. Recent progress in thin-film LN provides a great opportunity for this field: The ultra-low-loss nanophotonic waveguides and high-Q resonators [11] offer the possibility of realizing long-lived photonic states, and the strong EO effect in conjunction with the highly confined optical modes enables efficient photonic transition.

Recently, a photonic two-level system was achieved using a pair of evanescently coupled LNOI microring resonators [176], and fast index oscillation through the electro-optic modulation generates strong coupling between the two levels (Figure 18(a)). The Hamiltonian to describe such a system is $H = \omega_1 c_1^\dagger c_1 + \omega_2 c_2^\dagger c_2 + \Omega \cos(\omega_m t)(c_1^\dagger c_2 + h.c.)$, where $\omega_1$ and $\omega_2$ represent the eigen-frequencies of the two optical modes in the coupled resonators, $\Omega$ describes the coupling strength induced by microwave modulation, and $\omega_m$ is the microwave frequency. Based on this Hamiltonian, various phenomena such as Rabi-oscillation, Ramsey interference, Autler-Towers splitting, and Stark Shifts were demonstrated in this device (Figure 18(b-d)). In addition, the photonic molecule can be programmed into a bright-dark mode photon-pairs, allowing for photon storage and retrieval.

Coupled-resonator based modulator also allows efficient frequency shifting and beam splitting at GHz-scale. One approach to control the flow of light for a complete frequency conversion is proposed based on a generalized critical coupling condition [177]. The realization of photonic molecules in thin-film LN provides an opportunity to achieve such frequency shifters as well as frequency beam splitters, which are fundamental functionalities for controlling the frequency degree of freedom of light.

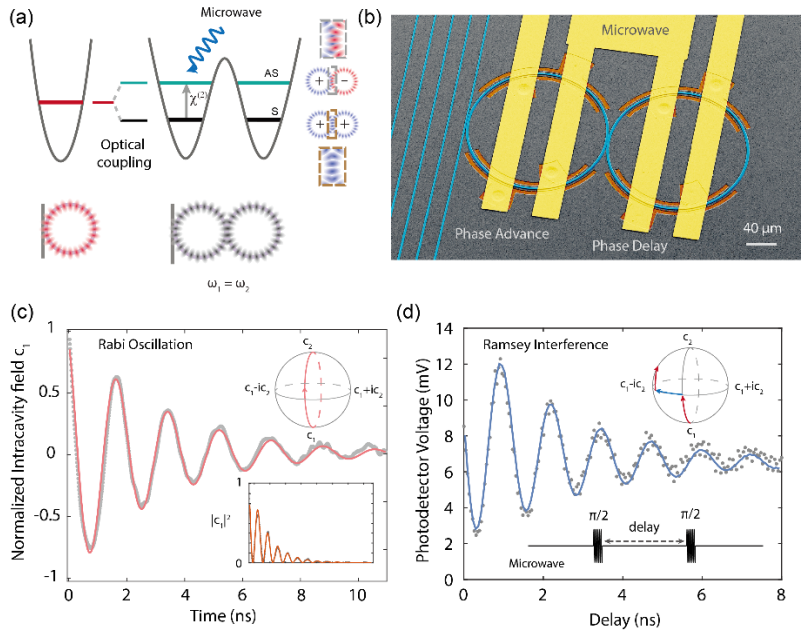

Figure 18: Realization of a photonic molecule using coupled resonators in thin-film lithium niobate [176]. (a) The concept of the photonic molecule, where hybrid modes are formed by a pair of evanescently coupled microring resonators. The transition between the two modes can be induced by electro-optic modulation. (b) Scanning electron micrograph of a couple-ring modulator that forms a photonic molecule. (c) Rabi oscillation and (d) Ramsey interference in the two-level system. Reproduced from [176], © The Author(s), under exclusive license to Springer Nature Limited 2018.

By engineering the density of states of, and coupling between, optical modes in LNOI waveguides and ring-resonators, an EO frequency shifter that works with CW optical input and continuous sinusoidal microwave drive was demonstrated, featuring a shift efficiency of ~99% and an on-chip device insertion loss of 0.45 dB [177]. Under the generalized critical coupling condition, light injected at one frequency can be efficiently coupled from one frequency mode to another without back conversion. Different from conventional single sideband modulators, the coupled-ring shifter allows bidirectional frequency exchange of photons (i.e., frequency swap operation). This bi-directionality allows the device to be reconfigured as a tunable frequency beam splitter, in which the splitting ratio and frequency are controlled by microwave power and frequency, respectively. Such frequency shifter and beam splitter allow coherent frequency control of photons, which are of particular interest to frequency-domain quantum computing [258]. In addition, extending the generalized critical coupling condition to multi-level system is also proposed [177], which can lead to cascaded frequency shifting beyond 100 GHz using only several tens of GHz microwave drive.

### 3.6. Cavity electro-optics

Combining electro-optically active optical resonators with low-loss microwave cavities – an approach known as cavity electro-optics [259–261] – can enable strong interactions between microwave and optical fields and create nearly quantum-limited modulation performance. Figure 19(a) illustrates a typical cavity EO device, such as those described in [178,179,262,263]. Two optical ring resonators form hybrid photonic molecule modes as previously described in Section 3.5. However, now the photonic molecule is driven by a microwave-frequency resonator (a lumped-element LC resonator in this case), with resonance frequency matched to the splitting between the hybrid optical modes. Additionally, this splitting can be tuned by a DC bias capacitor. The resonant enhancement of microwave power creates stronger optical modulation at the cost of a reduced bandwidth set by the linewidth of the microwave resonator. For example, one way to operate such a device is illustrated by the frequency-space diagram shown in Figure 19(b), where a strong pump laser was tuned to be resonant with the red optical mode at frequency $\omega_-$. An input microwave signal modulates the optical modes and induces resonantly enhanced sum-frequency generation into the blue optical mode at frequency $\omega_+$, while the difference frequency at $\omega_- - \omega_m$ was suppressed due to being far off resonance. With this configuration, both the optical pump and modulation sideband were resonantly enhanced. Not all cavity EO devices use this coupled-ring-resonator approach; other designs exploit neighboring longitudinal [264] or transverse [190,262,265] optical modes of a single optical resonator, but the principle of operation is the same.

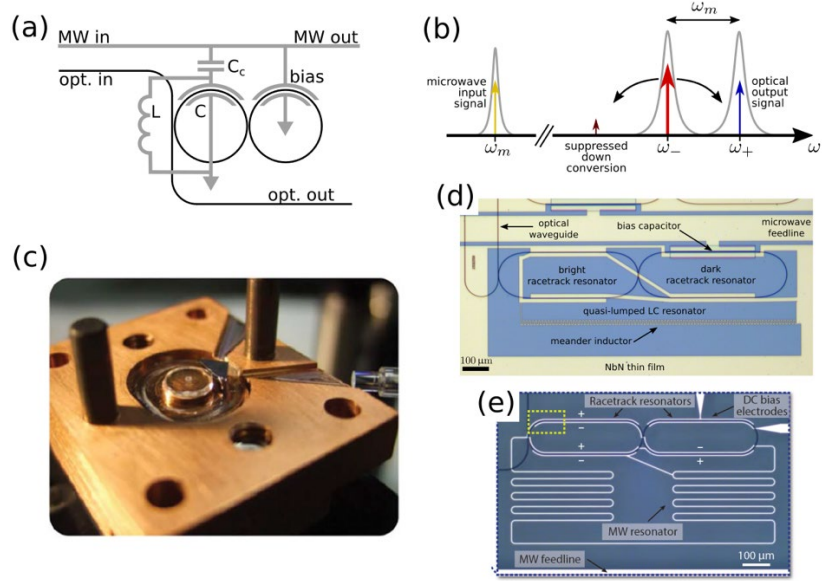

Figure 19: Cavity EO in lithium niobate. (a) Schematic diagram of a dual-ring cavity EO system. (b) Frequency-space diagram showing microwave-optical transduction based on single-sideband modulation. (c) A cavity EO transducer based on a polished LN whispering gallery mode resonator and a copper 3D cavity [262]. (d) and (e) Cavity EO transducers using thin-film lithium niobate resonators and superconducting LC resonators [178,179]. (c) Reproduced from [262], © 2016 Optical Society of America; (a, b, d) reproduced from [178] © 2020 Optical Society of America.; (e) reproduced from [179] © 2020 Optical Society of America.

The main motivation for the development of such modulators is to achieve an efficient transducer between microwave and optical fields to couple microwave-frequency superconducting qubits to optical networks. Such superconducting quantum devices have demonstrated rapid progress in recent years [266], which has encouraged interest in using them to create large-scale networks. However, the microwave-frequency photons used to encode information in these devices are susceptible to strong thermal noise and attenuation at room temperature, which limits the practical size of such networks to something that can fit within a single cryogenic environment. Quantum interconnects [267] based on optical fiber links offer a way around this restriction, because they provide relatively weak attenuation and negligible thermal noise due to the high optical carrier frequency. A key component of such an interconnect is a quantum transducer that can convert single photons between microwave and optical frequencies. Several approaches have been studied to create such a transducer, which are described in several comprehensive reviews [268,269]. An approach based on the piezoelectric interaction in lithium niobate is described in Section 5.3.1.

Cavity EO devices are particularly promising for this application because of their direct conversion mechanism (i.e. they do not rely on an intermediate form of energy, such as a mechanical excitation), which may enable higher efficiency and lower noise operation. The first cavity EO devices were based on bulk lithium niobate whispering gallery mode resonators, like that shown in Figure 19(c) [261,262,265]. These devices benefit from low-loss optical modes (quality factor up to $\sim 10^8$), but the large size of the devices mean that the EO interaction strength is comparatively weak. Thin-film LN offers relatively low loss ($Q$-factor up to $\sim 10^7$ [11]) and smaller optical modes, which can allow for a stronger EO interaction. Recently, cavity EO transducers in thin-film LN, shown in Figure 19(d) and (e) have demonstrated microwave-to-optical photon conversion efficiency as high as $2 \times 10^{-5}$ [178,179]. More recently, by mitigating photorefractive effect and using a much higher pump power, on-chip conversion efficiency of 1% has been achieved [270]. While this still falls well below the limit required for direct quantum transduction, near-unity efficiency is expected to be feasible for optimized devices. We note that several other material platforms have also been investigated for cavity EO devices [190,264].

*3.7. Synthetic dimension based on electro-optic modulation*

Synthetic dimension allows the construction of high-dimensional spaces beyond the apparent geometric dimensions. It is of particular interest to study, for example, topological physics. In photonic systems, spectral synthetic dimension can be controlled under temporal modulation, in which different frequency modes form a synthetic lattice in frequency domain, and time modulation provides coupling among the lattice points—a frequency crystal. Realizing such photonic synthetic dimensions requires the ability to simultaneously modulate a large number of optical cavities or waveguides. LNOI is a suitable platform to investigate this field and push the experimental realization forward.

The basic concept of the synthetic frequency dimensions is to use dynamic modulation of the optical index to couple optical modes with different frequencies [271]. A dynamic index modulation can create a time-dependent permittivity $\varepsilon(r,t) = \varepsilon_0(r) + \delta\varepsilon(r,t)$, where $\varepsilon_0$ and $\delta\varepsilon$ are the static permittivity and modulation-induced permittivity change, respectively. If a harmonic modulation is applied to a ring cavity or an optical waveguide (i.e., $\delta\varepsilon(r,t) = \delta(r)\cos(\omega t + \varphi)$), such a system can be described by the equation of motion $i\frac{d}{dt}a_n = ge^{i\varphi}a_{n-1} + ge^{i\varphi}a_{n+1}$, where $a_n$ represents the field amplitude at different frequencies $E = \sum_n a_n E_n e^{i\omega_m t}$ and $g$ is a coupling strength that relates to the strength of modulation [271]. Basically, time-dependent modulation couples different frequency modes, allowing for using different optical frequencies as lattice points of synthetic dimensions (Figure 20(a)).

Various theoretical work has been conducted on the topic of forming synthetic dimensions through time-modulation, including Bloch oscillation in synthetic dimension [273], photonic Weyl point [274], realizing photonic gauge field in a synthetic dimension [275], generating high-dimension in frequency space by folding the one-dimensional frequency crystal [276], combining the synthetic dimension of frequency with physical dimensions [277] or with other synthetic dimensions [278]. With the rapid progress on theory, experimental demonstrations of such synthetic dimensions have been recently explored in optical fiber systems. The band structure of a one-dimensional synthetic frequency crystal was theoretically investigated and experimentally probed [279]. Moreover, a combination of frequency dimension with the degree of freedom of clockwise-counterclockwise modes were also realized in the fiber system, which led to an observation of a wide variety of physics [280].

Integrated photonics on LNOI provides a powerful platform for experimental realizations of frequency crystals. The theoretical investigation and experimental realization of high-dimensional frequency crystals with dimensions up to four were recently demonstrated (Figure 20(b)) [272]. The density of states from one- to four- dimension were formulated and probed for such frequency crystals (Figure 20(c)). Coherent random walks (quantum random walks with non-interacting photons) were also experimentally measured in one- and two- dimensional crystals.

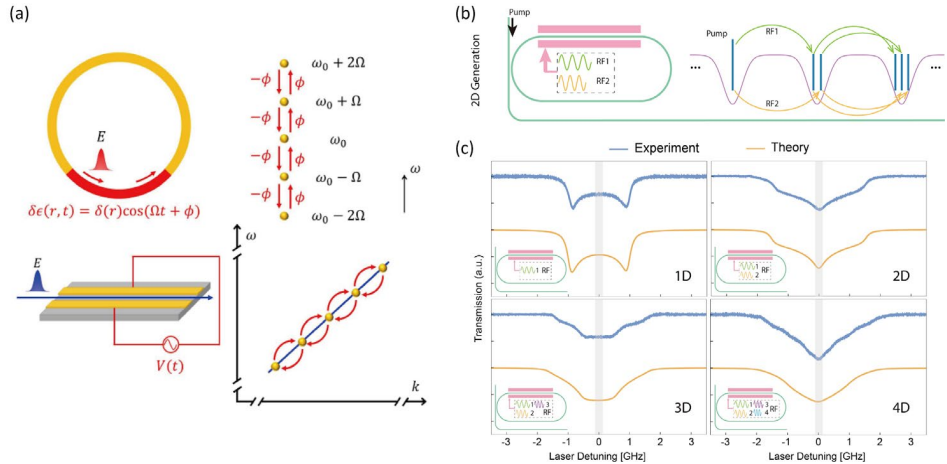

Figure 20: Realization of synthetic dimension [271] and high dimensional frequency crystals [272] based on electro-optic modulation. (a) A synthetic dimension can be achieved through EO modulation in a ring resonator or a waveguide. Different frequency modes are coupled by EO modulation and form a tight-binding lattice. (b) A concept of realizing high-dimensional frequency crystals in a single resonator. (c) Measurements of the density of the states of the frequency crystals with dimensionality ranging from one dimension to four dimensions. (a), Reproduced from [271], ©2018 Optical Society of America; (b), Reproduced from [272], © 2020 Optical Society of America.

In addition to the investigation of basic properties of the frequency crystal, thin-film LN is also a promising platform to explore topological photonics, a rapidly emerging field. The concept of topological insulator originates from condensed matter physics, in which a bulk insulator can have a conducting state localized on its edge. Such edge states are robust to perturbation and defects due to the non-trivial topology of the band structure of the material. This concept has been transferred to photonics and resulted in various new phenomena including photonic topological edge states, topological lasing, etc. There have also been several theoretical studies related to topological photonics in synthetic crystals [281–289]. We believe the integrated thin-film LN platform is well suited for experimental implementation of these proposals.

## 4. All-optical nonlinearity

LN possesses both second and third order nonlinearities and has a wide transparency window from ultraviolet to mid-infrared. The significant reduction of linear propagation loss (both at visible and near-infrared) and absence of multiphoton absorption (for wavelengths above 800 nm) are of interest for ultra-broadband nonlinear processes even at very modest pump powers. In this section, we review recent demonstration of harmonic generation, sum frequency generation, different frequency generation, optical parametric generation, and four wave mixing. Another unique potential offered by integrated LN photonic platform arises from the interplay between $\chi^{(2)}$ and $\chi^{(3)}$-based process, that can result in large effective third order nonlinearity (due to cascaded second-order nonlinearity), self-referencing, dispersive wave tailoring, and soliton compression.

At the heart of all nonlinear optical interactions is the phase-matching condition. On the thin-film LN photonic platform, significant progresses have been made on engineering group velocity, group velocity dispersion, and modal phase matching at wide range of wavelengths. More importantly, LN's ferroelectric property allows domain engineering (poling), which enables quasi-phase matching. Combined with powerful dispersion engineering and cavity integration, nanophotonic periodically poled LN (PPLN) could ultimately become a platform where a new branch of nonlinear science and applications will be born. We will cover dispersion engineering and phase matching in thin-film LN devices, and highlight several emerging applications in both classical and quantum fields.

## 4.1. Dispersion engineering

Precise engineering of both the group velocity (GV) and group velocity dispersion (GVD) is a crucial step for utilizing the nonlinear material properties of LN. In contrast to traditional LN waveguides (Figure 21(a)), a thin-film platform allows for the effective index of a waveguide mode to be tuned carefully by changing the waveguide geometry. Here, we discuss the effects of various device parameters on both GV and GVD engineering. We also provide context for the importance of dispersion engineering and how it can be used for various nonlinear photonic applications.

The GVD (or $\beta_2$) of a material at a given frequency $\omega_0$ can be written as $\beta_2 = \left(\frac{\partial^2 k}{\partial \omega^2}\right)_{\omega=\omega_0}$, where $k = \frac{2\pi}{\lambda} \cdot n_{\text{eff}}$ is a wavevector [290]. Dispersion curves can be generated by simulating the effective mode index ($n_{\text{eff}}$) of transverse-electric (TE) and transverse-magnetic (TM) modes of LN waveguides and numerically calculating the second derivative. Dispersion engineering on thin-film LN was first demonstrated in high-Q microring resonators with flexible GVD control [291] and has since been utilized on the platform for such applications as Kerr frequency comb generation [28] and supercontinuum generation [65].

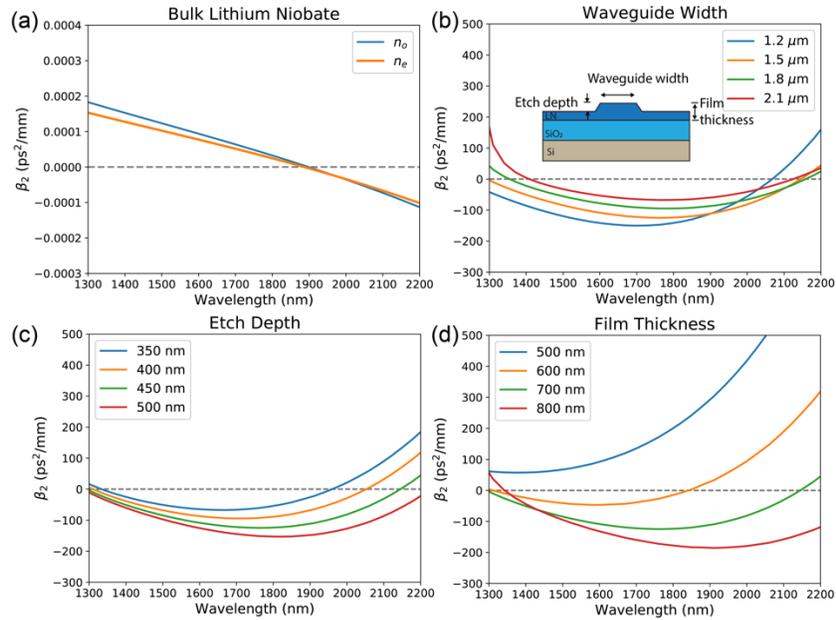

Figure 21: Illustration of geometric control of the waveguide dispersion. We compare the TE mode GVD for thin-film LN waveguides with different widths (b), etch depths (b), and initial film thicknesses (d) against LN's intrinsic material dispersion (a). Unless indicated otherwise, the simulation is run with a 700 nm thick, uncladded X-cut LN with etch angle of 60 degrees, 1.5 µm top width, and etch depth of 450 nm.

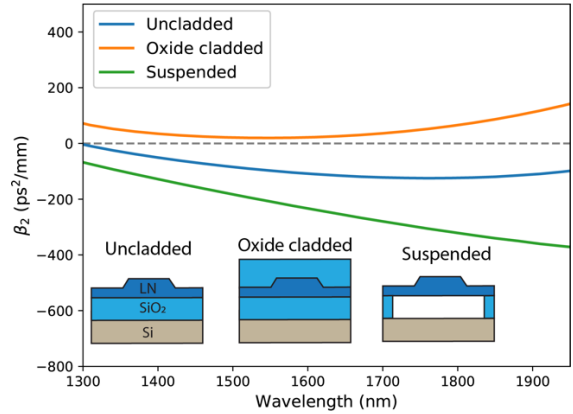

Figure 22: The effect of different cladding configurations on the TE mode dispersion of TFLN waveguides, including uncladded, 1 μm $SiO_2$ top cladding, and suspended/free standing structure. The LN waveguide geometry is the same as in the case of Figure 21.

We used finite-difference eigenmode (FDE) solver and numerical derivation to simulate dispersion curves for thin-film LN waveguides. The Sellmeier equations for both undoped and 5% magnesium oxide (MgO) doped congruently grown bulk LN were used [292]. We evaluate the effect of different geometric parameters, including waveguide width, etch depth, initial LN film thickness, and the etch angle on the dispersion curve. In Figure 21 (b)-(c) we show the simulated $\beta_2$ for the fundamental TE waveguide modes of an undoped X-cut thin-film LN waveguide on a $SiO_2$ substrate with different geometric variations, and compare it to that of the ordinary and extraordinary crystal axes of bulk undoped LN. it can be seen that in the case of thin-film LN, the GVD curve takes on a larger range of $\beta_2$ value than for bulk crystal. Additionally, the dispersion can be carefully tailored with small variations in the geometry, which is invaluable in applications where near-zero (GVD ~ 0) dispersion or anomalous dispersion (GVD < 0) is desired.

Beyond these changes to the waveguide geometry, other configurations of thin-film LN can be explored. For example, an oxide layer is often deposited on top of the LN waveguide as cladding, either to pull the optical mode further into the core of the waveguide or to make the structures more robust for additional fabrication procedures and post-process handling. The buried oxide layer beneath the LN film can also be removed to produce a suspended LN structure. These free-standing structures are important to leverage the piezoelectric properties of LN discussed later. The effect of the cladding on dispersion is summarized in Figure 22, which shows the dispersion curve of the TE mode in the case of a suspended, air cladded, and oxide cladded structure for the same LN geometry.

Group velocity dispersion is essential for various nonlinear processes. Anomalous GVD (GVD <0) is often required to achieve positive parametric gain via four-wave mixing for ultrabroadband optical spectral generation in light of a positive Kerr nonlinearity. The sign of GVD is engineered via tuning the waveguide dimensions at the operational wavelength. Some of the most well-known examples are Kerr microcombs and supercontinuum generation. LN material is also attractive for Pockels (electro-optic) microcombs where solitons can form at both fundamental frequency and its second harmonic frequency via the $\chi^{(2)}$ effect or $\chi^{(2)}$ and $\chi^{(3)}$ interactions. The combination of the GVD at both frequencies reveals different nonlinear dynamics as modelled in ref. [293]. It is also known that periodically poled LN (PPLN) can be used to generate an effective Kerr nonlinearity, the sign of which can be manipulated by choosing a positive/negative phase mismatch condition [294]. Therefore, thin-film PPLN platform offers several important knobs to engineer nonlinear optical interactions.

In addition to the second order derivative of the effective refractive index (related to GVD), the first order derivative (related to the group velocity) along with higher order terms can also be engineered to achieve certain targeted nonlinear features. For example, zero group velocity mismatch between the fundamental and SH wavelengths is ideal for broadband second harmonic generation. In addition, minimal temporal walk-off is also critical to achieve large effective $\chi^{(3)}$ and high conversion efficiency in cascaded $\chi^{(2)}$-based SCG via pulse pumping (Figure 23). Higher order dispersion has been intensively studied in integrated waveguides for modulation instability, soliton self-frequency shift, pulse generation and dispersive wave (DW) generation. For example, DW generation can be used to coherently enhance the spectral emission which is far away from the input light frequency. The position of the DW can be controlled via engineering the dispersion operator which consists of all the higher order dispersion terms (Figure 24). This has been utilized to generate visible and mid-infrared light via supercontinuum generation pumped by a mode-locked laser at the near infrared [65,296,297].

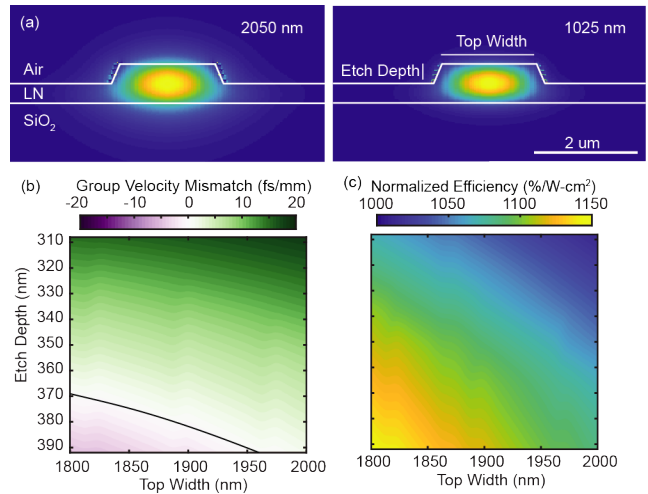

Figure 23: Group velocity engineering for second harmonic generation (SHG). (a) Waveguide cross-section and optical mode profiles at the fundamental (left) and second harmonic (right) wavelength. (b) Simulated group velocity mismatch as a function of waveguide top widths and etch depths. (c) Simulated normalized SHG efficiency. Reproduced from [295], © 2020 Optical Society of America.

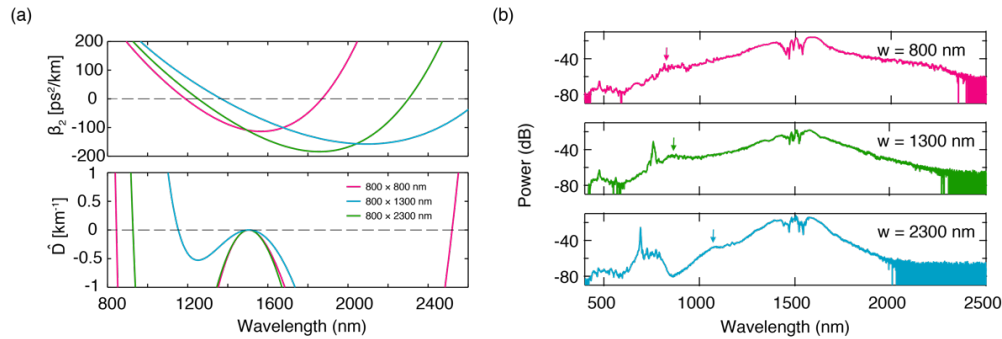

Figure 24: Engineering of higher order dispersion for dispersive wave generation. (a) Group velocity dispersion and dispersion operator of air-clad LN rib waveguides of three different top widths. (b) Dispersive wave radiation via supercontinuum generation in LN waveguides with corresponding top widths in (a). The position of the dispersive wave (indicated by the arrows) can be predicted at near zero crossings of the dispersion operator. Reproduced from [65], © 2019 Optical Society of America.

*4.2. Phase matching*

In $\chi^{(2)}$ mediated three wave mixing processes, momentum is generally not conserved for the three frequencies involved due to material and waveguide dispersion. The momentum mismatch, also known as the phase mismatch, for three wave mixing is $\Delta k = k_1 + k_2 - k_3$, where $k_i = n_{\text{eff}}(\omega_i)\,\omega_i/c$ is the propagation constant of the mode at frequency $\omega_i$, $n_{\text{eff}}(\omega_i)$ is the effective index of that mode, and $c$ is the speed of light. The phase matching condition, $\Delta k = 0$, must be met in order to achieve efficient energy transfer from the pump wavelength to the target wavelength(s) as light propagates through some length, $L$, of the nonlinear medium. When the phase matching condition is not met, energy will oscillate between the pump and target wavelength(s) with period $\Lambda = \frac{2\pi}{|\Delta k|}$, such that energy transferred to the target wavelength(s) is at most that accumulated over a propagation distance of $\Lambda/2$, regardless of the length of the nonlinear medium. Thus, in order to achieve efficient frequency conversion by taking advantage of the full propagation length through the nonlinear medium, one must engineer conditions under which phase matching is satisfied. Below are several ways in which phase matching can be achieved.

### 4.2.1. Birefringent phase matching

At the heart of the phase mismatch problem is that in the typical dispersion regime of $\frac{dn_{\text{eff}}}{d\omega} > 0$, $n_{\text{eff}}(\omega_3) > n_{\text{eff}}(\omega_2) > n_{\text{eff}}(\omega_1)$ for $\omega_3 > \omega_2 > \omega_1$ if the light at all three frequencies follow the same dispersion curve. In a birefringent material such as LN, this occurs when the light at all three frequencies occupies the same polarization and spatial modes. In birefringent phase matching, $\Delta k = 0$ can be achieved by having one or more of the modes involved occupy a different polarization.

Since LN is a negative uniaxial material, the material index for light polarized along the ordinary axis at a particular frequency is always higher than that polarized along the extraordinary axis. Thus, the inequality $n_{\text{eff}}(\omega_3) > n_{\text{eff}}(\omega_2) > n_{\text{eff}}(\omega_1)$ can be broken by having the the one or both of $\omega_1$ and $\omega_2$ be polarized along the ordinary axis.

The polarization configuration in which both $\omega_1$ and $\omega_2$ are polarized along the ordinary axis while $\omega_3$ is polarized along the extraordinary axis is typically known as "type-I", whereas the configuration in which $\omega_1$ is polarized along the ordinary axis while $\omega_2$ and $\omega_3$ are polarized along the extraordinary axis is known as "type-II". Both configurations are commonly used to phase match $\chi^{(2)}$ frequency conversion in bulk crystals, and detailed treatments of how these are implemented in practice can be found in most textbooks on nonlinear optics, such as Ref. [290].

Birefringent phase matching is rarely used in LN integrated photonics, because it precludes frequency conversion mediated by the largest component of the $\chi^{(2)}$ tensor, $\chi^{(2)}_{zzz}$, which is more than five times the second largest component of the $\chi^{(2)}$ tensor element. For interactions that make use of $\chi^{(2)}_{zzz}$, light at all of the three frequencies involved must be polarized along the extraordinary axis, i.e., "type-0" polarization configuration, which can only be phase matched using other methods.

### 4.2.2 Intermodal phase matching

One of the ways in which frequency conversion in the type-0 polarization configuration can be phase matched is through intermodal phase matching, in which the inequality $n_{\text{eff}}(\omega_3) > n_{\text{eff}}(\omega_2) > n_{\text{eff}}(\omega_1)$ is broken by having one of the modes at a different spatial profile, rather than polarization.

Intermodal phase matching is typically achieved by having light at $\omega_3$ occupy a higher order spatial mode, which will have a lower $n_{\text{eff}}$ than the fundamental mode due to a weaker confinement. Dispersion engineering via varying waveguide dimensions is employed to achieve intermodal phase matching for a specific set of frequencies. Intermodal phase matched second harmonic generation (SHG) in the type-0 polarization configuration has been demonstrated in both etched LN waveguides [117] and resonators [173] (see Figure 25). Because of symmetry, the overlap integral of even and odd modes will approach 0. As a result, fundamental and third-order modes are often used, such as that used in Figure 25. This restriction can be removed by breaking the symmetry of the nonlinear material (e.g., replacing half of waveguides with a different material) [298]. Additionally, intermodal phase matching in other polarization configurations (e.g. type-I) has also been demonstrated [299].

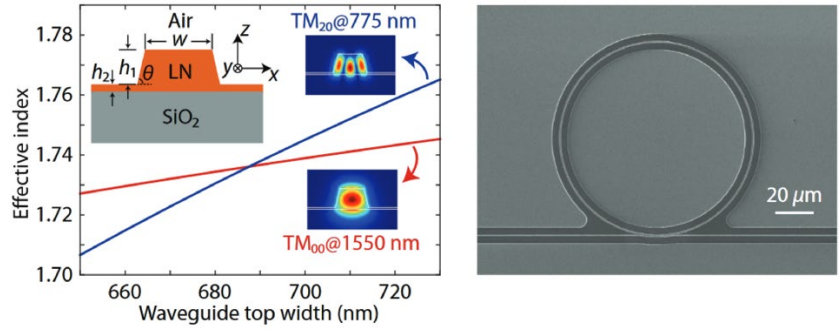

Figure 25: Intermodal phase matching in a microring resonator between the fundamental $TM_{00}$ spatial mode at the fundamental frequency and $TM_{20}$ higher order spatial mode at the second harmonic. Reproduced from [173], © 2019 American Physical Society.

4.2.3. Quasi-phase matching

While intermodal phase matching can enhance frequency conversion efficiency by allowing the use of the largest $\chi^{(2)}$ tensor component, it incurs a significant penalty from lower mode overlap between the higher order spatial mode(s) and the fundamental spatial mode(s) [31]. Thus, for optimal $\chi^{(2)}$ frequency conversion efficiency, two conditions must be satisfied: first, light at all three interacting frequencies must occupy the same polarization and spatial mode, and second, the process must be phase matched. However, these two conditions are in tension, since, as previously noted, phase matching requires that at least one of the interacting frequencies occupies a different spatial and/or polarization mode. One solution to this dilemma is to employ *quasi-phase matching* rather than to attempt true phase matching via either birefringent or modal phase matching.

In quasi-phase matching, $\Delta k \neq 0$ for the three-wave mixing process, allowing for the interacting frequencies to occupy the same polarization and spatial mode. However, the unidirectional/efficient power transfer between the pump and target wavelength(s) is recovered by inverting the sign of the relevant $\chi^{(2)}$ tensor component with period $\Lambda = \frac{2\pi}{|\Delta k|}$. Thus, the $\chi^{(2)}$ sign is inverted every $\Lambda/2$ of propagation distance, and prevents energy backflow into the pump. Equivalently, one can think of inverting $\chi^{(2)}$ sign as imparting an additional grating momentum $K = 2\pi/\Lambda$ that is equal and opposite to the intrinsic momentum mismatch $\Delta k$ to achieve momentum conservation.

***Periodic poling.*** Quasi-phase matching requires the periodic inversion of the sign of $\chi^{(2)}$ of the material. This can be achieved in ferroelectric materials, where the orientation of the ferroelectric domains (and thus $\chi^{(2)}$) can be inverted via the application of high voltage pulses in a process called periodic poling. Popular nonlinear optical materials that can be periodically poled include LN, lithium tantalate (LiTaO3), and potassium titanyl phosphate (KTP).

The techniques for fabricating bulk periodically poled lithium niobate (PPLN) are mature and have been commercialized. However, since bulk PPLN fabrication techniques rely on being able to pattern electrodes on the $\pm z$ faces of LN, techniques for PPLN fabrication on thin-film LNOI is somewhat more complicated, and have only matured in recent years.

The coercive field required for poling LN varies with material composition and doping. In general, congruently-grown undoped LN typically requires a higher coercive field for domain inversion [300–302], as compared to near-stoichiometric undoped LN [301]. Doping with MgO decreases the coercive field in both congruent and stoichiometric LN [300]. It has been reported that thin-film LN requires a higher coercive field than the corresponding bulk crystals [17,18,303,304]. It was thought that the increased coercive field could be caused by the bonding interface between the thin-film LN and buried $SiO_2$ [17], which hinders the domain inversion, or due to the $Li^+$ out-diffusion during the annealing process in LNOI fabrication [305], which reduces Li concentration and increases coercive field.

In Z-cut thin-film LNOI, there are two general methods for PPLN fabrication: poling before bonding LN to the oxide and substrate, and poling after bonding to the substrate. In the first case, bulk LN is poled with standard bulk LN poling techniques, and the poled wafer is then bonded onto the substrate and sliced using the same Smart-Cut technique for fabricating unpoled LNOI. This technique was used in one of the earliest demonstrations of thin-film PPLN [35] and several other demonstrations subsequently [169,306–308].

More recently, groups have demonstrated poling performed directly on thin-film Z-cut LN either by using an AFM tip to apply a very high local electric field for domain inversion [162], or by poling through the LN/oxide/handle wafer stack at an elevated temperature [168]. Feature sizes down to 100 nm has been demonstrated with the former method, while the latter method was used to fabricate poled microring structures for the demonstration of the highest second harmonic generation normalized conversion efficiencies to date.

In X-cut thin-film LNOI, PPLN fabrication was typically done with coplanar electrode gratings since electrodes cannot be accessed without etching the film. While developed initially for a hybrid platform with $SiN_x$ loaded waveguides on PPLN [17,18], the method has since been used to fabricate monolithic poled LN ridge waveguides. Poling in these cases is typically done before waveguide fabrication, though poling after waveguide fabrication has been demonstrated with $SiN_x$ loaded waveguides [309] and shallowly-etched LN ridges [309,310]. The latter case allows active monitoring of the poling process through the optical signal [310]. In-situ confocal second harmonic microscopy has also been applied in the former case to track domain growth over several pulses in close to real time [311]. Poling periods as short as 600 nm have been demonstrated on X-cut thin-film LNOI using coplanar electrodes [312].

Only linear periodically poled gratings are possible on X-cut films as the domain inversion is in the plane of the film, whereas in Z-cut films, the domain inverted regions can take arbitrary shapes (e.g. radial gratings) since the domain inversion occurs normal to the plane of the film.

There are three main methods to visualize poled domains: (1) selective etching, (2) piezoresponse force microscopy, and (3) second-harmonic microscopy.

Selective etching is a longstanding technique for verifying poled domain structures [313]. This method consists of etching with HF on exposed faces normal to the poling direction. Differently poled regions etch at different rates, resulting in a set of "hillocks" which indicate different domains and can be imaged conventionally. Figure 26(a) shows an example of selective etching of a poled waveguide [304], where a focused ion beam was used to mill a trench in the X-cut sample to expose the Z-faces, and an HF etch was applied. The left image shows an incompletely poled domains which do not reach the bottom of the LN film, while the right image shows completely poled domains. While this method does provide information on the three-dimensional structure of the poled domains (including depth into the crystal), its major drawbacks are its destructive nature and the amount of time required to expose the relevant faces, etch them, and then image them.

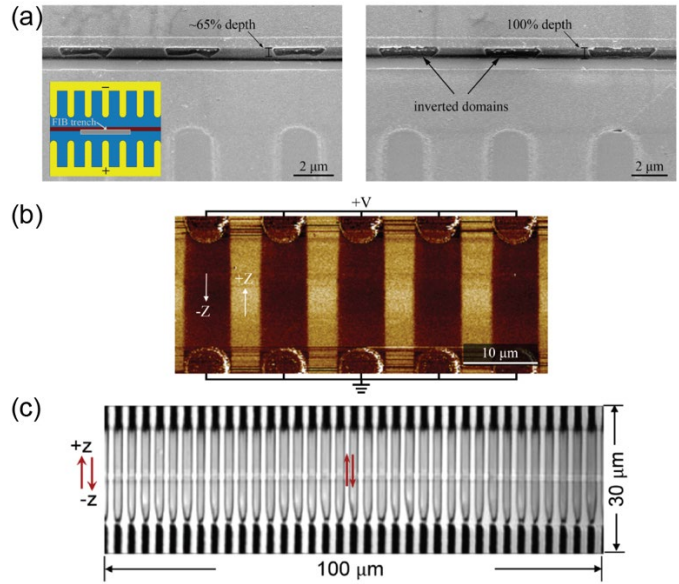

Figure 26: Characterization of poled domains in LN. (a) Selective etching [304], where focused ion beam was used to mill a trench in the X-cut waveguide to expose the Z-faces followed by HF etching. (b) Piezoresponse force microscopy [304]. This method can provide detailed information on the surface but without poling depth information. (c) Second-harmonic microscopy [314], where the domain boundaries appear as dark lines and the brightness of the poled region infers poling depth. The faint gray horizontal line across the image is a shallow-etched waveguide. (a)-(b) Reproduced from [304], © 2019 Optical Society of America under the terms of the OSA Open Access Publishing Agreement; (c) Reproduced from [314], © 2020 Optical Society of America under the terms of the OSA Open Access Publishing Agreement.

Piezoresponse force microscopy (PFM) is a non-invasive domain imaging technique. A probe is touched to the surface of a sample and an AC electrical field applied, inducing surface oscillations due to the converse piezoelectric effect [315]. Domains of different polarization have a different electromechanical response, providing the contrast mechanism. While this method gives high-resolution data on the shapes of poled domains at the surface of the material (see Figure 26(b)), it does not provide depth information, and image acquisition is relatively slow, on the order of tens of seconds.

Second-harmonic microscopy to evaluate poling quality has the advantage of speed while delivering adequate resolution. Second harmonic (SH) light from domains of opposite polarization is fully out of phase, resulting in destructive interference at the boundaries between domains and thus dark regions in the SH image. Due to resonance effects in the layer structure, the degree of contrast may depend on wavelength [316]. Figure 26(c) shows an example of a second-harmonic microscopy image of a periodically poled X-cut waveguide [314]. Detailed simulations of the SH signal given a geometry of poled domains have opened up the possibility of non-destructively inferring poling depth from SH microscope images, although the resolution is limited [34].

Indirect optical poling diagnostic methods have also been exploited by monitoring the output optical signal from the waveguides while poling. One way is to monitor phase modulation in the waveguide [309]. In principle, at 50% duty cycle, no phase modulation should be observed when an electric field is applied (through a separate set of electrodes than the poling electrodes) across the waveguide, along Z crystal axis. The average poling duty cycle can be inferred from the change in phase of the out-coupled light. This method can be utilized while poling and provides path-averaged information which can be complemented by examination of local poling quality using second-harmonic microscopy. Another in-situ monitoring technique is to inject CW telecom light into a waveguide and monitor the second harmonic signal at the out-coupled light as poling proceeds. This has been used to optimize poling schedules by e.g. poling and depoling a waveguide multiple times until the second harmonic signal was saturated [310].

### 4.2.4. Other phase-matching approaches

*Cyclic phase matching*. Cyclic phase matching (CPM) [317] is often used in ring or microdisk cavities to achieve type-I phase matching. When the crystallographic axis is not the symmetry axis, for example, on a X/Y-cut LN thin film, the effective nonlinearity and the refractive index of in-plane polarized light is changing as light propagates in the microring resonator ([317–319]). In a type-I second harmonic generation (SHG) process, there are four positions along the ring which the phase matching can be achieved. As compared to the type-I SHG when the c-axis is the symmetry axis, CPM can enable a much broader phase matching bandwidth (only certain discrete wavelengths would satisfy phase matching within this bandwidth) but at a sacrifice of conversion efficiency. In addition, the modulated effective nonlinearity is also needed to be included in the model, and the fact that SHG in different phase-matched regions can constructively or destructively interfere would lead to a wavelength-dependent gain profile [317]. SHG with a normalized conversion efficiency of 0.1%/mW is demonstrated in a double resonant X-cut LN whispering gallery microresonator [164], and CPM-based spontaneous parametric down conversion has been reported in a X-cut LN microdisk resonator [159]. A CPM-based process generally suffers from a lower conversion efficiency as compared to a QPM-based one given the same resonance condition.

It is also possible to achieve type-0 effective quasi-phase matching in such a cyclic fashion. For example, all participating waves can be in quasi-TE modes inside an X-cut microdisk [151]. As they circulate around the microdisk, they experience an effective nonlinear coefficient with a sign inversion every half cycle, mimicking quasi-phase matching in PPLN. At the same time, the effective indices for all participating waves vary at different azimuthal angles. Similar to cyclic phase matching, this produces a wide range of $\Delta k$ and results in a broader phase-matching bandwidth but with sacrificed efficiency as compared to standard QPM in PPLN.

*Metasurface-assisted phase-matching-free SHG.* Dielectric metasurface can be used to achieve phase-matching-free SHG [320]. This method utilizes silicon nanoantennas fabricated on the top surface of a LN waveguide to create a unidirectional phase gradient. When the pump light at $\omega_0$ in the fundamental TE mode is converted to its SH at $2\omega_0$ in the fundamental TE mode, the gradient metasurface couples the SH into its higher-order spatial modes. The unidirectional effective wavevector created by the dielectric metasurface prevents these SH signals from converting back to their fundamental TE modes, and the small mode overlap between these higher-order SH modes and fundamental pump modes prevents them from down-converting to $\omega$. Therefore, even in the absence of phase matching, the power transfer from the pump to SH is unidirectional, which results in broadband enhancement of SHG efficiency. A normalized conversion efficiency as high as 1,660% $W^{-1}cm^{-2}$ has been achieved at telecom wavelengths. The drawback of this approach is that collecting the higher-order SH modes is generally inefficient.

*Periodically grooved quasi-phase matching.* Besides periodic poling, periodically perturbing the waveguide width in the form of grooves can also achieve quasi-phase matching [117,321]. This change of the waveguide width induces periodic modulation in both the effective index and nonlinear coefficient along the propagation direction, providing a momentum "kick" that compensates for the phase mismatch. The added grooves, however, causes Bloch scattering loss. In general, deeper grooves result in higher nonlinear conversion efficiency due to tighter confinement, but cause larger Bloch loss at the same time.

### 4.3. Second-order nonlinear wavelength conversion

Second-order nonlinearity ($\chi^{(2)}$) enables frequency generation and wavelength conversion based on sum and difference frequency generation (SFG, DFG), second-harmonic generation (SHG), as well as optical parametric oscillation (OPO). Moreover, it can also be used to generate nonclassical states of light through spontaneous parametric down conversion (SPDC). These operations have been accomplished in bulk LN crystal or microscale waveguides, with much based on periodic poling. Here, we review recent demonstrations in thin-film LN waveguides and microcavities, especially with periodic poling, which are continuously pushing the nonlinear efficiency to higher records.

#### 4.3.1. Traveling-wave processes

The efficiency of the aforementioned parametric frequency conversion processes is greatly enhanced in nanophotonic PPLN waveguides compared to in bulk PPLN waveguides. This is because the normalized conversion efficiency of, for instance, SHG, scales as $\frac{P_2}{P_1^2} \propto (\frac{\zeta}{(A_1\sqrt{A_2})^{1/3}})^2$ where $\zeta$ is the mode overlap between the fundamental frequency and second harmonic modes, and $A_1$ and $A_2$ are the mode areas of the fundamental frequency and second harmonic, respectively; details on calculating $\zeta$, $A_1$, and $A_2$ can be found in Ref. [299]. Since mode areas $A_1$ and $A_2$ can be more than an order of magnitude smaller in nanophotonic PPLN waveguides compared to bulk PPLN, the conversion efficiency can be much higher in nanophotonic waveguides if all other parameters are held equal. Thus, nanophotonic PPLN waveguides allow one to achieve conversion efficiencies comparable to those of bulk PPLN with a much shorter PPLN section. This is desirable not only for its compactness, but also because the phase matching bandwidth will be broader, as the phase matching bandwidth is inversely proportional to interaction length.

Since the absolute (quasi-)phase matched SHG conversion efficiency in waveguides scales as $P_2 / P_1 \propto P_1/L^2$, where $P_2$ is the power of the generated second harmonic signal, $P_1$ is the power of the input fundamental frequency signal, and $L$ is the length of the PPLN waveguide, the figure of merit for (quasi-)phase matched SHG in waveguides is typically given in terms of the normalized conversion efficiency $\eta = \frac{P_2}{P_1^2 L^2}$, which has units of units of %/W·cm$^2$.

To date, ultrahigh normalized conversion efficiencies have been achieved in both SiN$_x$ loaded waveguides and monolithic LN ridge waveguides via type-0 quasi-phase matching. In the former case, high normalized conversion efficiency of 1160%/W·cm$^2$ was achieved in a 5 mm SiN$_x$ loaded waveguide on PPLN through engineering the waveguide cross section to limit leakage of the fundamental TE mode to TM slab modes [322]. In the latter case, ultra-high normalized conversion efficiency of 4600%/W·cm$^2$ was achieved in a 300 μm long LN ridge waveguide through the use of active monitoring of the second harmonic signal to achieve optimal poling [310]. Longer PPLN ridge waveguides up to 4 mm have also been demonstrated [29,32], albeit with lower normalized conversion efficiencies (<3000%/W·cm$^2$) due to non-optimal poling duty cycle (i.e. <50% duty cycle) and non-uniform poling. Figure 27 shows some typical poling setups for X-cut LNOI PPLN waveguides.

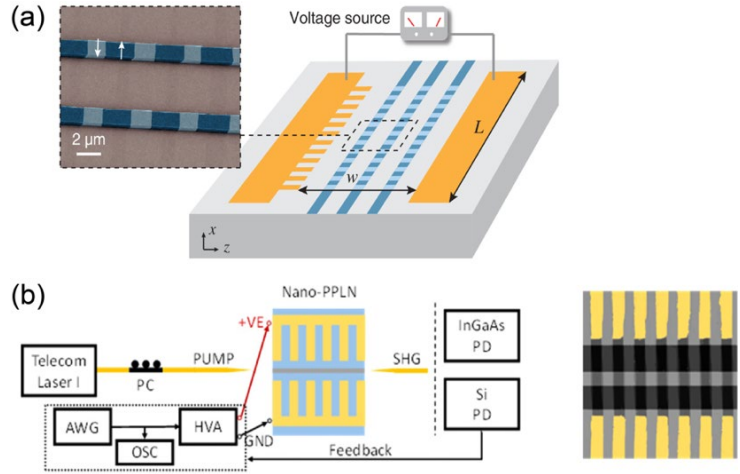

Figure 27: PPLN waveguide frequency converters. (a) Poling schematic for X-cut LN film before waveguide etching, and false color SEM of resultant etched waveguides with duty cycle of ~(39 ± 3)%. (b) Schematic for actively monitored poling of etched LN nanophotonic waveguide and false color AFM image of resultant 50% domain duty cycle. Reprinted with permission. (a) Reproduced from [29], © 2018 Optical Society of America; (b) reproduced from [310], © 2019 Optical Society of America.

Additionally, $\chi^{(2)}$ frequency conversion has also been demonstrated in unpoled LN ridge waveguides with type-I modal phase matching [299]. Such a configuration sacrifices conversion efficiency as it uses a smaller component of the $\chi^{(2)}$ tensor, and has lower mode overlap $\zeta$, but has a higher thermal tuning range than phase matching done in the type-0 polarization configuration. Improvements to conversion efficiency while retaining broad thermal tunability can likely be achieved by using type-I quasi-phase matching rather than type-I modal phase matching for higher mode overlap.

Table 5 summarizes some representative works on traveling-wave based second harmonic generation in thin-film LN waveguides.

Table 5. Representative works on traveling-wave based second-harmonic generation in thin-film LN.

| Device type | Crystal cut | Phase matching | Normalized conversion efficiency (%/W cm$^2$) | Year | Ref. |
|---|---|---|---|---|---|
| SiNx loaded waveguide | X | Type-0 QPM | 160 | 2016 | [17] |
| SiNx loaded waveguide | X | Type-0 QPM | 1,160 | 2019 | [322] |
| Etched waveguide | X | Metasurface assisted | 0.4 | 2017 | [320] |
| Etched waveguide | Z | Type-I intermodal | 22.2 | 2018 | [299] |
| Etched waveguide | X | Type-0 intermodal | 41 | 2017 | [117] |
| Etched waveguide | X | Type-0 (periodically grooved) | 6.8 | 2017 | [117] |
| Etched waveguide | X | Type-0 QPM | 2,200 | 2019 | [32] |
| Etched waveguide | X | Type-0 QPM | 2,600 | 2018 | [29] |
| Etched waveguide | X | Type-0 QPM | 4,600 | 2019 | [310] |

### 4.3.2. Resonant-enhanced processes

Optical cavities can strongly enhance these parametric conversion processes given the same optical pump power. Taking SHG in a ring resonator as an example, to achieve optimal conversion efficiency, the following conditions need to be satisfied: (1) Energy conservation and multi-resonance: $\omega_2 = 2\omega_1$ with both frequencies on resonance, where the subscripts 1 and 2 indicate fundamental and second harmonics; (2) momentum conservation: $m_2 = 2m_1$, where $m_{1/2}$ are the azimuthal mode numbers; (3) large effective second-order nonlinear susceptibility; (4) large mode overlap inside the nonlinear material; (5) small effective mode area; (6) small cavity length; (7) high intrinsic quality factors; and (8) critical coupling. Detailed derivation and mathematical formulas can be found in Refs. [173,323]. Ideally, one would prefer all participating waves to be in their fundamental transverse modes, polarized along the Z crystal axis to utilize $d_{33}(\chi_{zzz})$ and tightly confined in the resonator with maximum modal overlap. However, material dispersion often compromises phase matching. This problem can be alleviated by quasi-phase matching through periodical poling, where the momentum conservation requirement is relaxed to $m_2 = 2m_1 + M$ ($M = 2\pi R/\Lambda$ is the domain period number with $R$ being the ring radius and $\Lambda$ being the poling period). Most importantly, the key advantage of resonance-enhanced $\chi^{(2)}$ wavelength conversion relies on its capability of circulating the optical waves. In doubly resonant and critically coupled conditions, the conversion efficiency enhancement can scale as $\sim Q_1^2 Q_2$.

In microring resonators, the dispersion can be engineered similarly to that in waveguides by adjusting the waveguide width, etch depth and film thickness as well as crystal cut. Without domain engineering, modal phase matching is often adopted [172,173]. For example, in Z-cut microrings, the fundamental wave is usually at fundamental quasi-TM (TM$_{00}$) mode, and the second harmonics can be at higher order quasi-TM (TM$_{20}$) mode. This conversion thus utilizes the largest nonlinear coefficient $d_{33}$. Note that mode symmetries need to be conserved to ensure non-zero overlap. With this approach, Luo et al. demonstrated telecom to visible SHG with a conversion efficiency of 1.5%/mW as well as DFG at telecom wavelengths [173]. It is worth mentioning that temperature can also be used to effectively fine tune the phase matching [172].

Recent progress on PPLN microrings and racetracks have brought SHG conversion efficiency to an unprecedented level. In 2019, Chen et al. demonstrated a periodically poled racetrack resonator on X-cut MgO-doped LN thin films [30] (see Figure 28(a)). The poling is applied on one of the 300 μm straight waveguide sections. This configuration allows all waves to be in their fundamental quasi-TE modes and utilizes the largest nonlinear coefficient $d_{33}$. Quasi-phase-matched SHG with conversion efficiency of 230%/mW is achieved, corresponding to an efficiency of 10$^{-6}$ per single photon. The equivalent single-photon efficiency is calculated as $\eta_s = \eta E_{ph}/c$, where $E_{ph}$ is the pump photon energy and $\tau$ is the cavity lifetime. In the same year, Lu et al. demonstrated radially poled, Z-cut microring [31] (see Figure 28(b)). A single pulley waveguide was designed to efficiently couple both the infrared and visible frequency bands. Due to the fabrication limit on the poling period, TE$_{00}$-to-TM$_{00}$ SHG was adopted, utilizing the $d_{31}$ coefficient. Nonetheless, an ultra-high on-chip SHG efficiency of 250%/mW was achieved. More recently, the same group further improved the Z-cut PPLN microring and realized a poling period of 2.95 μm [168]. This allowed for QPM SHG with all waves in their fundamental quasi-TM modes to utilize $d_{33}$. As a result, record high SHG efficiency of 5,000%/mW was achieved in the low-power regime, reaching a single-photon nonlinearity approaching 1%.

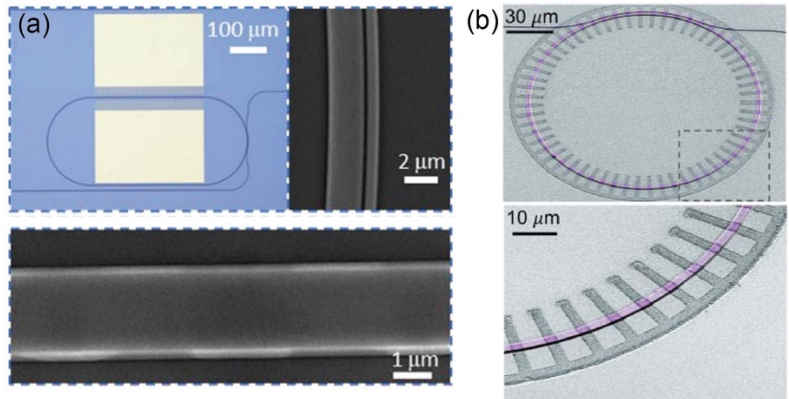

Figure 28: PPLN racetrack and microrings. (a) PPLN racetrack resonator in X-cut LNOI, where the poling was applied along the straight section of the racetrack. (b) Radially poled Z-cut microring resonator. (a), (b) Reproduced from [30], [31], © 2019 Optical Society of America.

Besides racetrack and microring cavities, on-chip whispering gallery resonators (including microdisks and wide ring-shaped resonators) and photonic crystal cavities have also been studied extensively for $\chi^{(2)}$ nonlinear wavelength conversion.

In high-Q microdisks, various $\chi^{(2)}$-nonlinear processes have been demonstrated, including second harmonic generation [141,153,158], sum-frequency generation [152,160], and spontaneous parametric down conversion [159]. Furthermore, effective third-order nonlinear phenomena have been observed as a result of cascaded $\chi^{(2)}$ processes. For example, third-harmonic generation has been demonstrated through cascaded SHG and SFG [150,151]; effective four-wave mixing ($\chi^{(3)}$ process) has been demonstrated as a result of cascaded SHG and DFG [161].

Microdisks support abundant resonance modes, differing by their polarization states, as well as azimuthal and radial mode numbers. It is thus convenient to satisfy the multi-resonance condition, where the wavelengths involved in the wave-mixing processes coincide with cavity resonances. However, efficient phase matching is nontrivial. Most $\chi^{(2)}$ processes demonstrated in Z-cut LN microdisks rely on modal phase matching [141,150,161], where the largest nonlinear tensor element $d_{33}$ are not used. In X-cut LN microdisks, cyclic phase matching is commonly involved [153,159,164], where the fundamentals are quasi-TE modes and second harmonic is at a quasi-TM mode. An effective quasi-phase matching can also be achieved in X-cut microdisks when all waves are at quasi-TE modes, interacting through $d_{33}$ but with an oscillating sign. Under this scheme, SHG with an efficiency of 9.9%/mW and THG with an efficiency of 1.05%/mW$^2$ were measured [151]. Recently, periodic poling has also been applied to Z-cut LN microdisks [162]. With a bottom layer acting as a poling electrode, piezoresponse force microscopy was used to achieve a poling period of ~2 μm. Using the piezoresponse force microscopy method, the same group further demonstrated dual-period poling with unit domain size as small as 90 nm in width, and observed SHG, THG, and even FHG [163]. However, the conversion efficiency remains moderate (0.05%/mW) due to limited Q and non-optimal mode overlap and confinement.

In addition, in high-Q ($3\times10^6$ at 976 nm), large (200 μm diameter) and wide (7 μm) ring-shaped whispering gallery resonators, a mixture of nonlinear processes were observed, including SHG of the pump light, cascaded stimulated Raman scattering, SHG of the Raman signal, and SFG between the pump and Raman signal [94]. Periodic poling was also applied to this Z-cut on-chip whispering gallery resonators, which showed improved SHG efficiency of 0.09%/mW [169], limited by the nonideal poling configuration and external coupling.

Photonic crystals support high quality factor (~$10^5$) while maintaining ultra-small mode volume ~ $(\lambda/n)^3$. They strongly enhance optical interactions and are ideal for nonlinear photonic applications. The first SHG in LN photonic crystals was demonstrated in a free-standing L3 LN photonic crystals fabricated from bulk LN substrate [180]. Subsequently, LNOI wafers were utilized to realize both 1D [181] and 2D photonic crystals [143] and to demonstrate SHG as well as SFG. So far, the conversion efficiency remains low (current record of 0.078%/W [143]), since the fundamentals and harmonics (or sum-frequency) are not on resonance simultaneously. In fact, it has been a big challenge to achieve multiple resonances simultaneously in such a different wavelength regime (i.e., IR and visible). Recent progresses on doubly resonance photonic crystals cavities may solve this problem [324], and lead to ultra-efficient nonlinear conversions or even single-photon nonlinearity. Another challenge is that photonic crystals are more susceptible to photorefractive effects due to the strong local field intensity. This problem will limit its use in classical applications but may be less prominent in the quantum regime, where only single or few photons are in the cavity.

In Table 6, we summarize recent works on cavity-based SHG, including the cavity type, crystal cut, phase-matching type, and conversion efficiency.

Table 6. Representative works on cavity-based SHG in thin-film LN.

| Device type | LN crystal cut | Phase matching | Conversion efficiency (%/W) | Year | Ref. |
|---|---|---|---|---|---|
| Microdisk | Z | Intermodal | 0.28 | 2017 | [150] |
| Microdisk | Z | Intermodal | 10.9 | 2014 | [141] |
| Microdisk | X | Cyclic | 0.36 | 2017 | [159] |
| Microdisk | X | Cyclic | 2.3 | 2018 | [153] |
| Microdisk | X | Cyclic | 110.6 | 2016 | [164] |
| Microdisk | X | Cyclic | 990 | 2019 | [151] |
| Ring-shaped WGM resonator | Z | Type-0 QPM | 0.9 | 2018 | [169] |
| Microdisk | Z | Type-0 QPM | 1.44 | 2020 | [162] |
| Microdisk | Z | Type -0 QPM | 51 | 2020 | [163] |
| Racetrack | X | Type-0 QPM | 230,000 | 2019 | [30] |
| Microring | Z | Type-I QPM | 250,000 | 2019 | [31] |
| Microdisk | Z | - | 1.35 | 2015 | [158] |
| Microring | Z | Type-0 QPM | 5,000,000 | 2020 | [168] |
| Photonic crystal L3 cavity | X | - | 0.012 | 2016 | [180] |
| 1D photonic crystal | X | - | 0.0004 | 2018 | [181] |
| 2D photonic crystal | X | - | 0.078 | 2019 | [143] |

4.3.3. Spontaneous parametric down conversion

In addition to demonstrations of classical $\chi^{(2)}$ wavelength conversion processes, spontaneous parametric downconversion (SPDC), an exclusively quantum process in which a pair of photons is generated from a single pump photon, has also been demonstrated on the thin-film LN platform in microdisks [159], PPLN waveguides [32–34,325] and PPLN ring [326]. High coincidence-to-accidental ratio (CAR) of 668 ± 7 has been achieved at a pair generation rate of 12 MHz in Ref. [34] in a 5 mm PPLN waveguide on X-cut LNOI. Photons generated via SPDC in PPLN waveguides generally exhibit strong energy-time entanglement, which can be leveraged to produce entangled photon pairs in multiple wavelength channels to enable frequency domain multiplexing; this has been demonstrated in a submillimeter PPLN waveguide with signal-idler pairs generated over a coarse wavelength division multiplexing grid separated by more than 120 nm [33]. Entangled photon pairs generated via SPDC can also be used as sources of heralded single photons, though the spectral purity of the heralded photon has yet to be demonstrated on thin-film LN [34]. For biphoton spectra featuring strong energy-time entanglement, it is expected that pulsed pumping and narrow band filtering at the cost of pair generation rate will be needed to generate spectrally pure heralded single photons. Alternatively, as demonstrated theoretically [327], spectrally pure heralded single photons can be achieved by dispersion engineering the joint spectral amplitude to be separable without filtering.

*4.4. Cascaded second-order nonlinearity*

In addition to the intrinsic cubic third order nonlinearity, phase-mismatched second order processes can induce an effective Kerr nonlinearity due to back-action of the second harmonic (SH) wave at $2\omega$ on the fundamental wave (FW) at $\omega$. It originates from the occurrence of two successive second order nonlinear processes: upconversion to SH at $2\omega$ via second harmonic generation (SHG) and then downconversion back to the FW at $\omega$, which results in a nonlinear phase shift proportional to $[\chi^{(2)}]^2$ on the FW [328]. In the presence of phase mismatch $\Delta k$ between the FW and SH, this effective third order nonlinearity $\chi_{\text{eff}}^{(3)}$ is approximately proportional to $-[\chi^{(2)}]^2/\Delta k$. More importantly, in lithium niobate, $\chi_{\text{eff}}^{(3)}$ can be tuned to either a positive or negative value by engineering the quasi-phase matching (QPM) condition in periodically poled lithium niobate waveguides ($\Delta k = k_{\text{SH}} - 2k_{\text{FW}} - k_{\text{QPM}}$). Soliton generation might be possible in the visible spectral range, where the normal GVD of the material is typically too large to be overcome by the waveguide dispersion on integrated platforms. In addition, at near phase matching condition, a giant $\chi^{(3)}$, which equals $\chi_{\text{Kerr}}^{(3)} + \chi_{\text{eff}}^{(3)}$, can be achieved to significantly reduce the pump power for third-order nonlinear frequency conversion, such as four wave mixing [150,329]. Jankowski *et al.* reported an achieved $\chi_{\text{eff}}^{(3)}$ ~200 times higher than the $\chi_{\text{Kerr}}^{(3)}$ of LN at 2050 nm using an integrated PPLN waveguide [295]. The nonlinear parameter $\gamma \sim \chi^{(3)}/A_{\text{eff}}$, where $A_{\text{eff}}$ is the effective optical mode area, defines the strength of a Kerr nonlinear interaction. The recent advancement on the thin-film LN nano-waveguides offers high optical confinement over the wavelength scale (~μm), leading to the $A_{\text{eff}}$ more than an order of magnitude smaller than the mode size of weakly confined reverse proton exchange (RPE) LN waveguides. Therefore, thin-film-LN-based cascaded second-order process opens up exciting opportunities, including realizing dynamical processes typically associated with $\chi^{(3)}$ nonlinearities using extremely low pump powers. Recent demonstration [295] reports a multi-octave spanning supercontinuum spectrum with a few pJ pulse energy.

This capability of tailoring the effective nonlinearity offers incredible flexibility for controlling nonlinear dynamics when combined with dispersion engineering of integrated photonics. For example, group velocity mismatching (GVM, $\Delta k'$) or the temporal walk-off between the FW and SH is the dominant factor of the SHG bandwidth ($\Delta\lambda \sim 1/|\Delta k'|L$, where $L$ is the PPLN waveguide length). Realization of cascaded second order processes in a group velocity matched regime ($\Delta k' \sim 0$) would tremendously boost both Kerr-based spectral broadening and the conversion bandwidth of SH light. Ref. [295] has shown efficient QPM SHG with less than 100 fJ pulse energy, and a 3-dB SHG bandwidth > 110 nm, an order of magnitude broader than using conventional PPLN technologies (Figure 29). In addition, there are still a rich group of nonlinear dynamics which can be explored in dispersion engineered PPLN waveguides, such as large nonlinear phase shift, laser mode-locking, soliton compression, spatial soliton waves, Cherenkov radiation, nonlinear mirror, GVM-induced Raman radiation, Kerr microcombs, and more [328,330–337]. Almost all the nonlinear interactions will benefit from the extra degree of freedom in control of GVM, PM, GVD at both FW and SH, and even higher order dispersion terms. For example, cascaded phase shifts through interaction with femtosecond pulses can be operated under both the stationary and nonstationary regime [330], which is determined by the GVM characteristic length, SHG coherence length, and the SH GVD length [338]. These regimes have been intensively studied for a cascaded quadratic soliton compressor [330,332] and frequency shifting using controllable Raman-like nonlinearities [337]. Further study of the many dynamical processes that can be accessed with engineering of the waveguide dispersion in conjunction with quasi-phase-matching is the subject of ongoing research.

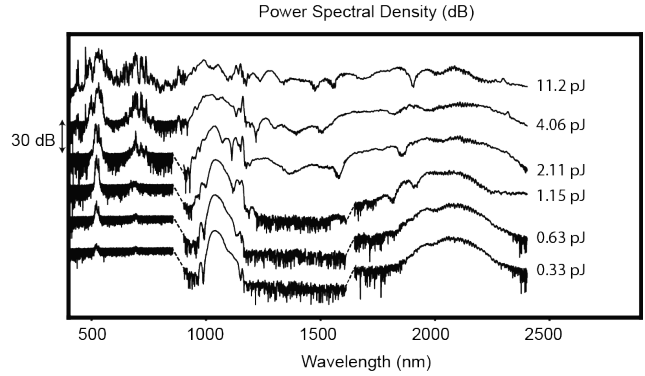

Figure 29: Supercontinuum spectra at various pulse energies, assisted by the cascaded second-order nonlinearity in a PPLN waveguide. Reproduced from [295], © 2020 Optical Society of America.

*4.5. Kerr comb*

Optical frequency combs have a wide range of applications, such as precision spectroscopy, frequency metrology, and optical communications. While they are traditionally produced through the pulse train emitted by a mode-locked laser, microresonator frequency comb generation, wherein the comb teeth are formed at the resonator mode frequencies via nonlinear processes in the cavity, is quickly maturing as a field. Most common is Kerr frequency comb generation, which makes use of the $\chi^{(3)}$ optical nonlinearity and high circulating optical power to generate comb lines through four-wave mixing (FWM). Locking to Kerr solitary waves (solitons), which balance third order nonlinearity and dispersion, as well as cavity loss and gain [339], can often produce octave-spanning combs for *f*-2*f* referencing.

Kerr frequency comb generation holds the promise of a compact, low footprint comb source for a wide range of applications, and when utilized carefully, the Kerr process can produce highly coherent low-noise spectra. Thin-film LN is of particular interest for chip-based frequency comb generation due to its capability of hosting an entire photonic circuit on a single chip (Figure 30(a)). Included in the photonic circuit are various components, such as electro-optic resonance tuning and on-chip self-referencing, that leverage LN's $\chi^{(2)}$ nonlinearity and EO property. These material properties usually are missing from other material platforms used for frequency comb generation, leading to the need for bulky off-chip components for full functionality of the frequency comb.

Kerr combs spanning 700 nm were demonstrated on the LNOI platform on X-cut lithium niobate wafers [28]. The waveguide geometry was dispersion engineered for anomalous dispersion in both the TE and TM waveguide modes, and frequency combs were successfully demonstrated on both. For the crystal cut used in this work, the TE mode was favorable, as the light is polarized along the Z-axis (polar axis) of the material, allowing for utilization of the large $r_{33}$ component of LN. However, we see in the TE comb spectrum (Figure 30(b)) many peaks from the strong Raman effect in this polarization, corresponding to scattering off of various phonon branches. The Raman effect is strongest when the light is polarized along the crystal's Z-axis [175] and competes with the four wave mixing process used to generate the comb teeth.

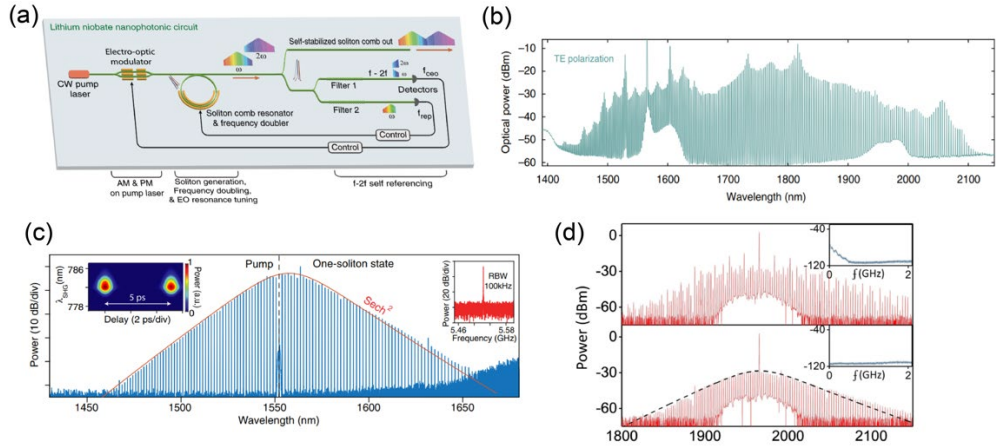

Figure 30: (a) Conceptual schematic of a full LN nanophotonic circuit for Kerr comb generation on chip, including soliton pulse formation, resonator tuning, frequency doubling, and f-2f self-referencing [27]. (b) Raman-Kerr combs demonstrated on X-cut LN [28] (c) Soliton comb on Z-cut LN at 1.5 μm [27] (d) Soliton Kerr combs on Z-cut LN generated at 2 μm in order to mitigate strong Raman gain [174]. (a), (c), (d) Reproduced from [27], [174] © 2019 Optical Society of America; (b) reproduced from [28], © The Author(s) 2019.

The combination of optimized $\chi^{(2)}$ performance and maximized Raman gain in the same polarization has caused a conundrum in functional Kerr comb generation, as the strong Raman effect can prevent soliton modelocking. To mitigate the interference of Raman, frequency combs can be generated along the non-polar axis of the material or around a longer optical wavelength where Raman effect is known to be weaker, or with a larger free spectral range (FSR) to avoid overlapping with the highest Raman gain wavelength. The former two solutions have been used to achieve stable soliton combs in the TE mode of Z-cut LN [27,174]. Furthermore, a soliton microcomb spanning > 200 nm at 1.5 μm (Figure 30 (c)) > 300 nm at 2 μm have been achieved by avoiding Raman scattering effects (Figure 30 (d)).

There are a number of challenges unique to Kerr frequency combs on the LNOI platform that must be considered before octave-spanning combs can be realized. Most importantly, a means of mitigating or bypassing the strong Raman effect along the crystal polar axis to achieve soliton formation remains to be demonstrated. Additionally, at high optical powers required for Kerr comb generation, the photorefractive and thermo-optic effects in LN act in opposition to each other, requiring careful consideration to the tuning procedures for locking onto the soliton state. Recently, 4/5 octave-spanning Kerr combs have been reported in a-Z-cut LN microresonator with a 335 GHz line spacing [340]. It is achieved by engineering the waveguide-ring coupling to raise the Raman oscillation threshold [341] and utilizing a relatively large FSR. There are also recent studies on controlling the photorefractive effect via the annealing technique and the cladding removal, which could help stabilize the cavity resonance at a high optical power [342].

### 4.6. Raman lasing

Raman spectroscopy is one of the common tools for investigating the structural information of lithium niobate through analysis of its vibrational transitions. It is also commonly used to obtain the information of bulk LN regarding crystallinity, surface quality, doping, strain, etc. [343–347] Raman scattering has various configurations depending on the propagation directions and polarization of both the pump and stokes light with respect the crystal axis of LN [348–351]. There are two branches of Raman-active phonon modes, which are A symmetry and E symmetry. The A mode is polarized along the Z crystal axis, and the E mode polarized in the X-Y plane [352]. There are both longitude and transverse modes in A and E symmetries. In addition, there exist both forward and backward scattering processes, and forward Raman scattering corresponds to a polariton effect where photon and phonon are strongly mixed and leads to electromagnetic radiation [353].

While Raman spectroscopy is an indispensable tool for material study, Raman oscillation plays an important role in nonlinear dynamics in LN crystals, and has been observed in LN microdisks and whispering gallery resonators [153,354]. A detailed study of Raman oscillations is reported in monolithic X-cut LN microring resonators and racetrack cavities pumped by a CW laser [175]. The dominant stimulated Raman scattering is found at a counter-propagating direction with respect to the pump, and several scattering configurations are explored (Figure 31(a)). The strongest Raman gain is found when the polarization of the pump is parallel to the crystal axis, where both the electro-optic (EO) and second order nonlinear effects are also the largest. This potentially leads to a strong energy transfer from the pump to the Stokes light if high optical power is used for other parametric oscillation such as second harmonic generation or in EO-based LN devices. Figure 31(b) shows the competition between four wave mixing and Raman oscillations in a CW pumped LN microresonator [174].

We note that there exist discrepancies of Raman oscillation strength among the work in thin-film LN devices [27,174,175]. The variation in the position and linewidth of Raman peaks could be attributed to the difference in fabrication processes of the thin-film LN-based devices, such as annealing [355], reactive ion etching, electron beam lithography, the Li/Nb concentration ratio of the wafer etc. The study of stimulated Raman scattering in the thin-film LN waveguides can be of importance to building LN-based nonlinear photonic circuits.

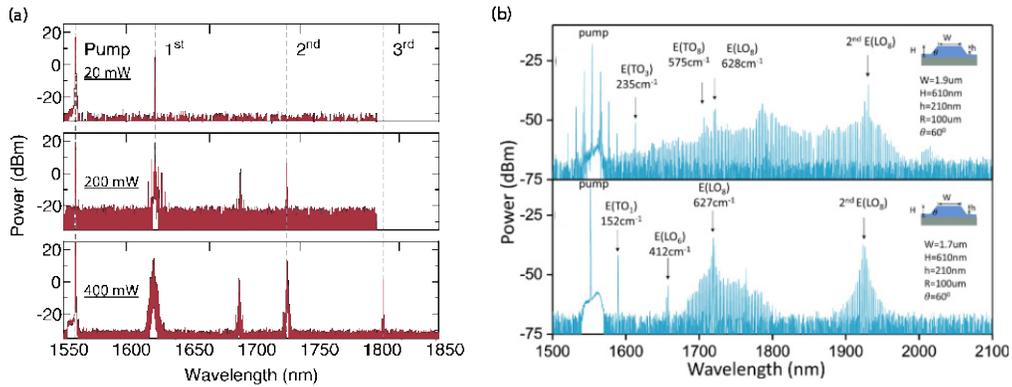

Figure 31: Raman oscillations in LN microresonators. (a) Three Raman spectra at different pump powers of 20, 200, and 400 mW [175]. The polarization of both the Stokes and pump light are along the LN polar Z-axis. The Raman oscillation is found to be dominant in the backward propagating direction with respect to the pump. (b) Optical spectrum of Raman–Kerr combs generated in a LN microresonator, where the corresponding phonon modes are labelled [174]. No evidence of soliton formation is observed while scanning the pump laser across the resonance. (a) Reproduced from [175], © The Authors(s) 2020, licensed under a Creative Commons Attribution 4.0 International License; (b) reproduced from [174], © 2019 Optical Society of America.

### 4.7. Supercontinuum generation

Supercontinuum generation is a nonlinear optical interaction where an input optical pulse undergoes enormous spectral broadening in a nonlinear waveguide due to various nonlinear processes including self-phase modulation, cross-phase modulation, four-wave mixing, and dispersive wave generation [356]. Over the past decade, there has been tremendous development of integrated photonic platforms that enable supercontinuum with moderate pump powers [357–365]. In particular, the integrated platform is well suited for the generation of a coherent broadband optical spectra, as the high material nonlinearity along with the tight optical confinement allows for efficient supercontinuum generation with sufficiently short pulses using a short waveguide length [366–370]. The generation of a coherent, octave-spanning spectrum is particularly critical for applications in time and frequency metrology and optical clocks where a fully-stabilized frequency comb source is required. Many of these comb sources rely on the stabilization of the repetition rate and the carrier-envelope offset of a mode-locked laser [371]. The carrier-envelope offset frequency ($f_{CEO}$) can be stabilized by using a self-referencing technique based on *f-2f* interferometry which requires the generation of a coherent, octave-spanning spectrum [372–374]. While the on-chip supercontinuum technology has become mature, the self-referencing has been performed using bulk optics. In recent years, there have been demonstrations of $f_{CEO}$ detection using on-chip *f-2f* interferometry which leverages simultaneous supercontinuum and second-harmonic generation in a single waveguide [375–377].

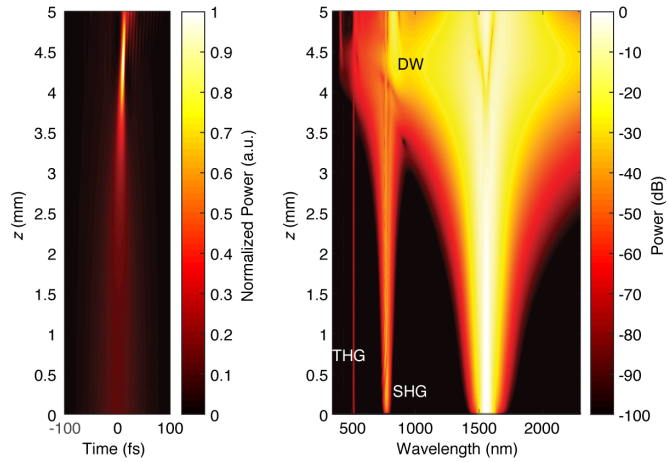

Figure 32: Numerical modeling of pulse propagation dynamics in a 0.5-cm-long LN waveguide. Plot shows temporal (left) and spectral (right) evolution. The waveguide cross section is 800 x 1250 nm. DW, dispersive wave; SHG, second harmonic generation; THG, third harmonic generation. Reproduced from [296], © 2020 Optical Society of America.

The LN platform allows for on-chip *f-2f* interferometry due to the large intrinsic $\chi^{(2)}$ and $\chi^{(3)}$ nonlinearities. Moreover, through dispersion engineering of the waveguide a strong second harmonic signal along with a broadband coherent supercontinuum can be simultaneously generated [65,378]. Coherent two-octave-spanning supercontinuum generation and $f_{CEO}$ detection through the spectral overlap of the second harmonic component with the generated supercontinuum spectrum was achieved in a single LN waveguide [65]. Numerical modeling of the pulse propagation [296] dynamics in the LN waveguide showed that the system can be modeled using a reduced scalar nonlinear envelope equation with both quartic and cubic nonlinearities [379–381]

$$\left[\frac{\partial}{\partial z} - i \sum_{n \geq 2} \frac{\beta_n}{n!} \left(i \frac{\partial}{\partial t}\right)^n + \frac{\alpha}{2}\right] E(z,t) = i \frac{\omega_0}{2 n_0 c \varepsilon_0} \left(1 + i \tau_{sh} \frac{\partial}{\partial \tau}\right) P_{NL}(z,t) \qquad (6)$$

where $\beta_n$ corresponds to the n$^{th}$ order dispersion coefficient, $\alpha$ is the linear propagation loss, $\omega_0$ corresponds to the pump frequency, $\tau_{sh} = 1/\omega_0 - \partial[ln(n(\omega))]/\partial\omega|_{\omega=\omega_0}$ is the optical shock time, and $P_{NL} = \varepsilon_0[\chi^{(2)}E^2 + \chi^{(3)}E^3]$ is the total nonlinear polarization with contribution from only non-negative frequencies, and $\tau$ is the local time in the moving frame. This equation enables modeling of the dynamics in LN waveguides spanning multi-octave bandwidths through the combined $\chi^{(2)}$ and $\chi^{(3)}$ nonlinearities, taking into account the birefringent nature of LN in solving effective refractive index and nonlinearity [382]. Figure 32 shows the simulated pulse propagation dynamics in a 0.5-cm-long dispersion-engineered LN waveguide based on 800-nm-thick X-cut LN thin film [296]. Here, the pump was a 90 fs pulse train centered at 1560 nm with a pulse energy of 107 pJ. As the pulse propagates in the waveguide, the second harmonic and third harmonic components were immediately generated. The main pulse undergoes spectral broadening through self-phase modulation. For $z > 4$ cm, a dispersive wave component is generated near 860 nm which continues to blue shift due to phase matching and approaches the second harmonic component. Dispersive wave generation is a phase-matched nonlinear process where energy from the pump is transferred across a zero-GVD point, and is commonly used to extend the overall generated spectral bandwidth [383–385]. The position of the dispersive wave can be approximated by considering the zero-crossing of the dispersion operator $\widehat{D} = \sum_{n \geq 2} \frac{\beta_n(\omega_0)}{n!}(\omega - \omega_0)^2$. An integrated LN $f$-$2f$ interferometer is demonstrated for stabilizing the $f_{CEO}$ of a mode-locked fiber laser [296]. The waveguide output was directly sent to a silicon avalanche photodiode for $f_{CEO}$ detection and a feedback loop was implemented for stabilization. Figure 33 shows (a) the generated supercontinuum spectrum along with (b) the single-sideband phase noise of the measured $f_{CEO}$ beatnote. This approach drastically simplified the conventional approach for f-2f interferometry, which consists of a bulk periodically-poled lithium niobate (PPLN) with a temperature controller and a delay line for compensation of the temporal walk-off between the $f$ and $2f$ spectral components, and offers promise towards the realization of a fully-integrated frequency comb source.

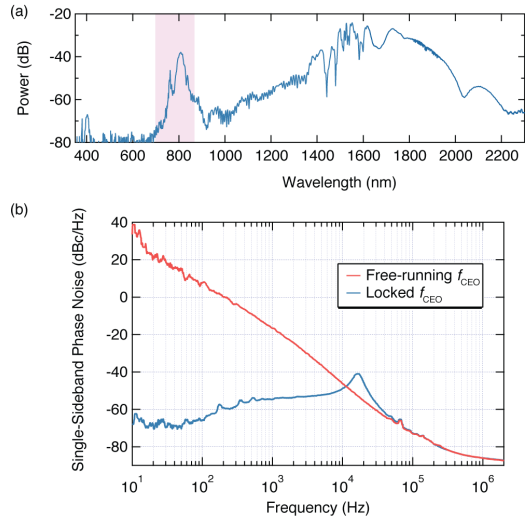

Figure 33: On-chip $f$-$2f$ interferometer. (a) Experimental supercontinuum spectrum. The region shaded in pink is used for $f$-$2f$ interferometry. (b) Single-sideband phase noise of the $f_{CEO}$ for free-running (red) and locked (blue) states. Reproduced from [296], © 2020 Optical Society of America.

Furthermore, the presence of both $\chi^{(2)}$ and $\chi^{(3)}$ nonlinearities allows for the extension of the spectral bandwidth even further through a combination of supercontinuum via dispersive wave generation, harmonic generation, and difference frequency generation (DFG), where an ultrabroadband supercontinuum generation spanning 0.35 to 4.1 μm with 240 pJ pulse energy [297]. The spectral components in the visible were generated through SHG and sum-frequency generation, along with cascaded SHG which allows for spectral generation down to the UV regime. In addition, the spectrum extends into the mid-infrared (mid-IR) regime via DFG. The spectral generation of UV to the mid-IR in a single LN device offers promise towards the development of an integrated ultrabroadband source for spectroscopy applications.

Lastly, there has been a demonstration of a novel approach for broadband SCG in an integrated PPLN waveguide via cascaded nonlinearities [295]. This approach utilizes quasi-phase matched $\chi^{(2)}$ interactions to enable broadband SHG with 3-dB bandwidths >110 nm and effective self-phase modulation with nonlinearities that are 200 times larger than the intrinsic $\chi^{(3)}$ nonlinearity in LN. Such large nonlinearities allow for the generation of a supercontinuum spectrum spanning 2.5 octaves with pulse energies as low as 10 pJ. Further details are given in Section 4.4.

## 5. Piezo-optomechanics

Optomechanics [386] provides an efficient interface between light and elastic waves, via the photoelastic and moving boundary interactions. The recent development of piezo-optomechanics further links elastic waves and microwaves on thin film piezoelectric materials. Such integrated piezo-optomechanical devices utilize electrically-driven acoustic waves to deflect, modulate, and frequency control the light and demonstrate superior performance than conventional bulk acousto-optic components.

An ideal material for piezo-optomechanics would possess strong piezoelectricity, photoelasticity, as well as low mechanical and optical losses. Commonly used piezoelectric materials include ZnO, AlN, GaAs, and LN, which are non-centrosymmetric materials. While all these materials have similar photoelasticity, and have been used to demonstrate high-Q optical cavities, LN shows the highest piezoelectric coefficient.

Other piezoelectric materials with stronger piezoelectricity and/or photoelasticity, such as PZT and $TeO_2$, could further improve the efficiency of piezo-optomechanical devices, should high quality thin film with low optical loss be available. $TeO_2$ is widely used in bulk acousto-optic modulators (AOMs) for its excellent properties for acousto-optics, but nanofabrication of $TeO_2$ is challenging and needs further development [387,388].

Heterostructures, which use one material for piezoelectric coupling and another material for optimized optomechanical coupling, have also been explored for piezo-optomechanics. Acoustic waves generated on LN have been used to measure the elasto-optic coefficients of the arsenic trisulfide thin film overlaying on the LN [389]. Piezo-optomechanical coupling between superconducting qubits and optical cavities using AlN on Si has been achieved [390], benefiting from the well-developed silicon optomechanics platform.

Heterogeneous integration of LN thin film could also benefit piezo-optomechanics, and this is discussed further in Section 5.2. For example, taking advantage of the high acoustic velocity in diamond (19 km/s in bulk diamond, vs 8 km/s in bulk LN), LN thin film on diamond could achieve mechanical frequencies of >40 GHz. Piezoelectrically generated acoustic waves in LN could efficiently drive the spin qubits of color centers in diamond. Point defects in SiC have also been explored for acoustic wave-spin interactions [391].

### 5.1. Piezoelectricity and photoelasticity

Since the piezo-optomechanical coupling strength depends on both the piezoelectric and photoelastic tensors, the design of piezo-optomechanical devices is challenging. Control and manipulation of elastic waves in thin-film LN could be even more complicated. Fortunately, LN surface acoustic wave (SAW) technology has been developed for several decades for its use in mobile communication systems [392], which provides valuable experience and insight into the design of piezo-optomechanical devices.

Due to the anisotropy of LN, SAW devices use a variety of crystal orientations such as X, Z, 42°-Y, 128°-Y. While optical fields in LN with X or Y polarization have the same ordinary refractive index, the acoustic phase velocities in the X, Y, and Z directions are all different. The choice of crystal orientation significantly changes the performance of piezo-optomechanical devices. While most published work selected the crystal orientation of LN thin film based on its commercial availability, the optimal device orientations might be difference for piezoelectric and optomechanical couplings, and thus piezo-optomechanical devices can have two parts with different orientations and then be connected using an acoustic waveguide [170]. This suggests that there is room for further investigation, and a comprehensive study of LN crystal orientations for piezo-optomechanical devices would be highly valuable to the community.

Determined by its point group of 3m, the piezoelectric coefficient tensor $d_{ij}$ of LN is given by

$$\begin{bmatrix} 0 & 0 & 0 & 0 & d_{15} & -2d_{22} \\ -d_{22} & d_{22} & 0 & d_{15} & 0 & 0 \\ d_{31} & d_{31} & d_{33} & 0 & 0 & 0 \end{bmatrix}, \qquad (7)$$

and the photoelastic tensor is given by,

$$\begin{bmatrix} p_{11} & p_{12} & p_{13} & p_{14} & 0 & 0 \\ p_{12} & p_{11} & p_{13} & -p_{14} & 0 & 0 \\ p_{31} & p_{31} & p_{33} & 0 & 0 & 0 \\ p_{41} & -p_{41} & 0 & p_{44} & 0 & 0 \\ 0 & 0 & 0 & 0 & p_{44} & p_{41} \\ 0 & 0 & 0 & 0 & p_{14} & p_{66} \end{bmatrix}. \qquad (8)$$

The experimentally measured values are included in Table 1.

Since the piezoelectricity of LN is strong, the secondary electro-optic effect, which is caused by the electric field generated by the elastic wave via piezoelectricity, plays a significant role in the overall coupling from microwaves to light. The contribution of this secondary electro-optic effect can be more than one third of the overall coupling.

### 5.2. Microwave acoustic modes

Several approaches based on LN have been pursued to implement piezo-optomechanical systems. General considerations include efficient piezoelectric excitation of acoustics waves, large optomechanical coupling, and low acoustic wave propagation loss. Figure 34 illustrates various acoustic waves used in LN piezo-optomechanical systems.

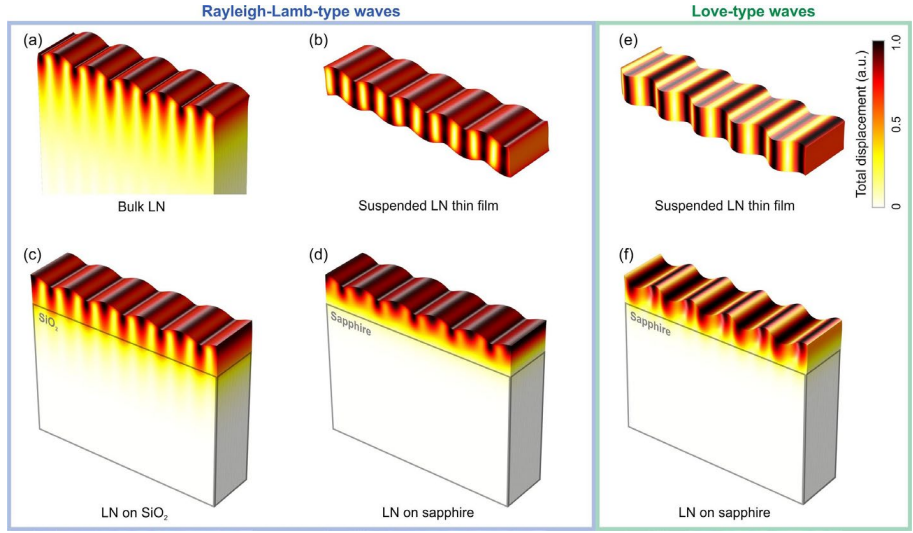

Figure 34: Total displacement profiles of typical acoustic waves used in LN piezo-optomechanics. There are (a-d) Rayleigh-Lamb type and (e)(f) Love-type shear waves on (a) bulk LN substrate, (b)(e) suspended LN thin film, (c) LN on SiO$_2$, and (d)(f) LN on sapphire.

### 5.2.1 Surface acoustic wave on bulk LN

SAWs on bulk LN substrates are low-loss propagating elastic waves confined to the surface of LN, which can be efficiently generated by interdigital transducers (IDTs). The propagation loss is typically 1-7 dB/mm at GHz frequencies at room temperature, and about one order of magnitude lower at cryogenic temperatures [393,394]. However, bulk LN substrates cannot confine optical modes. Thus, Ti indiffused layers, deposited higher optical index materials, or material-on-oxide layers are needed to guide the optical light on the chip.

An acousto-optic frequency shifter has been built employing Ti indiffused optical waveguides on bulk LN substrate [395]. Since there is almost no acoustic contrast between the region with and without Ti indiffusion, the optical waveguides defined by Ti indiffusion do not reflect SAW or scatter SAW into bulk acoustic waves. This is important for building integrated acousto-optic frequency shifters or non-reciprocal optical devices, in which the unwanted SAWs would generate other unwanted sidebands or degrade the isolation. However, the optical contrast formed by Ti indiffusion is also limited, and thus the optical mode size is large, for example, 10 μm. For high frequency applications, with the acoustic frequency in GHz range, the acoustic wavelength is <1 μm which results in reduced interaction between the optical and the acoustic waves, due to the limited mode overlap. This limits the efficiency of the acousto-optic devices.

To improve interaction between optical and acoustic waves, overlay layers on bulk LN can be used to guide the light, while the generation and propagation of the SAW are still mainly in the LN substrate. Such devices do not rely on LNOI wafers, and feature a more robust and compatible fabrication that avoids etching of LN. The optomechanical coupling happens either in the overlay optical waveguides, whose strain fields are driven by the SAW on LN substrate, or in the LN substrate by the evanescent optical fields.

This configuration provides an opportunity to achieve overall efficient piezo-optomechanical coupling by using materials with high photoelastic coefficients as top optical waveguides. Such structures have also been used to measure the photoelastic coefficients of the optical waveguide material [389]. However, it is challenging to form a high-Q acoustic resonators using this approach, since the top optical waveguide inevitably scatters the SAW into bulk acoustic waves, especially when the acoustic wavelengths are comparable to the size of optical waveguides.

### 5.2.2 Surface acoustic wave on LNOI

As both the optical and acoustic indices (inverse phase velocities) of LN are higher than the underneath silicon dioxide layer, thin-film LN on oxide can simultaneously confine optical and acoustic modes. Compared to suspended thin-film LN, LNOI can handle larger RF powers. For example, through similar IDT structures, up to ~100 mW microwave power is delivered to the suspended LN thin films, while more than 1 W microwave power is delivered when thin-film LN is laying on the oxide layer [396].

With optical waveguides defined by etching thin-film LN, waveguide-based acousto-optic modulators have been demonstrated (Figure 35(f)(g)) [398–400]. However, similar to the overlay waveguides on bulk LN substrate, the optical waveguide structure defined on the LN thin film can scatter the thin film acoustic waves, especially when the acoustic wavelength becomes comparable to the structure feature sizes (e.g. etch depth). Thus, such devices usually work at sub-GHz acoustic frequencies and have limited $Q$ and $fQ$ product for the acoustic resonance.

Ti-indiffused waveguide on thin-film LN could potentially be used to reduce the scattering of the acoustic waves while providing the optical waveguide structure. Alternatively, optical confinement in lateral dimension could be eliminated (while maintaining vertical confinement provided by thin-film LN), thus allowing the light to propagate as a 2D "free-space" beam (a slab-waveguide mode). Recently developed integrated LN acousto-optic frequency shifter (AOFS) [396,401] employ such "free-space" regions and achieve high purity frequency shifting.

LNOI also has some limitations for piezo-optomechanics. Since the acoustic contrast between LN and oxide is small, thin-film LN on oxide only confines a few acoustic wave modes, for example the Rayleigh-like modes. However, Rayleigh modes are not the most effective modes for optomechanical coupling due to their strain profile. The top of the thin film sees a tensile strain and the bottom sees a compressive strain, and the induced optical index changes would be partially cancelled out in the overlap integration with the optical mode. Also, oxide is not a good thermal conductor, and thus limits the stability of devices under high RF power inputs.

5.2.3 LN on Sapphire

Since the acoustic velocity of sapphire is much higher than LN, LN on sapphire supports more interesting acoustic modes than LN on oxide [394]. For example, X cut LN on sapphire supports shear waves (Love waves) propagating in the crystal Y direction. Unlike the aforementioned Rayleigh mode, shear modes create strain from top to bottom of the thin film with the same sign, and thus the induced optical index change adds constructively in the overlap with the optical mode. This particular shear acoustic mode, whose major strain component is $e_{YZ}$ (index 4 in the reduced representation), employs the high piezoelectric coefficient $d_{24} = d_{15}$, and high photoelastic coefficient $p_{14}$. It also features a higher acoustic velocity than the Rayleigh-like modes, and thus increases the acoustic frequency of the devices.

However, sapphire is a hard material and challenging for fabrication. Currently, the major limitation is the availability of high-quality LN on sapphire wafers. Bubbles formed between the thin-film LN and sapphire substrate during the bonding process significantly impact the usable area of LN on sapphire wafer, to less than 50%. However, we believe that these difficulties will be overcome in near future as more demand for high quality LN on sapphire wafers will drive the technological advancements. Molecular beam epitaxy growth of LN on sapphire has been demonstrated [402], and it opens another pathway for high quality LN on sapphire, although the current film quality is still not comparable with ones prepared by the Smart-Cut.

5.2.4 Suspended thin-film LN

As acoustic waves do not radiate into vacuum, suspended LN thin films are more flexible in design and could achieve higher acoustic frequency and larger Q factors. Such suspended structures can be fabricated using LNOI, in which the oxide is removed by a wet etch with HF or using HF vapor etching to preserve the mechanical integrity of the structure. Alternatively, LN on silicon could be used, where $XeF_2$ gas etching would be used to remove Si and release LN devices. $XeF_2$ gas etching does not etch LN and, being a dry process, does not suffer from surface tension issues that are common in liquid-phase approaches.

Suspended LN disks support both optical whispering gallery modes and mechanical modes (Figure 35(a)), and have been used to achieve optomechanical coupling [166], but at a relatively low mechanical frequency of 100s of MHz. Optomechanical crystals [142,170,182] have also been developed on suspended LN. Periodic structures provide bandgaps for both light and acoustic waves, while resonators are formed by introducing defect modes within the bandgap, for example, by introducing a perturbation to the periodic unit cell. Optomechanical crystals feature small optical and acoustic mode volumes and large overlaps between the two modes. However, piezoelectrically driving such small acoustic modes is challenging. It can be done using electrodes directly deposited on the optomechanical crystals [182] or by employing an inter-digitated transducer (IDT) on a suspended acoustic waveguide then guided to the crystals [170]. Due to the small size of electrodes, the demonstrated microwave-to-mechanical conversion efficiency is ~0.1%. Efforts have also been made to guide the acoustic wave from large efficient IDTs to small sized waveguides [403].

On the other hand, efficient microwave-to-mechanical conversion has been demonstrated on suspended LN slabs with optical waveguides [167]. The microwave return loss from the acoustic resonator can be less than -10 dB [404]. The optical waveguides do not cause acoustic losses in the suspended structures. Thus, devices can achieve higher acoustic frequencies (in the GHz range) and acoustic Q factors. Also, optical ring resonators defined by etched optical waveguides benefit from higher Q factors than photonic crystals. However, the optomechanical coupling rate is compromised due to the large mode volumes.

Some of the concerns for suspended thin-film LN are robustness and high RF/optical power handling. About 1 W optical light has been successfully coupled from lensed fibers to suspended optical waveguides, and only about 100 mW RF power has been applied to the suspended LN acoustic cavities. Higher powers could damage the suspended LN structures, probably due to the localized heating.

### 5.3. Applications of microwave acoustic modes

Various applications of acousto-optic devices have been discussed [405]. Recent developments in LN piezo-optomechanics have resulted in demonstrations of state-of-the-art devices (Figure 35) for microwave to optical conversion, optical frequency shifting, optical gyroscopes [399], and nonreciprocal optical transmission. Here, we highlight a few applications that have attracted the most attention.

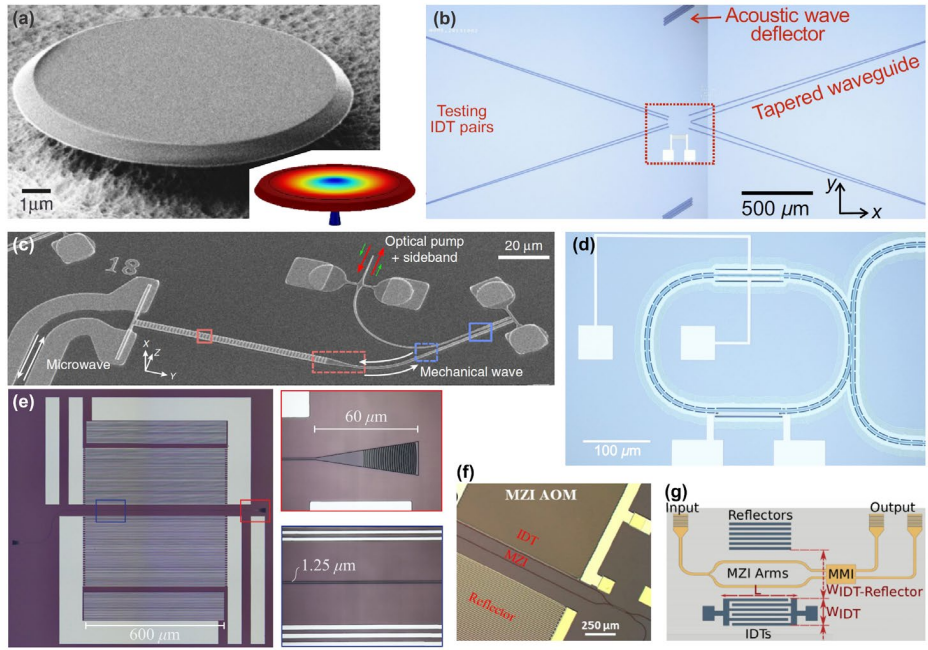

Figure 35: LN optomechanical devices. (a) Suspended optomechanical disk cavity; (b) Acousto-optic frequency shifter on LNOI; (c) Suspended piezo-optomechanical cavity; (d) Acousto-optic modulator on LNOI; (e) Acousto-optic waveguide on LN-on-sapphire; (f)-(g) Hybrid LN acousto-optic modulators. Reproduced from (a) [166] © 2016 The Author(s), (b) [396] © 2020 Optical Society of America, (c) [170] © 2020 The Author(s), (d) [167] © 2019 Optical Society of America, (e) [397] © 2020 The Author(s), (f) [398] © 2019 Chinese Laser Press, (g) [389] © 2019 IEEE.

### 5.3.1 Microwave-to-optical conversion

Compared to LN cavity electro-optic implementations, LN piezo-optomechanical implementations [167,170] demonstrate higher microwave-to-optical conversion efficiencies by taking advantage of higher $Q$ factors and smaller mode volumes of the mechanical resonances. The state of the art implementation demonstrates a single microwave photon to optical photon conversion efficiency of 5% using piezo-optomechanical crystals (Figure 35(c)) [170].

Piezo-optomechanical microwave-to-optical convertors are categorized into two groups -- one employs a mechanical resonance and an optical resonance, and the other employs an extra microwave electromagnetic resonance to facilitate the microwave to mechanical conversion. Due to the strong piezoelectricity, LN devices demonstrate efficient microwave to mechanical conversion even without an electromagnetic resonator. Using IDTs, >80% microwave power has been successfully delivered to the suspended LN slabs (Figure 35(d)) [404].

In order to achieve microwave-to-optical conversion with unitary efficiency, there are a few strategies that could be considered. While LN features strong piezoelectricity, the optomechanical coupling rate of LN devices is smaller than that in the start of the art silicon devices. This is due to the lower photoelastic coefficients of LN compared to silicon, and the limitations in LN fabrication. Hybrid devices that use LN and silicon have the potential for higher microwave to optical conversion efficiencies, and such hybrid devices using AlN and silicon have been recently demonstrated [390], though their efficiency is still limited.

Similar to double-ring cavity electro-optics (Section 3.5 and 3.6), employing a second optical resonance can enhance the microwave to optical conversion efficiency by the ratio between the microwave frequency and the optical linewidth. Such improvements are significant for high Q optical cavities with linewidths of tens of MHz. However, there is still a need to match the mechanical resonant frequency and the differential frequency between the optical resonances, which can be done, for example, by DC electro-optic tuning.

### 5.3.2 Integrated acousto-optic modulation and frequency shifting

Commercial bulk acousto-optic modulators typically operate at acoustic frequencies below hundred megahertz, though devices featuring gigahertz frequencies have been available with compromised performance. Meanwhile, integrated acousto-optic modulators have been demonstrated featuring even higher acoustic frequencies in gigahertz regime. Among the integrated devices, LN acousto-optic frequency shifters (AOFS) [396,401] demonstrate frequency shifting efficiencies that are 10-100 times higher than those of the pioneering integrated AOFS developed on thin film AlN (Figure 35(b)). However, the frequency shifting efficiencies are only a few percent, which is still quite low compared to the mature commercial bulk AOFS. There are a few challenges for integrated AOFS. First, the size of acousto-optic interaction regions for the integrated devices are limited to ~100 μm, while bulk devices can feature a few centimeters. Second, the travelling acoustic waves in integrated devices are typically generated by IDTs, whose small feature size limits the RF power at about 1 W, while bulk AOFS can handle 10 W of RF power.

Several efforts are being made to improve the integrated realization of standard acousto-optic modulator and frequency shifters. One is to employ optical or acoustic resonance. Another is employing the LN-on-sapphire platform (Figure 35(e)) [397] for better acoustic confinement and better substrate thermal conductivity to handle high RF power.

### 5.3.3 Nonreciprocal optical devices

Since the acoustic velocity is about 5 orders of magnitude smaller than the electromagnetic wave velocity, acoustic waves at microwave frequencies have wave numbers, $k$, that are 5 orders of magnitude larger than their EM counterparts. Through acousto-optic interactions, a travelling acoustic wave can thus result in a strong nonreciprocity between light travelling in opposite directions. Non-magnetic acousto-optic isolators or circulators are being developed, which are compatible with magnetic fields and desired for quantum systems. Such acousto-optic nonreciprocal devices have been developed on thin film AlN [406,407]. However, the achieved efficiency is still much lower than the state-of-the-art magnetic bulk optical isolators and circulators.

## 6. Heterogeneous integration

Significant efforts have been made on heterogeneous integrations of thin-film LN with other materials. On one hand, many mature photonic integrated circuit (PIC) platforms, such as Si and $SiN_x$, lack $\chi^{(2)}$ nonlinearity and/or the capability of quasi-phase matching. Integrating thin-film LN through flip-chip or wafer bonding can complement their functionality while leveraging their existing fabrication/manufacturing processes. On the other hand, thin-film LN is a promising platform for quantum information processing, but the material itself does not offer memories, detectors and deterministic sources. Furthermore, integrated lasers, amplifiers, linear detectors, as well as driving electronics are desirable in some classical applications, but they are not provided natively by LN. Therefore, many works have been focused on integrating these missing elements on thin-film LN photonic circuits. In this section, we review hybrid modulators, detectors, single-photon emitters, rare-earth-ion doping, and give some perspectives on III-V integration on thin-film LN.

### *6.1. Hybrid modulators*

EO modulators based on thin-film LN have been discussed generally in Section 3.2 as well as 3.3, with emphasis on monolithic, dry-etched approaches. Since hybrid EO modulators represent a considerable portion in the literature and have ignited significant interests [39,100], we dedicate this subsection to summarize the rationale, approach, and current progresses in hybrid modulators.

High-performance and scalable EO modulators are critical components for applications ranging from optical communications to quantum signal distribution. Current all-silicon optical modulators, based on plasma dispersion effect and free-carrier absorption, are pursued mainly due to the excellent scalability and low cost offered by their CMOS compatible fabrication processes [408]. However, compared to EO modulators, their performance is largely limited in terms of bandwidth, extinction ratio, insertion loss, temperature stability, and working spectrum.

Hybrid LN/Si modulators can combine the scalability of silicon photonics with the excellent EO performance of LN. This approach relies on bonding of bare LN thin films on pre-patterned Si photonic circuits. It was first implemented in hybrid ring modulators [111–113] by bonding pieces of LN thin films prepared from ion slicing onto Si ring resonators. Later, LN/SOI hybrid traveling-wave modulators with 3 dB EO bandwidth exceeding 100 GHz were demonstrated [23]. The key design point is to transition from the Si waveguide mode to the Si/LN supermode, and maximize confinement inside LN for the modulation region (see Figure 36(a)). Bonding of dry-etched thin-film LN modulators on Si photonic circuits were also achieved, where the splitters/combiners and grating couplers of the MZM were all Si, but the phase modulation sections, including LN ridge waveguides and metal electrodes, were in LN thin films [22]. In this approach, full mode transition from Si waveguide to thin-film LN waveguide is possible (see Figure 36(b)), which offers stronger field confinement in the LN compared to the bare-film-bonding approach. More recently, hybrid LN (unpatterned)/$SiN_x$ MZM have also been demonstrated, which is CMOS-compatible and is aimed for future integration with SiP photonics [202]. Theoretical studies of CMOS-compatible hybrid modulators were also being carried out [227]. We expect such hybrid integration to soon flourish and make LN a popular component in larger integrated systems.

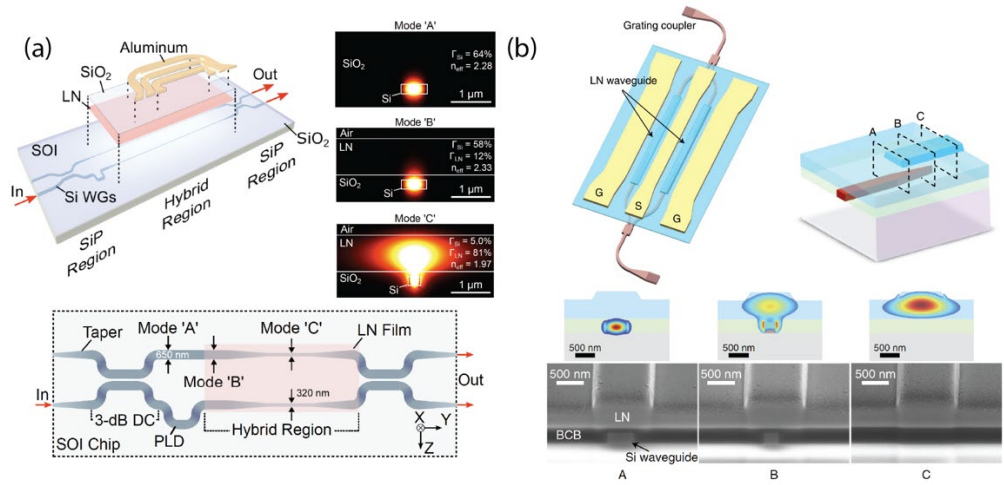

Figure 36: Hybrid LN/SOI modulators. (a) Bare LN thin film bonded on pre-patterned SOI waveguides [23]. In the hybrid region, the Si waveguide is designed to have large confinement in LN. (b) Dry-etched thin-film LN modulator bonded on SOI waveguides [22]. In this approach, the optical mode can transition completely from the Si waveguide to the LN waveguide. (a), Reproduced with permission from [23], © 2018 Optical Society of America; (b), reproduced with permission from [22], © The Author(s), under exclusive license to Springer Nature Limited 2019.

Another class of hybrid modulator uses LNOI as the base substrate, and adopt rib loading to form waveguides to avoid LN etching (see Section 2.3.3). This approach has been successfully implemented using $SiN_x$, $Ta_2O_5$, chalcogenide glasses, and Si (as summarized in Table 3 and Table 4). Though rib-loaded modulators have not been able to fully utilize CMOS-fabrication facilities, it does reduce some process compatibility issues associated with LN etching, and can reach similar performances as monolithic ones [409]. Beyond standard EO modulators used for telecommunication, integrated EO Fourier transform spectrometer has also been demonstrated based on $SiN_x$-loaded thin-film LN modulators [410].

### 6.2. Detectors

On-chip photon detection is crucial for scalable photonic information processing. Waveguide-integrated detectors are highly desirable—they eliminate the loss from fiber-to-chip couplings and are compact for large array integration [411]. For quantum applications, integrating single-photon detectors directly on LN waveguides would allow single-photon generation (through SPDC or integrated solid-state emitters), manipulation (through EO modulations/switching), and detection all on the same chip, promising a fully integrated quantum photonic processor.

Superconducting nanowire single-photon detectors (SNSPDs) and transition-edge sensors (TESs) are two leading single-photon detection technologies, especially at infrared wavelengths [412–415]. They are suitable for heterogeneous integration with photonic waveguides as they only require a single layer of material. On bulk LN substrate, SNSPDs based on polycrystalline NbN have been studied since 2012 [416,417]. On Ti in-diffused LN waveguides, integrated SNSPDs based on amorphous WSi as well as W TESs have both been demonstrated [418,419].

Very recently, direct integration of SNSPDs on thin-film LN waveguides have been successfully demonstrated [420,421] by overcoming a number of fabrication challenges. Figure 37(a) shows a scanning electron micrograph of a hairpin NbN SNSPD integrated on a LNOI waveguide. Conventionally, SNSPDs are patterned before waveguide fabrication, because the superconducting nanowires are highly sensitive to constrictions as well as inhomogeneities and are better done on a clean, flat substrate. However, thin-film LN waveguide fabrication involves aggressive dry etching and wet cleaning, which will damage the superconducting nanowires. It is, therefore, more suitable to pattern the SNSPD after waveguide fabrication, which makes the lithography step for SNSPDs particularly challenging. Furthermore, both demonstrations employed specially tailored deposition processes for the superconducting thin film to avoid excessive heating, which is incompatible to LN thin films [423,424]. From our own experience, on bulk LN, pyroelectric effect tends to cause electrostatic discharge and damage the nanowires under rapid temperature change. But this effect does not seem to happen on LNOI waveguides. Besides fabrication yield, a noticeable problem is that the partially etched LN waveguide makes the absorption rate of the TE mode small (~ 0.1 dB/μm), which requires the nanowire length to be hundreds of micrometers long to reach near-unity on-chip efficiency. Future incorporation of photonic crystal mirrors/cavities [425,426] or micro-rings [427] may be used to enhance the absorption. Besides direct lithographical integration, another viable alternative is to use flip-chip transfer [428], which circumvents the challenging fabrication and improves the yield.

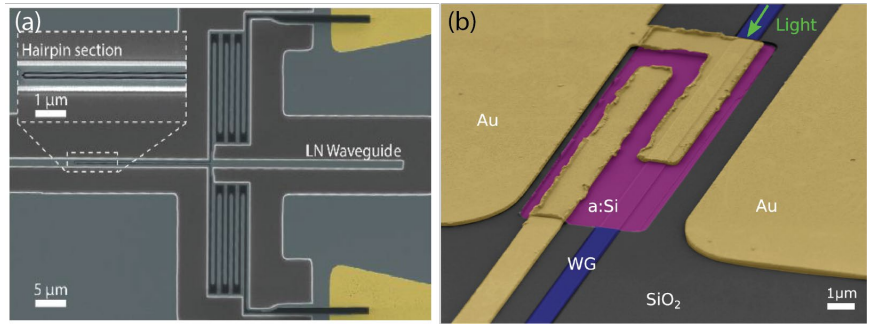

Figure 37: Heterogeneously integrated photodetectors on LNOI waveguides. (a) NbN superconducting nanowire single-photon detector on monolithically etched LNOI waveguide [421]. (b) Amorphous Si photodetector on LNOI waveguide [422]. (a), Reproduced from [421], © The Authors; (b) reproduced from [422] with the permission of AIP Publishing.

Besides single-photon detectors, integration of classical detectors is equally important and is particularly useful for applications such as large-scale neural networks [429]. On Ti in-diffused LN waveguides, GaAs detectors have been demonstrated [430]. On the thin-film platform, metal-semiconductor-metal photodetectors based on amorphous silicon have been reported, featuring a responsivity of 22 mA/W to 37 mA/W in the 635 nm to 850 nm wavelength range [422] (see Figure 37(b)). Note that the integrated SNSPDs [420,421] and amorphous-Si detectors [422] discussed here can be fabricated by direct material deposition and patterning on LNOI waveguides. But conventional semiconductor detectors, such as Si or III-V *p-n*/*p-i-n* or avalanche photodiode, will likely require flip-chip bonding. As an example, we will discuss III-V integration in Section 6.5.

## *6.3. Single-photon emitters*

Similar to integrated single-photon detectors, on-chip single-photon sources are another critical component for photonic quantum information processing. While heralded single-photon sources can be produced through SPDC processes, their probabilistic nature poses significant challenges for scaling up.

Solid-state quantum emitters and quantum dots in 2D materials are excellent sources of on-demand single photons due to their small spatial footprint and good emission properties. Recently, several works have demonstrated the coupling of quantum dot emitters to integrated on-chip silicon-based photonic platforms [431–433]. The main drawback of silicon based optical materials is the absence of an effective mechanism for cryogenic compatible EO modulation (both phase and amplitude) in these materials. In 2018, direct integration of efficient single-photons sources in the telecom band with LN photonic platform was demonstrated [434] with moderate coupling efficiency of 40.1%. InAs quantum dots embedded in an InP nanobeam were transferred to a LN photonic circuit by pick-and-place technique using focused ion beam (Figure 38) [431]. Future integration of various on-chip passive and active components in LN platform e.g., phase and amplitude modulators, beam splitters and filters will enable demonstration a full reconfigurable integrated source of single photons on demand. Another promising platform to explore the deterministic generation of single photons are quantum dots and color centers in 2D materials. Recently, 2D $Wse_2$ flake was placed [435] on a facet of Ti-indiffused Mach Zehnder interferometer in bulk LN for coupling of single photons to integrated photonic structures. Though it's currently on bulk LN waveguides, it can be extended to the thin-film platform. These studies pave the way for the realization of on-chip, integrated and multi-qubit photonic quantum photonics circuits.

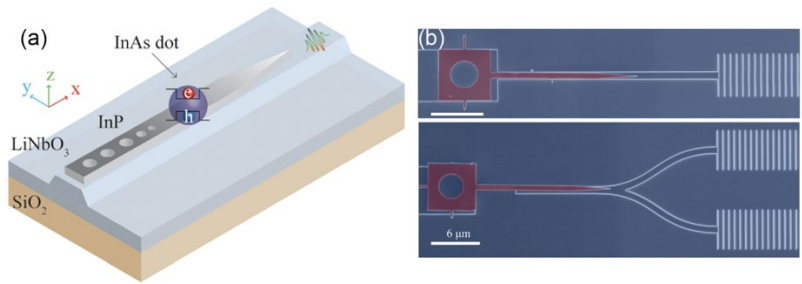

Figure 38: Integrated single-photon emitters on LN waveguides. (a) Schematic of InAs quantum dot embedded in a InP nanobeam sitting on top of a LNOI waveguide. (b) Scanning electron micrographs of the fabricated device, where the nanobeam with emitter is pick-and-placed on a grating-coupled waveguide and beamsplitter for photon correlation measurement. Reproduced from Aghaeimeidodi et al., Appl. Phys. Lett. **113**(22), 221102 (2018) with the permission of AIP Publishing [434].

### 6.4. Rare-earth ions

Rare-earth-ion-doped (REI-doped) crystals have been used for several optical applications, going back decades. These include lasers, amplifiers, scintillators, sensors, as well as classical and quantum signal processing [436,437]. To exploit their properties within LN-based applications, REI-doped LN in bulk and waveguide geometries have been developed and utilized over several years [438]. However, it is until only this past year that REIs, namely $Tm^{3+}$, $Er^{3+}$, and $Yb^{3+}$ have been investigated in thin-film LN [134,439]. This is partially due to the fact that bonded LN on insulator has only recently became commercially available and that previous studies of REIs in modified crystals—ranging from polycrystalline to surface-doped bulk samples—revealed some non-ideal properties from those in crystals grown from the melt (e.g. using a Czochralski process) [440], in particular for the case of REI-doped LN [441]. We now summarize the results of Refs. [134,439], which were investigated mainly for quantum science applications. Note that similar studies have also been performed in Ref. [442].

In the work of Ref. [439], a smart-cut process was used to bond a 1% Tm-doped film of X-cut LN to $SiO_2$ on LN (see Figure 39(a)). The relatively low doping concentration is suitable for quantum processing, which often requires suppression of REI-REI interactions to reduce decoherence. Partially-etched grating couplers and single-mode waveguides at 794 nm were created and cladded in LN while wavelength and polarization of the optical mode was optimized for interaction with the $^3H_6 \rightarrow \,^3H_4$ optical transition of $Tm^{3+}$. Waveguides were also fabricated with orthogonal polarization for comparison.

The sample was cooled to 3.6 K along with a bulk crystal for comparison. Broadband absorption spectra were measured that is consistent with a reference bulk crystal. Importantly, the polarization-dependence of the transition dipole moment was also consistent despite variations in light polarization owing to their racetrack waveguide design, indicating that the smart-cut process did not add residual strain compared to the bulk crystal. A luminescence spectrum was also measured using the thin film and bulk crystal, showing comparable transition linewidths (0.7 nm) between these samples and absence of background fluorescence from the thin film. This indicates that the smart-cut process does not create any additional fluorescent centers that absorb at the 773.4 nm excitation wavelength used here. Time-resolved luminescence further revealed consistent fluorescence lifetimes of 157 μs in both the thin film and bulk samples, suggesting no additional non-radiative decay pathways were induced by the smart-cut. Time-resolved spectral hole burning revealed hole widths and depths that were comparable to that of bulk Tm:LN measured under similar conditions, accounting for power broadening and spectral diffusion, and with two-orders of magnitude lower laser power than Ti-indiffused waveguides owing to the small (0.07 μm$^2$) transverse area of the optical mode.

Despite these results, the uniformly doped wafer may not be ideal for low-loss integration at 794 nm. To potentially overcome this, and to operate in the desirable telecom C-band, $Er^{3+}$ ions were implanted into etched off-the-shelf Z-cut LNOI. A similar spectroscopic study of the 1532 nm $^4I_{15/2} \rightarrow \, ^4I_{13/2}$ transition of $Er^{3+}$ was performed in Ref. [134]. Here, grating couplers as well as micro-ring and cm-long waveguides were created and optimized to guide single mode TE-polarized light at 1532 nm, which interact strongly with the $Er^{3+}$ transition. See Figure 39(b).

A thin cladding layer was deposited on the chip before implantation of $Er^{3+}$ ions (energy of 350 keV, fluence of $1.14 \times 10^{14}$ ions/cm$^2$) likely to reduce channeling, with SRIM simulations indicating 50 nm penetration depth into the LN film. Post-implantation annealing was carried out for 5 h at 350 °C to overcome implantation damage and recover transmission. Specifically, loaded $Q$s of 200,000 were measured at 1532 nm, owing to the absorption of Er, with loaded $Q$s of up to 1 million off resonant from the $Er^{3+}$ transition at 1515 nm (average $Q$ of 500,000).

The sample was cooled to 2.8 K, and resonance fluorescence was performed revealing a fluorescence linewidth of 170 GHz, comparable to the literature value of 180 GHz, with the difference owing to imperfect wavelength calibration. The sample was further cooled to 1.7 K, and time-resolved excitation revealed a fluorescence lifetime of 3.2 ms, which is somewhat larger than a 2.0 ms population lifetime of the $^4I_{13/2}$ level measured in a bulk sample by spectral hole burning, with the difference to be investigated in future work. Supporting cavity-ion coupling, they also measure a reduction in fluorescence lifetime of 0.6 ms (corresponding to a Purcell enhancement of 3) of ions in the ring compared to the waveguide, consistent with the measured $Q$s, optical mode volumes and field overlap with implanted ions.

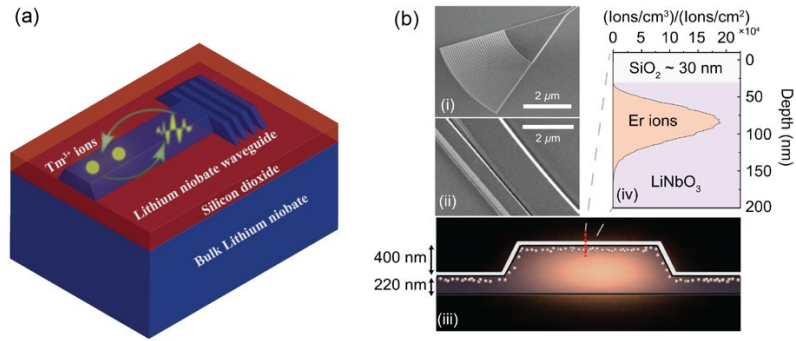

Figure 39: Rare-earth ions in LNOI integrated photonics. (a) Illustration of rare-earth ions ($Tm^{3+}$ indicated) in etched thin-film LN on SiO2/LN. (b) SEM images of a grating coupler (i) and bus-resonator coupling section (ii) on thin-film LN. (iii) Schematic cross-section of a thin-film LN

waveguide (transverse-electric mode) implanted with rare-earth ions. (iv) SRIM simulation of the implantation depth of $Er^{3+}$ in $SiO_2$-coated LN under a 350 kV acceleration voltage and a flux of $1.14\times10^{14}$ ions/cm$^2$. (a) Reproduced from [439], © 2019 American Chemical Society; (b) Reproduced from [134], with permission from AIP Publishing.

Overall, these studies predominantly address the low-temperature spectra and fluorescence properties of REIs and are a step towards future REI-based quantum applications in thin-film LN. We note however that the bonding process used to realize thin-film LN may cause variations in the crystal properties, in particular near surfaces, in ways that are not revealed by the previous studies and more measurements (e.g. coherence, sub-level properties, spectral diffusion) are needed.

On the classical end, very recently, on-chip waveguide amplifiers [443,444] and microdisk lasers [445–447] were demonstrated using $Er^{3+}$-doped thin-film LN. In these works, $Er^{3+}$-doped LN thin films were ion-sliced from uniformly doped bulk crystals. In Ref. [443], a 3.6 cm-long spiral LNOI waveguide was fabricated using photolithography assisted chemo-mechanical etching technique (see Section 2.3.2), and a maximum net gain of 18 dB was observed in the small-signal-gain regime around 1530 nm. In Ref [444], a dry-etched ridge waveguide was used, showing a 5 dB net gain with 5 mm length. In Refs. [445–447], high-Q microdisks were used as cavities and lasing at telecom wavelengths was observed. Strong photorefractive effect was observed in these microdisk lasers, especially in the presence of a short-wavelength pump. This effect led to a pump power-dependent emission wavelength [445] and reduced pump efficiency [446]. Mitigation of the photorefractive effect, e.g. by co-doping MgO, is needed in future developments. These preliminary demonstrations show the potential of using REI doping to achieve gain and lasing on LNOI platform.

## 6.5. III-V materials

Integrated active photonic components, including lasers, amplifiers, and photodetectors, have been developed extensively in III-V materials with state-of-the-art performance. The integration of LN electro-optics and III-V components could open opportunities for optical communications, microwave photonics, and light detection and ranging (LIDAR). Such opportunities have been foreseen by several on-going research programs supported by government funding agencies worldwide.

The integration of III-V with LN components could benefit from previous experiences in integration with other materials, such as LN on Si [448], as well as III-V on silicon-on-insulator (SOI) [449–451] and $Si_3N_4$ waveguides [452]. As LN is inert to most chemicals used in III-V fabrication, the III-V stack layers can be transferred to LN and then fabricated without damaging the LN waveguides. This approach could enable fabricating photodetectors directly on the LN waveguides. Alternatively, it is possible to transfer LN thin film to the III-V substrates, and then fabricate the LN devices. The transfer process could be similar to transferring LN thin film on SOI substrates [448]. For integration using overlaying materials, optical tapers are usually used to couple the light between different materials [448–452].

On the other hand, pick-and-place tools could enable direct placement of fabricated III-V components on the LN chips. This approach has been used to transfer III-V lasers on silicon photonic circuits [453]. In additional to the optical tapers, efficient butt/end-fire coupling between III-V lasers and LN waveguides can be achieved by proper mode-size matching and accurate facet alignment. This approach could be more suitable for integrating high power III-V lasers or optical amplifiers, as the III-V components can directly sit on the substrates (e.g. Si) with high thermal conductivity.

## 7. Challenges and Opportunities

The field of integrated photonics on thin-film LN is advancing rapidly, promising many exciting opportunities in telecom, datacom, microwave photonics, and quantum photonics. To enable these, however, several important challenges need to be addressed. In this section we discuss these challenges and outline several exciting near- and long-term opportunities.

### 7.1. Charge carrier effects

Mobile charge carriers in electro-optic materials can create a wide variety of optical and electro-optical effects, including photoconductivity [454], photorefractivity [455], and dielectric relaxation [456]. These are long-standing problems with LN material, and often limit the stability and power handling capability of LN devices. Recent measurements [183,342,457,458] suggest that these effects are stronger in LNOI than in comparable bulk devices, possibly due to the tighter optical confinement and higher crystal defect concentration in LNOI. These charge carrier effects typically manifest as bias or detuning drift when operating electro-optic devices in LNOI, with timescales typically on the order of milliseconds to hours or longer [457]. While such drifts are an important practical challenge for realization of highly stable LNOI devices, charge carrier effects may also be useful for *in situ* device tuning [459] or to create strong all-optical nonlinearities [183].

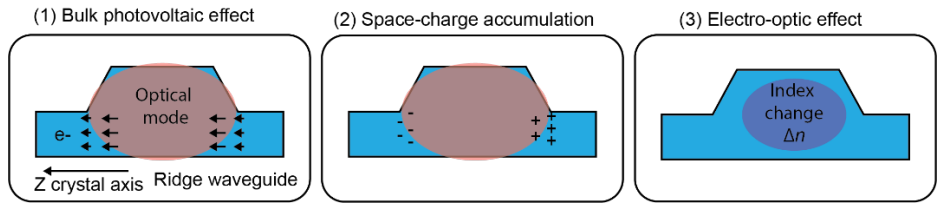

Figure 40: A schematic depiction of the mechanism of photorefraction in LNOI ridge waveguides. (1) Optical power excites charge carriers, which move preferentially along the Z crystal axis due to the bulk photovoltaic effect in lithium niobate, (2) excited charge carriers migrate and reach a steady-state spatial distribution, and (3) the resulting space-charge field perturbs the refractive index via the EO effect.

Photorefractive materials experience refractive index changes when exposed to optical power due to the excitation of charge carriers from crystal defect sites. Figure 40 shows a schematic depiction of this mechanism inside a LN ridge waveguide. More details on the photorefractive effect can be found elsewhere, for example in references [460,461].

The photorefractive effect provides both opportunities and challenges for the LNOI platform. Recent work has shown that the photorefractive effect in LNOI is stronger and responds at faster timescales than in bulk devices, especially in high-quality factor resonant photonic devices [457,458]. Indeed, resonance tuning of photonic crystal cavities in LNOI has been demonstrated using optical powers corresponding to just a few photons inside the cavity [183]. Photorefractive effects are particularly prominent in LNOI devices operating at visible wavelengths [15]. Such strong power-dependent detuning can make it more challenging to operate LNOI devices at high powers or where stable phase tuning is required [168,178,179]. Efforts to reduce photorefractive effects in LNOI have shown that photorefractive shifts can be reduced by annealing and using uncladded waveguides [342], but further work is required to understand the impact of material properties and device geometry on photorefraction in LNOI devices. Other methods of mitigation used in bulk LN devices include transition metal doping [462] and operating at elevated temperature [463].

Dielectric relaxation is a process by which charge carriers in a dielectric material can migrate to shield an applied electric field. For the simple case of a uniform material with dielectric constant $\epsilon$ and resistivity $\rho$, this relaxation occurs at a rate given by the dielectric relaxation time $\tau_r = \epsilon\rho$. It is well known that such relaxation can lead to drift in the bias voltage required for electro-optic devices [456,464–466]. To our knowledge, these effects have yet to be studied directly in LNOI devices, but recent works have observed bias drift consistent with dielectric relaxation [467]. Thermo-optic tuning can be used as a drift-free method of bias control [16,468], but this requires tradeoffs in power dissipation and tuning bandwidth. We believe that understanding and mitigating dielectric relaxation in LNOI is a key challenge for creating stable electro-optic devices and networks in this platform.

*7.2. Quantum photonics*

Optical photons have many properties of interest for realization of quantum technologies: they exist under ambient conditions, are generally impervious to environmental noise, can travel long distance, and can be generated, manipulated and detected relatively easily. While many important demonstrations have already been made using discrete optical components, including recent demonstration of photonic quantum computational advantages [469], the future quantum applications will require integration of large number of different functionalities on the same photonic platforms [470]. Such integrated quantum photonics platforms will need to: (i) be ultra-low loss in order to preserve fragile quantum states; (ii) enable precise control of the temporal and spectral profiles of photons; (iii) allow fast and low-loss optical switches to route quantum information; (iv) be able to operate in visible wavelength range, where many single-photon sources and quantum memories operate, as well as telecom wavelength range, where low-loss optical fibers exist; (v) feature strong nonlinearities for efficient frequency up- and down-conversion to bridge these wavelength ranges, as well as enable efficient quantum transduction and entangled photon pair generation; (vi) allow integration of photodetectors and operating electronics. Thin-film LN photonic platform can simultaneously meet nearly all of these requirements and thus is an excellent candidate for integrated quantum photonics.

Among the most popular and arguably most advanced schemes for on-chip processing of single-photon quantum states are path-encoded schemes. In this approach, optical quantum information is encoded in the position (waveguide) in which a photon is present [470]. Quantum operation can be performed using basic building blocks including low-loss optical waveguides, beam splitters and phase shifters. With growing complexity of the optical circuits [471], additional challenges related to fabrication variation arises, which motivates active tuning of photonic devices [472]. As an example, beam splitter can be achieved by Mach-Zehnder interferometers, which can be tuned to the desired operation point, which increases the number of active devices and optical phase shifters, but allows for compensating for defects and fabrication imperfections [473]. Among path-encoded quantum photonic realizations, the use of thermal phase shifters is very popular, as they can be implemented in many material platforms such as Si and $SiN_x$ and are very compact in size. However, thermal heaters draw electrical current and with increasing number of thermal phase shifters, the total power consumption also increases. Additionally, with denser packing of devices, thermal crosstalk and cooling becomes another issue that needs to be considered. Finally, thermal tuning is slow which limits the operation speed of quantum photonic network.

The low propagating loss achieved in thin-film LN is well suited for quantum applications, and the electro-optic effect allows low-loss phase shifters, which could drastically reduce the power-consumption compared to the use of thermal phase shifters.

The use of electro-optic switches and phase shifters enables additional functionalities going further than reducing the power consumption and heat generation of thermal phase shifters. Indeed, the ability to perform fast modulation of light makes thin-film LN a very promising platform for time- and frequency-encoded quantum operations. Energy-time entangled quantum states carry quantum information in temporal and spectral modes. The ability to process quantum information in the time domain has been widely used for quantum communication, where the temporal modes can propagate long distances in optical fibers [474], and frequency-domain quantum processing enables parallel and multiplexed operations [258]. Time-domain quantum processing employs fast switches, optical delay lines as well as unbalanced interferometers as fundamental building blocks [475], while frequency-domain processing requires spectral filtering, as well as frequency shifting achieved by e.g. electro-optic modulation [476]. Thin-film LN could be suited for both time-domain as well as frequency-domain quantum processing [477], as fast switches, interferometers and electro-optic frequency shifters can be achieved in a compact and low-loss platform.

Figure 41 shows some basic resources and elements for photonic quantum computing that are suited for the thin-film LN platform. For example, arrays of integrated PPLN SPDC sources can be multiplexed in time with the help of photon detection in combination with fast EO switches [478,479] (see Figure 41 (a)). Similarly, these probabilistic sources can also be multiplexed in frequency domain (a broadband source sliced into multiple frequency bins) using EO frequency shifters to produce quasi-deterministic single-photon source [480] (see Figure 41 (b)). Tunable EO-based frequency shifters as described in Section 3.5 can be conveniently used to realize gates for frequency-encoded qubits (Figure 41 (c)). These can serve as the basis for frequency-domain quantum computing [477] (Figure 41 (d)).

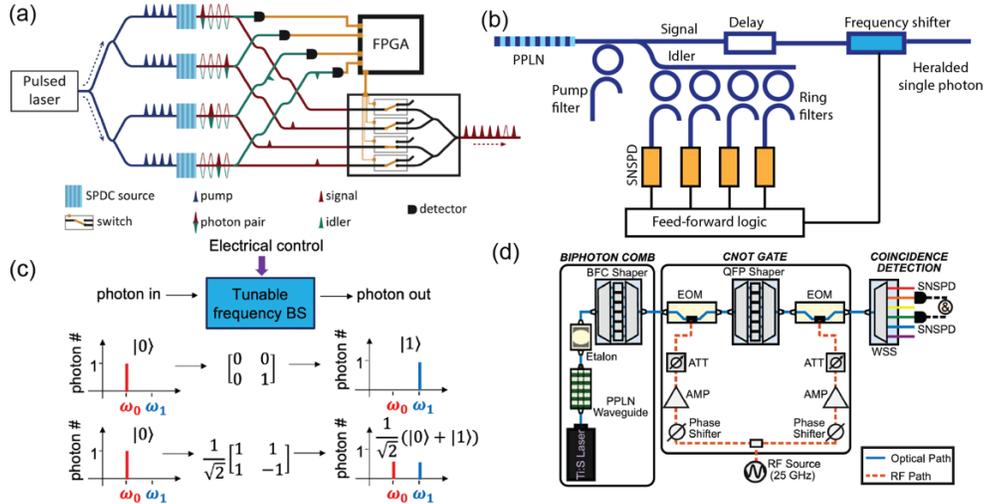

Figure 41: Basic resources and elements for photonic quantum computing that are well suited for thin-film LN integrated photonic platform. (a)Time-multiplexed single-photon sources based on SPDC and fast-optical switches [478]. (b) Frequency-multiplexed single-photon sources. (c) Encoding qubit in frequency domain and basic single-qubit gate based on tunable frequency shifters. (d) General purpose quantum frequency processor (QFP) can be implemented using EO modulators and wave shapers [477]. A tunable frequency beamsplitter (BS) can greatly simplify the scheme and reduce the required resources. All these essential components and their subunits, such as PPLN SPDC sources, fast switches, frequency shifters, single-photon detectors, filters, wave shapers, can be implemented on the thin-film LN platform. (a) reproduced from [478] © Author(s) 2019, licensed under a Creative Commons Attribution (CC BY) license; (d) reproduced from [477] © Author(s) 2019, licensed under a Creative Commons Attribution 4.0 International License.

### 7.3. Nonlinear photonics

The thin-film LN platform provides exciting new directions for the realization of novel frequency comb generation schemes, deterministic generation of chimera states and band solitons in addition to Kerr and Pockels solitons, study of co- and counter-propagating mode coupling, and high dimensional topological effect. These directions exploit the large strong $\chi^{(2)}$ and $\chi^{(3)}$ optical nonlinearities and the phase modulation capability of LN.

Stable pulse trains known as solitons can be utilized within a microresonator to produce broad and coherent frequency comb spectra. Because it has both $\chi^{(2)}$ and $\chi^{(3)}$ optical nonlinearities, thin-film LN is a promising platform for both Kerr (produced via $\chi^{(3)}$) [339] and Pockels (produced via $\chi^{(2)}$) [481] soliton generation. Furthermore, the strong EO effect in LN shows promise for the experimental demonstration of other stable and more exotic resonator states, including chimera-like states with purely local coupling [482], corresponding to a chaotic state circulating in the resonator with no expansion or contraction, and band solitons [483], dispersionless and coherent structures occurring in modulation instability region under high modulation strength and anomalous dispersion.

However, the role of strong Raman gain observed in LN needs to be understood and engineered since it may play an important role in the overall nonlinear dynamics of the resonator. The combined effects of Raman scattering, four-wave mixing, and electro-optic modulation can result in a broad spectrum of small EO combs around lines formed by Raman and Kerr effects [484]. Such a broad, stable frequency comb with low repetition rates could be used for applications such as optical frequency synthesizer and high-resolution comb spectroscopy. Additionally, the strong Raman oscillation in LN manifests in the counterpropagating direction with respect to the pump field [175], effectively coupling the clockwise and counterclockwise modes in the resonator. The role of Raman and the dynamics of counterpropagating EO fields remains largely unstudied, and the thin-film LN platform presents a promising avenue for exploring more complex nonlinear dynamics.

### *7.4. Magnetic-free non-reciprocity and optical isolation*

Lorentz reciprocity is a fundamental principle in electromagnetics. It holds true in linear, time-invariant medium with symmetric permittivity and permeability tensors. However, non-reciprocal devices, such as isolators and circulators, are critical for signal routing and blocking, as well as laser protection. The conventional method to realize non-reciprocity is to use magnetized materials, whose permittivity is nonsymmetric. In integrated photonics, heterogeneous integration of magneto-optical materials has been pursued. But it is generally difficult to fabricate, incompatible with large scale integration, and associated with high insertion loss.

In LN, there are multiple approaches to achieve magnetic-free non-reciprocity and on-chip optical isolation.

First, reciprocity can be broken through optical nonlinearity, such as wavelength conversion [485]. LNOI PPLN waveguides are well suited for this approach. The key principle is based on the fact that wave-mixing processes are high directional due to the stringent phase matching requirements. Recently, such an optical isolator was achieved based on difference frequency generation in PPLN waveguides (see Figure 42(a)) [486]. In this work, at the presence of an optical pump, signal light propagating in the same direction as the pump will undergo efficient difference frequency generation and pass a set of properly arranged frequency filters. Back propagating waves, however, will not experience efficient DFG and will be blocked by the filters. As the fabrication precision, quality, and yield of LNOI PPLN waveguides continue to improve, we expect the conversion efficiency and insertion loss to reach a level to satisfy the requirement of practical applications. Future works may include integrating filters directly on-chip, or involving different polarization modes to facilitate on-chip filtering.

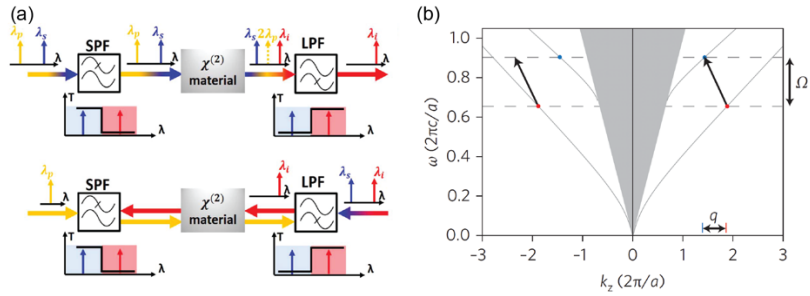

Figure 42: Magnetic-free optical isolation. (a) Optical isolator based on $\chi^2$ wavelength conversion [486]. (b) Nonreciprocal photonic interband transition through time modulation, which can lead to optical isolation [487]. (a) Reprinted from [486] © 2020 The Optical Society; (b) reprinted with permission from [487] © 2009 Springer Nature.

The second way to break reciprocity is to break time-reversal symmetry through EO modulation [488]. In fact, a simple traveling-wave modulator is naturally non-reciprocal—it only modulates efficiently in one direction. This can be utilized to implement optical isolators. For example, in a traveling-wave phase modulator, a proper driving voltage can be chosen so that when the light and microwave are co-propagating, the optical carrier frequency will be completely converted into sidebands and then filtered, while the reverse would not be modulated at all.

Another elegant approach for implementing optical isolation through EO modulation is to introduce an interband photonic transition. The two photonic bands can be even/odd and TE/TM modes in a single waveguide, or symmetric/asymmetric modes in coupled waveguides [489–491]. In the case of indirect interband transition, as illustrated in Figure 42(b) [487], input photon will be frequency up/down-converted only in one direction due to momentum/phase mismatch. It is worth mentioning that when photons gain/lose energies of microwave fields, they also gain/lose the phase of microwave field. Using this property, it is possible to construct an effective magnetic field using a pair of such modulators driven by microwave fields with different phases [285], which could lead to optical isolation as well.

Similar to the case of EO modulation, acousto-optic modulation is also able to break time-reversal symmetry and achieve non-reciprocity by using phonons to scatter the input photons. Photons modulated by acoustic waves will absorb or emit phonons and therefore achieve similar transitions as the EO modulation process. Compared to EO modulation, the advantage of acousto-optic approach is the large momentum of the phonons due to the short wavelength of acoustic waves, which can lead to larger phase mismatch and thus better suppression of undesired transitions. This non-reciprocal process of acousto-optic modulation has been demonstrated in AlN [492] and Si [493], and can be readily adapted to integrated LN systems.

*7.5. Microwave photonics*

Microwave photonics utilizes optical components to generate, transmit, and process high frequency RF signals, achieving tasks that are difficult using conventional electronics. Specifically, upconverting microwave signals to optical domain allows low-loss signal transport and flexible spectral tailoring over ultra-wide bandwidth. A generic microwave photonics system consists of laser sources, optical modulators, optical signal processors, and photodetectors. Central to a microwave photonic system are high-performance EO components that have large bandwidth, good linearity, negligible insertion losses, and can handle large optical power. Traditional microwave photonics are based on discrete fiber-optic components that combine bulk LN EO modulators, high power III-V lasers and detectors. This approach is not scalable due to the large footprint and increased insertion losses when daisy-chaining. In the past decade, significant effort has been devoted to integrated microwave photonics [494,495]. However, existing material platforms, namely SOI, $SiN_x$ and InP, are facing particular challenges, especially in realizing high-performance EO components. In particular, $SiN_x$ does not allow optical modulation; Si suffers from two-photon absorption and does not offer low-loss and high-bandwidth modulators; InP enables various active and passive components, but its waveguides have relatively high losses and modulator has limited linearity.

LNOI is well suited for microwave photonics. It not only has transparency window and ultralow linear loss, but also offers EO modulators with performances (> 100 GHz, few-volt half-wave voltage, < 0.5 dB on-chip insertion loss, and ~ 100 dB $Hz^{2/3}$ SFDR) that to date few other material platforms can compete with. Furthermore, it allows narrow resonant filters (MHz) with EO and thermal tunability. In addition, it offers broadband frequency comb sources for multi-band operation, and magnetic-free isolators to facilitate integration of on-chip lasers. However, the LN platform lacks native light sources, amplifiers, and detectors. To realize a fully integrated microwave photonic transceiver, heterogeneously integration III-V (e.g., InP and InGaAs) photonic components is a promising direction (see Section 6.5). Figure 43 shows a proposed microwave photonic receiver (based on the concept in Ref. [496]) that integrates III-V lasers, amplifiers and detectors with thin-film LN photonics. In this proposed receiver, beating between two EO frequency combs of different line spacings (referred as signal and reference combs) slice a 50-GHz microwave signal into multiple low-bandwidth channels which can be detected in parallel using a de-multiplexer (DEMUX) followed by an array of photodetectors. A high bandwidth microwave signal can be decomposed and processed by employing broadband and flat-top EO combs, which can be realized by cascading an intensity modulator with three phase modulators (as described in Section 3.4.1). With such effort of multi-function/material integration, we expect thin-film LN to play a major role in future integrated microwave photonics.

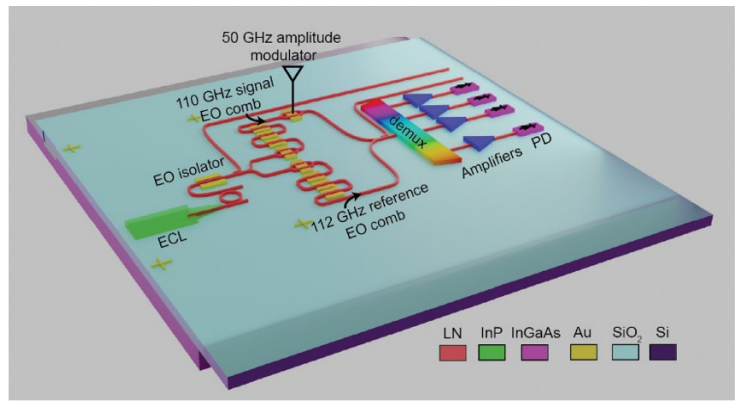

Figure 43: A proposed integrated microwave photonic receiver. It combines LNOI passive (low-loss waveguides, high-Q resonators, arrayed waveguides) and EO (modulators, isolators, frequency combs) devices with heterogeneously integrated III-V lasers and high-speed detectors. This figure is adapted from a grant proposal funded under the DARPA LUMOS program. ECL: external cavity laser; PD: photodetector.

*7.6. Wafer-scale lithium-niobate photonic integration*

Among the key advantages of silicon photonics is the compatibility with CMOS fabrication techniques, which enable large-scale and low-cost production of integrated photonic components at high volumes. Consequently, a large variety of foundries are already offering commercial tape-outs of silicon photonic chips with guaranteed performances, which significantly reduces the cost and enables broad access to a very advanced technology platform. Unfortunately, LN is not compatible with current CMOS or silicon photonics processes due to material contamination issues, mainly caused by lithium diffusion or residues from the etching process. Additionally, low roughness etching of LN, required for low-loss optical waveguides, has only been achieved very recently, and is not yet broadly available. For this reason, almost all thin-film LN devices were achieved using chip-scale proof-of-concept realizations, often relying on electron-beam lithography for pattern definition.

In order to drastically increase the innovation speed, foundry access to wafer-scale LN photonics would be highly desirable. There are different paths to advance thin-film LN from chip-scale demonstrations to wafer-scale processes, mainly focusing on either heterogeneous and monolithic integration. In the approach of heterogeneous integration, LN is bonded to other materials, such as $SiN_x$ [110], to enable wafer-scale processing. Recently it was proposed that LN could be bonded as the final step in standard CMOS silicon photonics processing, where Si waveguides as well as electrodes are defined in a scalable process before LN is added. This approach eliminates the risk of contamination and makes LN compatible with standard Si photonics foundry processes [227]. In such an approach it is however not possible to pattern or process the LN thin films, which could lead to a reduction in performance.

Alternatively, monolithic wafer-scale processing of thin-film LN has been shown to also be a possibility [497]. While LN is not compatible with CMOS processes, it can be processed at large scales and low cost using existing MEMS or even dedicated future lithium-niobate photonics foundries. The advantage of the monolithic approach is reduced process complexity as no bonding to other materials is required and potentially higher performance could be achieved as the LN can be etched and processed directly without process limitations. Wafer-scale processing of 4-inch and 6-inch monolithic thin-film LNOI wafers has been recently demonstrated [497] using scalable deep-UV lithography, see Figure 44. Wafer-scale LN etching was achieved, resulting in final film thickness variation of the etched wafer of below 6 nm, limited by both film thickness uniformity of the initial wafer as well as the etch. While these thicknesses and etch variations are much larger than currently achievable in silicon photonics, they are already sufficient for large-scale and low-cost manufacturing of high-performance LN photonic chips, and show that scalable wafer-processing is possible.

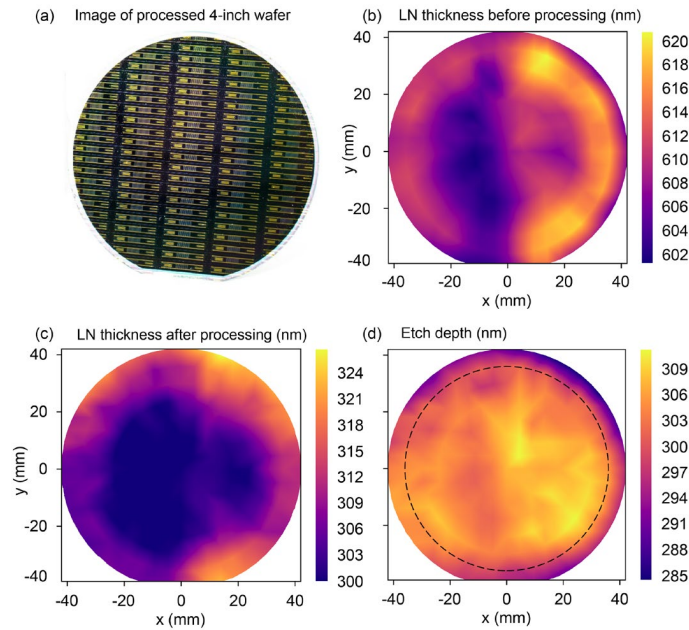

Figure 44: Wafer-scale LNOI fabrication. (a) Image of a processed LNOI wafer including metal electrodes. (b) Thickness of the LN thin film before processing. (c) Film thickness after etching. (d) Etch depth variation across the wafer, showing below 6 nm variation. Reprinted with permission from [497], © 2020 the Optical Society.

Due to the high potential and performance of thin-film LN, as well as the already demonstrated ability to perform heterogeneous integration as well as monolithic wafer-scale processing, we hope and expect that in the future foundry services will become available, which would significantly accelerate the development of this highly promising photonics platform.

## 8. Conclusion

In this Review, we attempted to cover the various aspects of integrated thin-film LN photonics as comprehensively as possible—from the materials and passive optical components to active components based on electro-optics, all-optical nonlinearity, and piezo-optomechanics, as well as heterogeneous integration. We also identified some challenges that need to be overcome to realize the exciting opportunities that lie ahead for this material platform.

Overall, thin-film LN integrated photonics, building on half-century legacy of discrete components realized in LN crystals, is rapidly revolutionizing a broad range of applications, from its traditional uses in telecommunications and nonlinear optics, to emerging fields such as quantum photonics, cavity electro-optics, and piezo-optomechanics. Within the past few years, thin-film LN modulators have significantly outperformed their widely used bulk counterparts in power consumption, bandwidth, and size, which are key metrics for next generation optical communication systems. Among many promising new material platforms, such as polymer, plasmonic or barium titanate, while LN is neither the smallest nor possessing the highest EO efficiency, its well-balanced material properties and adequate integrability with decades of industry proven operation could results in rapid deployment. Such potential in volume applications is key to catalyze LN technology development and could enable scaling up and cost reduction for LN thin-film platform at the foundry-level. Besides modulators, thin-film LN is breeding a new collection of powerful integrated nonlinear optical devices. Thin-film nonlinear wavelength converter and down-conversion sources have shown orders of magnitude higher efficiencies than that using diffusion-based waveguides. In addition, frequency combs, supercontinuum sources, Raman lasing, microwave-to-optical transduction, and acousto-optic modulators with impressive metrics have all been demonstrated on this platform. Furthermore, integration with superconducting detectors, quantum-dot single-photon emitters, and rare-earth ions are promising fully integrated quantum processors. Moreover, investment and involvement from industry and national facilities are pushing integrated thin-film LN photonics from laboratory proof-of-concept demonstrations to high-yield, wafer-scale productions.

Future optimization in material preparation (e.g., less expensive and higher quality wafers or use of stoichiometric LN), device fabrication (etching, poling, doping, annealing), and system integration (optical and RF packaging) will immediately improve device performance. Studies of photorefraction, charge accumulation, power handling, and loss mechanisms are urgently needed and essential for many applications including achieving extreme optical nonlinearities and operating EO devices at high power, in harsh environments and with long-term stability.

The immediate future of thin-film LN is clear and bright. Its versatility and complexity are fascinating and intriguing. We expect it to soon become a strong contender and indispensable player in the league of integrated photonics, and more importantly, make an impact on our daily life.

## Acknowledgements

We thank Smarak Maity (Harvard), Prashanta Kharel (HyperLight), Cheng Wang (City University of Hong Kong), Yoshitomo Okawachi (Columbia), and Marc Jankowski (Stanford) for helpful discussion and critical reading of the manuscript. We acknowledge collaborations with Cheng Wang, Martin M. Fejer, Joseph Kahn, Alex Gaeta, Michal Lipson, Xi Chen, Peter Winzer, Nathalie Picque, Oliver King, Ronald Esman, Shanhui Fan, Keji Lai, Carsten Langrock, Brendon Buscaino, James Leatham, Kevin Luke, Lingyan He, and Tianhao Ren. We also acknowledge collaboration and support from NanoLN (Hui Hu) and Covesion (Corin Gawith). Relevant research conducted at Harvard is sponsored by NSF, ONR, AFOSR, DARPA, DOE, ARL, ARO, Raytheon, Nokia Bell Labs, Rockwell Collins, and Google; and device fabrication is performed at the Harvard University Center for Nanoscale Systems, a member of the National Nanotechnology Coordinated Infrastructure Network. D. Z. is supported by the Harvard Quantum Initiative (HQI) postdoctoral fellowship. N.S. acknowledges support by the Natural Sciences and Engineering Research Council of Canada (NSERC), the AQT Intelligent Quantum Networks and Technologies (INQNET) research program, and by the DOE/HEP QuantISED program grant, QCCFP (Quantum Communication Channels for Fundamental Physics), award number DE-SC0019219. Finally, we thank everyone who contributed to the field of integrated thin-film lithium niobate photonics, without whom this review would not have been possible.

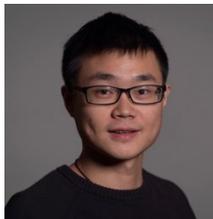
**Di Zhu** is a postdoctoral fellow at Harvard University. He received his Ph.D. (2019) and M.Sc. (2017) from Massachusetts Institute of Technology, and B.Eng. (2013) from Nanyang Technological University, all in Electrical Engineering. His current research interests include integrated quantum photonics, applied superconductivity, and nanoscale electromagnetics. His PhD thesis on superconducting nanowire single-photon detectors was recognized by the MIT Jin-Au Kong thesis award. He was a recipient of the inaugural Harvard Quantum Initiative Postdoctoral Fellowship and A*STAR National Science Scholarship.

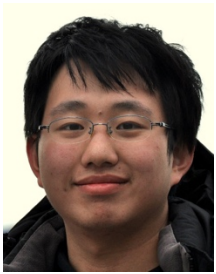
**Linbo Shao** is a Postdoctoral Fellow at Harvard University. He received his Ph.D. in engineering science (2019) and M.S. in applied physics (2014) from Harvard university, B.S. in microelectronics (2014) from Peking University. He works on integrated acousto-optics and electro-acoustics on the lithium niobate platform and color centers in diamond.

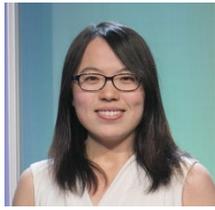

**Mengjie Yu** is a postdoctoral fellow in the John A. Paulson School of Engineering and Applied Sciences at Harvard University. She received her Ph.D. in Electrical Engineering in 2018 from Cornell University. Her research focuses on integrated photonics in silicon, silicon nitride, and lithium niobate platforms, and her interests include nonlinear frequency conversion, frequency combs, mid-infrared photonics, and optical coherent computing. Mengjie Yu is the 2020 OSA Ambassador. She was the winner of the 2016 Maiman Student Paper Competition and the 2016 Emil Wolf Student Paper Competition, and a finalist of the 2020 Tingye Li Innovation Prize. She was the Caltech 2019 Young Investigator Lecturer. She has published 32 peer-reviewed journal papers and 44 conference papers and is a referee for 15 peer-reviewed journals. Currently, she serves as chair of the OSA Integrated Optics Technical Group.

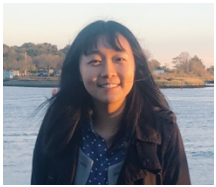

**Rebecca Cheng** is a graduate student in Applied Physics at Harvard University. She received her Sc.B. in Mathematical Physics from Brown University in 2018. She is currently studying integrated nonlinear photonics in Professor Marko Lončar's group.

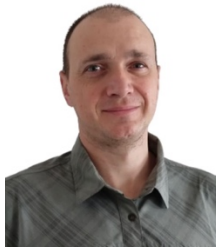

**Boris Desiatov** received his BS in Physics and Mathematics in 2006 and an MS and a Ph.D. in Applied Physics from Hebrew University of Jerusalem, Israel in 2015. After a postdoctoral fellowship at Harvard University, he joined Bar Ilan University as Assistant Professor of Electrical Engineering. His research is focused on integrated photonic devices, quantum information, quantum optics, electro-optical systems, and nonlinear optics.

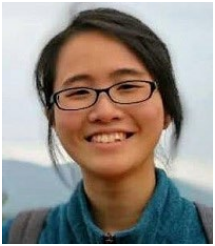

**CJ Xin** is a PhD student in Applied Physics at Harvard University. With the Lončar Group since 2018, she has been working on quantum frequency conversion in the thin-film lithium niobate on insulator photonic platform. She is interested in developing scalable sources of single photons and entangled photon pairs for quantum information processing.

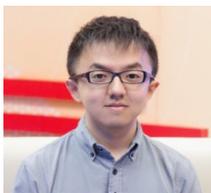

**Yaowen Hu** is a graduate student in the Department of Physics at Harvard University. He received his Bachelor of Science in Physics at Tsinghua University in 2018. During his undergraduate studies, he worked on coupling superconducting qubits with spin ensembles. He was awarded the highest honor and the Students of the Year award of Tsinghua University, both in 2017. He is currently working in Prof. Marko Lončar's group. His main interests are photonic computing, nonlinear photonics, and topological photonics.

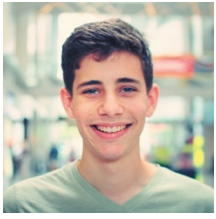
**Jeffrey Holzgrafe** is a Ph.D. candidate in Applied Physics at Harvard University. He received his M.Phil. in Physics from the University of Cambridge in 2017, supported by a Marshall Scholarship, and his B.S. in Engineering from Olin College in 2015, supported by a Goldwater Scholarship. Jeffrey is currently studying integrated photonics and quantum transduction in Marko Lončar's Laboratory for Nanoscale Optics.

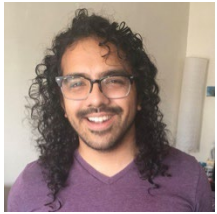
**Soumya Ghosh** is a PhD student in Physics at Harvard University, currently supported by an NSF Graduate Research Fellowship. He received his Sc.B in Mathematical Physics from Brown University, and is studying integrated nonlinear photonics in Marko Lončar's Laboratory for Nanoscale Optics.

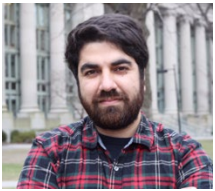
**Amirhassan Shams-Ansari** is a Ph.D. student in Electrical Engineering at Harvard University. He started his Ph.D. by exploring nonlinear optical phenomena in single-crystalline diamond platform. Over the last few years, he is mainly working on studying the thin-film lithium niobate platform's fundamental limitations, developing integrated electro-optics frequency combs, and implementing LN integrated devices in practical applications such as gas-phase spectroscopy.

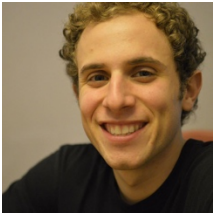
**Eric Puma** received his B.A. in Physics from Pomona College (2014) and his M.S. in Applied Physics from Harvard University (2020) where he is currently a PhD candidate in the group of Marko Loncar. His work focuses on the advancement of the lithium niobate on insulator platform for integrated quantum photonics. Formerly, he worked in the group of Frank Koppens at ICFO (2014-2017) developing graphene optoelectronics. Outside the lab, Eric is a musician and cyclist, serving as director of the Harvard World Music Collective (2018-2019) and as co-founder of the Caravan Project.

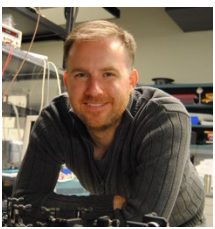
**Neil Sinclair** is a postdoctoral fellow in the group of Marko Lončar at Harvard University working on projects in nanophotonics and is co-appointed at the California Institute of Technology to develop quantum networks under Maria Spiropulu. Neil has worked on quantum entanglement theory in the group of Shohini Ghose during his B.Sc. studies at the University of Waterloo. For his graduate work, he developed quantum science experiments using rare-earth-ion-doped crystals under Wolfgang Tittel at the University of Calgary.

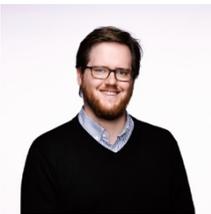
**Christian Reimer** is a physicist and entrepreneur working in the fields of nonlinear optics, integrated photonics and quantum optics. He received graduate degrees from the Karlsruhe Institute of Technology in Germany, Heriot-Watt University in Scotland, and the National Institute of Scientific Research in Canada. He then worked as a postdoctoral fellow at Harvard University, before becoming Co-Founder and Head of Product of HyperLight Corporation. HyperLight, a Venture-Capital funded start-up out of Harvard University, is specialized on integrated lithium niobate technologies for ultra-high performance photonic solutions.

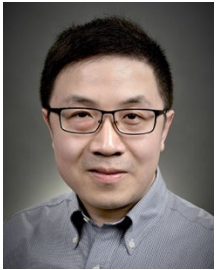

**Mian Zhang** is co-founder and CEO of HyperLight, a VC backed startup company located in Cambridge, MA, focusing on commercialization of ultrahigh performance lithium niobate integrated photonics technology. He has pioneered the technical development of thin-film lithium niobate photonic platform during his postdoc at Harvard. Mian obtained his PhD from Cornell University in 2015. Prior to that, he received his Bachelor's degree from University of Bristol in UK.

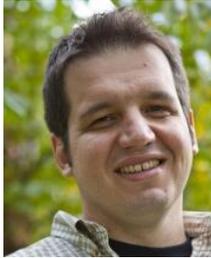

**Marko Lončar** is Tiantsai Lin Professor of Electrical Engineering at Harvard's John A Paulson School of Engineering and Applied Sciences (SEAS), as well as Harvard College Professor. Lončar received his Diploma from University of Belgrade (R. Serbia) in 1997, and his PhD from Caltech in 2003 (with Axel Scherer), both in Electrical Engineering. After completing his postdoctoral studies at Harvard (with Federico Capasso), he joined SEAS faculty in 2006. Lončar is expert in nanophotonics and nanofabrication, and his current research interests include quantum and nonlinear nanophotonics, quantum optomchanics, high-power optics, and nanofabrication. He has received NSF CAREER Award in 2009 and Sloan Fellowship in 2010. In recognition of his teaching activities, Lončar has been awarded Levenson Prize for Excellence in Undergraduate Teaching (2012), and has been named Harvard College Professor in 2017. Lončar is fellow of Optical Society of America, and Senior Member of IEEE and SPIE.